\documentclass[reqno,12pt]{article}
\usepackage{amsfonts,amssymb,
amsmath,amscd, bm, mathrsfs, %bbold,
mathrsfs,delarray,subfigure}
\usepackage{hyperref,cite}
\usepackage{xypic}
\usepackage[dvips]{color}
\usepackage[dvips]{graphicx}
\DeclareMathOperator{\ad}{ad}
\DeclareMathOperator{\Ad}{Ad}

\DeclareMathOperator{\id}{id}

\DeclareMathOperator{\Hom}{Hom}
 
\DeclareMathOperator{\Fun}{Fun}
\DeclareMathOperator{\Map}{Map}

\DeclareMathOperator{\Gau}{Gau}
\DeclareMathOperator{\OGau}{OGau}

\DeclareMathOperator{\End}{End}
\DeclareMathOperator{\Aut}{Aut}
\DeclareMathOperator{\OAut}{OAut}

\DeclareMathOperator{\tr}{tr}

\DeclareMathOperator{\GL}{GL}

\DeclareMathOperator{\CS}{CS}
\DeclareMathOperator{\CE}{CE}

\DeclareMathOperator{\CCE}{CCE}
\DeclareMathOperator{\ZCCE}{ZCCE}

\numberwithin{equation}{subsection} 
\numberwithin{subsection}{section} 

\font\sansserif=cmss12
\font\scriptsansserif=cmss12 at 7 truept
\font\scriptscriptsansserif=cmss10 at 5 truept
\def\sans{\fam=14}
\newcommand{\mathsans}[1]{{{\sans #1}}}

\font\euler=eusm10 at 12.8 truept
\font\scripteuler=eusm7
\font\scriptscripteuler=eusm5 
\newcommand{\bfdot}{{\boldsymbol{\,\cdot\,}}}

\newcommand{\ul}[1]{{\underline{#1}}{}}

\begin{document}

\hrule\vskip.5cm
\hbox to 14.5 truecm{June 2014 \hfil DIFA 14}
%\hbox to 14.5 truecm{Version 2 \hfil}
\vskip.5cm\hrule
\vskip.7cm
%\begin{large}
\centerline{\textcolor{blue}{\bf 4--D SEMISTRICT HIGHER}}   
\centerline{\textcolor{blue}{\bf CHERN--SIMONS THEORY I}}   
%\end{large}
\vskip.2cm
\centerline{by}
\vskip.2cm
\centerline{\bf Emanuele Soncini and Roberto Zucchini}
\centerline{\it Dipartimento di Fisica ed Astronomia, Universit\`a di Bologna}
\centerline{\it V. Irnerio 46, I-40126 Bologna, Italy}
\centerline{\it I.N.F.N., sezione di Bologna, Italy}
\centerline{\it E--mail: emanuele.soncini@studio.unibo.it, zucchinir@bo.infn.it}
\vskip.7cm
\hrule
\vskip.6cm
\centerline{\bf Abstract} 
\par\noindent
We formulate a $4$--dimensional higher gauge theoretic Chern--Simons theory.
Its symmetry is encoded in a semistrict Lie $2$--algebra equipped 
with an invariant non singular bilinear form. We analyze the gauge invariance of the theory
and show that action is invariant under a higher gauge transformation up to a higher winding number.
We find that the theory admits two seemingly inequivalent 
canonical quantizations. The first is manifestly topological, it does not require a choice of any
additional structure on the spacial $3$--fold. The second, more akin to that of ordinary Chern--Simons theory,
involves fixing a CR structure on the latter. Correspondingly, we obtain two sets of semistrict higher 
WZW Ward identities and we find the explicit expressions of two higher versions of the WZW action. 
We speculate that the model could be used to define $2$--knot invariants of $4$--folds. 
%\vskip.4cm
\vfil\eject

\tableofcontents

\vfil\eject

%\include{highgau1}
%\include{highgau2}
%\include{highgau3}
%\include{highgau4}
%\include{highgau5}
%\include{highgau6}

%%%%%%%%%%%%%%%%%%%%%%%%%%%%%%%%%%%  highgau1.tex

\section{\normalsize \textcolor{blue}{Introduction}}\label{sec:intro}

\vfil
\hspace{.5cm} 
{\it Higher gauge theory} is a generalisation of ordinary gauge theory 
where gauge potentials are forms of degree $p\geq 1$ and, correspondingly, 
their gauge curvatures are forms of degree $p+1\geq 2$.  
It is thought to govern the dynamics of higher--dimensional extended objects. 
See ref. \cite{Baez:2010ya} for a readable, up--to--date 
review of this subject and extensive referencing. 

\vfil
The origin of higher gauge theory %, in its Abelian variant, 
can be traced back to the inception of supergravity. 
Higher gauge theory has subsequently found application in string theory in the study of 
$D$-- and $M$--branes \cite{Polchinski:1998rr,Becker:2007zj,Johnson:2003gi} 
as well as loop quantum gravity and, in particular, spin foam models
\cite{Baez:1999sr,Rovelli:2004tv}. 
Nowadays, the pursuit of higher gauge theory is motivated especially
by its potential to provide a Lagrangian description of the 
$N=(2, 0)$ superconformal $6$--dimensional field theory
governing the effective dynamics of $M5$--branes \cite{Fiorenza:2012tb,Lavau:2014iva}.

\vfil
From a mathematical perspective, higher gauge theory is intimately related to higher
algebraic structures, such as 2--categories, 2--groups \cite{Baez5,Baez:2003fs} 
and strong homotopy Lie or $L_\infty$ algebras \cite{Lada:1992wc,Lada:1994mn} 
and higher geometrical structures such as gerbes \cite{Brylinski:1993ab,Breen:2001ie}.
A state of the art exposition of these matters highlighting their
manifold relationships to various physical issues can by found in \cite{Schreiber2011,Gruetzmann:2014ica}. 

\vfil
Higher gauge theory can be formulated as a categorification of
ordinary gauge theory by codifying higher gauge symmetry into 
algebraic structures arising from the categorification of ordinary Lie
groups, weak or coherent Lie $2$--groups 
\cite{Baez:1998,Baez:2002jn,Baez:2004in,Baez:2005qu}.
This approach has been adopted in a large body of literature 
which would be impossible to summarise rendering full justice to all
contributions. We shall limit ourselves to note that until quite
recently most studies on the subject were limited to the case where 
the structure $2$--group is strict. Though every coherent $2$--group
is categorically equivalent to a strict $2$--group, categorical
equivalence is not a sufficiently fine notion for gauge theory:
it does not translate into any viable form of field theoretic equivalence. 
The study of higher gauge theory with non strict structure $2$--group
was first undertaken in the very broad context of $\infty$--Lie theory in refs. 
\cite{Sati:2008eg,Fiorenza2011,Fiorenza:2011jr}. An alternative approach to the topic 
was followed in refs. \cite{Ritter:2013wpa,Jurco:2014mva}.

%\vfil\eject

\subsection{\normalsize \textcolor{blue}{The scope and the plan of this paper}}\label{subsec:scope}

\hspace{.5cm} 
The present paper is devoted to the study of a model of 
non strict $4$--dimensional  higher Chern--Simons gauge 
theory which, in our hope,  may
have application in the study of $4$--dimensional topology just as 
the ordinary Chern--Simons theory does in $3$ dimensions. 
This paper employs  a version of non strict higher
gauge theory, called {\it semistrict}, first developed by one of the
authors in ref. \cite{Zucchini:2011aa}, which we shall outline next. 

Consider a gauge theory on a space time manifold $M$ whose symmetry 
is codified by a Lie algebra $\mathfrak{g}$. (We shall neglect global issues
here.) A connection is then a $\mathfrak{g}$--valued $1$--form
$\omega\in\Omega^1(M,\mathfrak{g})$. A gauge transformation is map
$\gamma\in\Map(M,G)$, where $G$ is a Lie group integrating $\mathfrak{g}$.
The gauge transformed connection ${}^g\omega$  is then given by
\begin{equation}
{}^g\omega=g(\omega-\sigma_g)
\label{intro0}
\end{equation}
where $g=\Ad\gamma$ and $\sigma_g=\gamma^{-1}d\gamma$. Note now
that $g\in\Map(M,\Aut(\mathfrak{g}))$ and $\sigma_g\in
\Omega^1(M,\mathfrak{g})$ and that 
\begin{subequations}
\label{intro1}
\begin{align}
&d\sigma_g+\frac{1}{2}[\sigma_g,\sigma_g]=0,
\vphantom{\Big]}
\label{intro1a}
\\
&g{}^{-1}dg(x)-[\sigma_g,x]=0, \qquad x\in\mathfrak{g},
\vphantom{\Big]}
\label{intro1b}
\end{align}
\end{subequations}
In the above relations, any reference to the group $G$ has disappeared:
everything is expressed in terms of $\mathfrak{g}$--valued forms 
and $\Aut(\mathfrak{g})$--valued maps. In this way, we have dodged the
technical task of integrating $\mathfrak{g}$ to $G$. In ordinary
gauge theory, this problem is not particularly difficult, but its
counterpart in semistrict higher gauge theory instead is. The basic proposal 
of ref.  \cite{Zucchini:2011aa} is extending this formulation to a higher 
gauge theory on $M$ whose symmetry is codified by a semistrict Lie
$2$--algebra $\mathfrak{v}$. Semistrict higher connections and gauge 
transformations are defined in terms of $\mathfrak{v}$--valued forms 
and $\Aut(\mathfrak{v})$--valued maps. An exposition of this framework
with new results not originally given in \cite{Zucchini:2011aa} is
provided in sect. \ref{sec:semigau}.

The gauge theoretic framework outlined in the previous paragraph has 
limitations: it can only work in perturbative Lagrangian field theory. 
Its adequacy for the analysis of parallel transport, a basic problem
in gauge theory, is not clear. Further, as it is well--known, 
relevant non perturbative effects are related to the center $Z(G)$ of $G$, 
information about which is lost in $\Aut(\mathfrak{g})$. It is nevertheless 
computationally efficient and directly generalisable to semistrict higher 
gauge theory.

Chern--Simons theory is a $3$--dimensional 
topological field theory of the Schwarz
type.  (See. ref.\cite{Marino:2005sj} for a recent review
of the model and exhaustive referencing). It was 
first formulated in 1989 by E. Witten in ref. \cite{Witten:1988hf}. 
Witten succeeded to show that many topological
knot and link invariants discovered by topologists earlier, 
such as the HOMFLY and Jones polynomials, could be
obtained as correlation function of Wilson loop operators in
Chern--Simons theory. 
He also proved that the Chern--Simons partition function is a 
topological invariant of the underlying $3$--manifold. 
Multiple connections with the $2$--dimensional  WZW model were also 
found \cite{Elitzur:1989nr}. In 1992, Witten also showed  that Chern--Simons theory is intimately
related to the topological sigma models of both $A$ and $B$ types
\cite{Witten:1992fb}.
This paper is a modest attempt to extend Chern--Simons theory 
to $4$ dimensions in the framework of semistrict higher gauge theory
with the hope of achieving a field theoretic expression of $2$--knot
and link invariants of $4$--manifolds and unveiling $3$-dimensional
higher analogs of WZW theory. 
In sect. \ref{sec:4dchern}, we describe a higher $4$--dimensional
Chern--Simons model whose symmetry is encoded in a balanced 
semistrict Lie $2$--algebra equipped 
with a invariant non singular bilinear form. We analyse in detail its gauge invariance 
and perform its canonical quantization. 

Finally in the appendices, we collect various results on $2$--groups
and Lie  $2$--algebras and their automorphisms which are scattered in
the literature in order to define our terminology and notation and for 
reference throughout the text. 

%\vfil\eject

\subsection{\normalsize \textcolor{blue}{Outlook and open problems}}\label{subsec:outlook}

%\vfil
\hspace{.5cm}Our study is divided roughly in two parts. 

%\vfil
The first part of the paper is devoted to the analysis of the gauge invariance 
of higher Chern--Simons theory. 
We find that, analogously to ordinary Chern--Simons theory, 
the higher Chern--Simons action is invariant under a higher gauge transformation up 
to a higher winding number only. Full gauge invariance of the quantum theory requires that 
the winding number be quantized in appropriate units. In all the examples 
which we have been able to work out in detail, the winding number actually vanishes, 
but we cannot prove its quantization in general and we are forced to assume 
it as a working hypothesis. This is a first aspect of the theory that requires further investigation.

%\vfil
The second part of the paper deals with quantization. Several approaches to the problem of quantization 
are possible in principle. Perturbative quantization based on a straightforward 
extension of Lorenz gauge fixing involves the choice of a background metric on the base manifold 
as well as the introduction of Faddeev--Popov ghost and ghost for ghost fields. In the presence of a metric
we cannot maintain gauge covariance without resorting to gauge rectifiers whose existence and interpretation 
is still problematic \cite{Zucchini:2011aa}. We are left with canonical quantization.
We find that the theory admits two apparently inequivalent canonical quantizations. 
We obtain correspondingly two sets of higher WZW Ward identities and we
find the explicit expressions of two higher versions of the gauged WZW action.

%\vfil
The canonical quantization of the first kind is manifestly topological in that it does not require a choice of any
additional structure on the spacial $3$--fold. That of the second kind involves fixing a CR structure on the latter. 
This is more akin to ordinary Chern--Simons theory's canonical quantization. 
CR spaces are in fact in many ways the closest $3$--dimensional 
analog of Riemann surfaces. The unitary equivalence of the quantization associated with distinct CR structures 
is an open problem necessitating a non trivial extension of the analysis of ref. \cite{Axelrod:1989xt}. 
Furthermore, the relationship between the the topological and CR quantizations 
remains elusive. 

It is necessary to clarify a point on the higher WZW actions emerging in the process of 
canonically quantizing our higher Chern--Simons theory. They encode the gauge covariance of the 
relevant wave functionals and, so, are determined by the Ward identities these obey and by a 
cocycle conditions extending the familiar Polyakov--Wiegmann relation. Presently,  
however, we have no evidence 
that they are related to some kind of $3$--dimensional sigma model as the ordinary 
gauged WZW action, although this remains a distinct possibility. 
In this respect it may be more useful to consider the restriction of
the higher Chern--Simons action to flat connection configurations
expressed as gauge transformation of the trivial connection on the
same lines as\cite{Elitzur:1989nr}. This is left to future work.

The solution of the questions raised in the preceding paragraphs
requires a more fundamental theory of higher gauge transformation than 
that employed in the present paper. Until recently, this was available
only for the strict case \cite{Baez:2004in,Baez:2005qu}.
Promising new results in this direction can be found in ref. \cite{Jurco:2014mva} . 

\vfil\eject 

\section{\normalsize \textcolor{blue}{Semistrict higher gauge theory}}\label{sec:semigau}

\hspace{.5cm} In this section, we shall illustrate the local aspects of semistrict higher gauge 
theory. Since we aim to the construction of higher Chern--Simons gauge theory
as a higher counterpart of ordinary one, we neglect bundle theoretic global issues altogether. 
Part of the material presented here has been already expounded in \cite{Zucchini:2011aa}, 
which the reader is referred to for further details and motivation, but also new results
are given. 

Before proceeding further, it is useful to recall the general philosophy underlying our approach, 
which was already alluded to in the introduction.
% to the reader's benefit. 
In an ordinary gauge theory with symmetry Lie algebra $\mathfrak{g}$, fields are $\mathfrak{g}$--valued 
forms and gauge transformations of fields are expressed in terms 
of $\Aut(\mathfrak{g})$--valued maps and  $\mathfrak{g}$--valued forms.
The theory, at least in its local aspects, can be formulated to a significant extent relying on the Lie algebra 
$\mathfrak{g}$ only. 
In the same way, in our formulation, in a semistrict higher gauge theory with symmetry Lie $2$--algebra
$\mathfrak{v}$, the fields are $\mathfrak{v}$--valued forms and 
gauge transformations of fields are expressed in terms 
of $\Aut(\mathfrak{v})$--valued maps and $\mathfrak{v}$--valued forms.
The theory, then, is formulated in terms of the Lie $2$--algebra $\mathfrak{v}$ only
analogously to the ordinary case. We present the semistrict theory
characterising it as much as possible as a higher version of the
ordinary one. 

Just as the gauge symmetry of ordinary gauge theory organizes in an
infinite dimensional group $\Gau(N,\mathfrak{g})$, the gauge transformation group, 
that %the gauge symmetry 
of semistrict higher gauge theory organizes as an infinite dimensional
strict $2$--group $\Gau(N,\mathfrak{v})$, the higher gauge
transformation $2$--group. The $1$-- and $2$--cells 
of $\Gau(N,\mathfrak{v})$ correspond respectively to gauge and
gauge for gauge transformations. The notion of gauge for gauge
transformation we adopt  is however more general than that customarily 
found in the literature encompassing also transformations of gauge
transformations which do not necessarily leave the action on higher 
gauge connections invariant unless further restrictions are imposed. 

The basic notions of Lie $2$--group and $2$--algebra theory are recalled in the appendices,
where our notation is also defined. All algebraic structures considered below are real and all
fields are smooth, unless otherwise stated.

%\vfil\eject 

\subsection{\normalsize \textcolor{blue}{Lie $2$--algebra gauge theory, 
local aspects}}\label{subsec:linfdoub}

\hspace{.5cm} In ordinary as well as higher gauge theory, fields propagate on a fixed $d$--fold $M$.
To study the local aspects of the theory, we assume that
$M$ is diffeomorphic to $\mathbb{R}^d$. On such an $M$, 
a {\it field of bidegree $(m,n)$} is any element of the space 
$\Omega^m(M,E[n])$ of $m$--forms on $M$ with values in $E[n]$, where $E$ is
some vector space. %, which we assume to have ghost number $0$. 
% related to $\hat{\mathfrak{v}}_0$, $\hat{\mathfrak{v}}_1$.

{\it Ordinary gauge theory}

In an ordinary gauge theory with structure Lie algebra $\mathfrak{g}$
(cf. app. \ref{sec:linfty}), fields are generally 
drawn from the spaces $\Omega^m(M,\mathfrak{g}[n])$. 
The main field of the gauge theory is the {\it connection} $\omega$, which is a bidegree $(1,0)$ field.
$\omega$ is characterized by its {\it curvature} $f$, %which is 
the bidegree $(2,0)$ field given by 
\begin{equation}
f=d\omega+\frac{1}{2}[\omega,\omega].
\vphantom{\Big]}
\label{xfcurv}
\end{equation}
$f$ satisfies the standard {\it Bianchi identity}
\begin{equation}
df+[\omega,f]=0.
\label{xfBianchi}
\end{equation}
The connection $\omega$ is flat if the curvature $f=0$. 

The {\it covariant derivative} of a field $\phi$ is given by the well--known expression 
\begin{equation}
D\phi=d\phi+[\omega,\phi]
\vphantom{\Big]}
\label{xfcovder}
\end{equation}
and %The covariant derivative so defined 
satisfies the standard Ricci identity 
\begin{equation}
DD\phi=[f,\phi].
\label{xfRicci}
\end{equation}
The Bianchi identity \eqref{xfBianchi} 
obeyed by $f$ can be written compactly as %\hphantom{xxxxxxxxxxxxxxxxxxx}
\begin{equation}
Df=0. 
\label{xcpfBianchi}
\end{equation}

{\it Semistrict higher gauge theory}

In semistrict higher gauge theory with structure Lie $2$--algebra $\mathfrak{v}$
(cf. app. \ref{sec:linfty}), fields organize %as a rule 
in {\it field doublets} 
$(\phi,\varPhi_\phi)\in\Omega^m(M,\mathfrak{v}_0[n])\times \Omega^{m+1}(M,\mathfrak{v}_1[n])$,
where $-1\leq m\leq d$. 
If $m=-1$, the first component of the doublet vanishes.
If $m=d$, the second component does. The doublets of this form are said to have bidegree
$(m,n)$. Above, we attached a suffix $\phi$ to $\varPhi_\phi$ to indicate that $\varPhi_\phi$
is the partner of $\phi$ in the doublet, not to mean that $\varPhi_\phi$ 
depends on $\phi$ in any way. This allows us to concisely denote the doublet $(\phi,\varPhi_\phi)$
simply as $\phi$ in many instances. 

In higher gauge theory of this type, there is a distinguished field doublet, the {\it connection doublet} 
$(\omega,\varOmega_\omega)$
of bidegree $(1,0)$. Associated with it is the {\it curvature doublet} $(f,F_f)$ of 
bidegree $(2,0)$ defined by the expressions
\begin{subequations}
\label{fFcurv}
\begin{align}
&f=d\omega+\frac{1}{2}[\omega,\omega]-\partial\varOmega_\omega,
\vphantom{\Big]}
\label{fcurv}
\\
&F_f=d\varOmega_\omega+[\omega,\varOmega_\omega]-\frac{1}{6}[\omega,\omega,\omega].
\vphantom{\Big]}
\label{Fcurv}
\end{align}
\end{subequations}
From \eqref{fFcurv}, it is readily 
verified that $(f,F)$ satisfies %the appropriate generalization of 
the {\it Bianchi identities}
\begin{subequations}
\label{fFBianchi}
\begin{align}
&df+[\omega,f]+\partial F_f=0,
\vphantom{\Big]}
\label{fBianchi}
\\
&dF_f+[\omega,F_f]-[f,\varOmega_\omega]+\frac{1}{2}[\omega,\omega,f]=0
\vphantom{\Big]}
\label{FBianchi}
\end{align}
\end{subequations}
%The name adopted here is justified by the analogy of \eqref{fFBianchi}
analogous to  the Bianchi identity \eqref{xfBianchi} of ordinary gauge theory. 
The connection $(\omega,\varOmega_\omega)$ is flat if the curvature components $f=0$ and $F_f=0$. 

Let $(\phi,\varPhi_\phi)$ be a field doublet of bidegree $(p,q)$. %\pagebreak  
The {\it covariant derivative doublet} of $(\phi,\varPhi_\phi)$ is
the field doublet $(D\phi,D\varPhi_\phi)$ of bidegree $(p+1,q)$ given by
\footnote{$\vphantom{\dot{\dot{\dot{\dot{x}}}}}$ The covariant derivative doublet 
of $(\phi,\varPhi_\phi)$  should be properly written as $(D\phi,D\varPhi_{D\phi})$.
We shall write it as $(D\phi,D\varPhi_\phi)$ for simplicity.} 
\begin{subequations}
\label{fFcovder}
\begin{align}
&D\phi=d\phi+[\omega,\phi]+(-1)^{p+q}\partial\varPhi_\phi, \hspace{4cm}
\vphantom{\Big]}
\label{fcovder}
%\\
\end{align}
\begin{align}
&D\varPhi_\phi=d\varPhi_\phi+[\omega,\varPhi_\phi]-(-1)^{p+q}[\phi,\varOmega_\omega]
+\frac{(-1)^{p+q}}{2}[\omega,\omega,\phi].
%\vphantom{\ul{\ul{\ul{\ul{\ul{x}}}}}}
\vphantom{\Big]}
\label{Fcovder}
\end{align}
\end{subequations}
The sign $(-1)^{p+q}$ is conventional,  since the relative sign of $\phi$, $\varPhi_\phi$
cannot be fixed in any natural manner. From \eqref{fFcovder}, we deduce easily the
appropriate version of the Ricci identities, 
\begin{subequations}
\label{fFRicci}
\begin{align}
&DD\phi=[f,\phi],
\vphantom{\Big]}
\label{fRicci}
\\
&DD\varPhi_\phi=[f,\varPhi_\phi]-[\phi,F]-[\phi,\omega,f].
\vphantom{\Big]}
\label{FRicci}
\end{align}
\end{subequations}
The explicit appearance of the connection component $\omega$ in the right hand side of \eqref{FRicci}
is a consequence of the presence of a term quadratic in $\omega$ in \eqref{Fcovder}.

The above definition of covariant differentiation is yielded 
by the request that the Bianchi identities \eqref{fFBianchi} 
be expressed as the vanishing of the covariant derivative
doublet $(Df,DF_f)$ of the curvature doublet $(f,F_f)$ 
\begin{subequations}
\label{cpfFBianchi}
\begin{align}
&Df=0,
\vphantom{\Big]}
\label{cpfBianchi}
\\
&DF_f=0
\vphantom{\Big]}
\label{cpFBianchi}
\end{align}
\end{subequations} 
as it is the case for the Bianchi identity of ordinary gauge theory, 
eq. \eqref{xcpfBianchi}.

%\vfil\eject 

\subsection{\normalsize \textcolor{blue}{The $2$--group of higher gauge transformations}}
\label{subsec:linfgautrsf}

\hspace{.5cm} Just as gauge transformations play a basic role in ordinary gauge theory, 
higher gauge transformations play a similar basic role in higher gauge theory.
In this section, following the approach of ref. \cite{Zucchini:2011aa}
already outlined in the introduction,
we shall review the main properties of higher gauge transformations
highlighting the way they generalize ordinary ones. To this end, we shall 
slightly extend the notion of the latter.

{\it Ordinary gauge transformations}

In ordinary gauge theory, symmetry is codified in a Lie algebra $\mathfrak{g}$. 
A {\it gauge transformation} is a pair of:

\begin{enumerate}

\item a map $g\in\Map(M,\Aut(\mathfrak{g}))$ (cf. app. \ref{sec:linftyauto}),

\item a flat connection $\sigma_g$, \hphantom{xxxxxxxxxxxxxxxxxxxxx}
\begin{equation}
d\sigma_g+\frac{1}{2}[\sigma_g,\sigma_g]=0,
\label{x1linfdgloba}
\end{equation}

\end{enumerate}
related to $g$ through the condition \hphantom{xxxxxxxxxxxxxxxxxx}
\begin{equation}
g^{-1}dg(\pi)-[\sigma_g,\pi]=0
\label{x1linfdglobb}
\end{equation}
(cf. app. \ref{sec:linfty}). We shall denote the gauge transformation by $(g,\sigma_g)$ or simply by $g$, 
having in mind that now $\sigma_g$ is not determined by $g$ but participates with $g$
in the transformation. Further, we shall denote by $\Gau(M,\mathfrak{g})$ the set of 
all such extended gauge transformations.

The definition of gauge transformation given above is more general than the one commonly quoted 
in the literature. If $G$ is a Lie group exponentiating $\mathfrak{g}$ and $\gamma\in\Map(M,G)$,
then the pair $(\Ad\gamma,\gamma^{-1}d\gamma)$ is a gauge transformation in the sense just defined.  
However, not every gauge transformation $(g,\sigma_g)$ is of this form. 

{\it Ordinary gauge transformation group} %\pagebreak 

$\Gau(M,\mathfrak{g})$ is an infinite dimensional Lie group, 
the (extended) {\it gauge transformation group} of the theory. 
The composition and inversion and the unit of $\Gau(M,\mathfrak{g})$
are defined by the relations \hphantom{xxxxxxxxxxxxxx}
\begin{subequations}
\label{x4linfdglob}
\begin{align}
&h\diamond g=hg, 
\vphantom{\Big]}
\label{x4linfdglobz}
\\
&\sigma_{h\,\diamond \,g}
=\sigma_g+ g^{-1}(\sigma_h),
\vphantom{\Big]}
\label{x4linfdgloba}
\\
&g^{-1_\diamond}=g^{-1},
\vphantom{\Big]}
\label{x6linfdglobz}
\\
&\sigma_{g^{-1_\diamond}}=-g(\sigma_g),
\vphantom{\Big]}
\label{x6linfdgloba}
\\
&i=\id_{\mathfrak{g}},
\vphantom{\Big]}
\label{x5linfdglobz}
\\
&\sigma_i=0,
\vphantom{\Big]}
\label{x5linfdgloba}
\end{align}
\end{subequations}
where $g,h\in\Gau(M,\mathfrak{g})$ and, in \eqref{x4linfdglobz}, \eqref{x6linfdglobz}, \eqref{x5linfdglobz},
the composition, inversion and unit in the right hand side are those of $\Aut(\mathfrak{g})$
thought of as holding pointwise on $M$(cf. eqs. \eqref{mor3tlinalga0}, \eqref{mor3/2tlinalgd0},
\eqref{mor3tlinalgg0}). 

{\it Higher gauge transformations}

In semistrict higher gauge theory, symmetry is codified in a Lie $2$--algebra $\mathfrak{v}$. 
A {\it higher $1$--gauge transformation} consists of the following %set of 
data.
%\vspace{.1cm}
\begin{enumerate}

\item a map $g\in\Map(M,\Aut_1(\mathfrak{v}))$ (cf. app. \ref{sec:linftyauto});

\item a flat connection doublet $(\sigma_g,\varSigma_g)$, 
\begin{subequations}
\label{1linfdglob}
\begin{align}
&d\sigma_g+\frac{1}{2}[\sigma_g,\sigma_g]-\partial\varSigma_g=0,
\vphantom{\Big]}
\label{1linfdgloba}
\\
&d\varSigma_g+[\sigma_g,\varSigma_g]-\frac{1}{6}[\sigma_g,\sigma_g,\sigma_g]=0;
\vphantom{\Big]}
\label{1linfdglobb}
\end{align}
\end{subequations}

\item an element $\tau_g$ of $\Omega^1(M,\mathfrak{aut}_1(\mathfrak{v}))$  
satisfying %the ``Maurer--Cartan equation'' 
\begin{align}
&d\tau_g(\pi)+[\sigma_g,\tau_g(\pi)]-[\pi,\varSigma_g]+\frac{1}{2}[\sigma_g,\sigma_g,\pi]
\vphantom{\Big]}
\label{2linfdglob}
\\
&\qquad\qquad\qquad\qquad\qquad\qquad
+\tau_g([\sigma_g,\pi]+\partial\tau_g(\pi))=0. 
\vphantom{\Big]}
\nonumber
\end{align}
\end{enumerate} %xxxxxxxxx
(cf. app. \ref{sec:linfty}) $g$, $\sigma_g$, $\varSigma_g$, $\tau_g$ are required to satisfy a number of relations. %\pagebreak 
If $g=(g_0,g_1,g_2)$ (cf. app. \ref{sec:linftyauto}), then one has $\vphantom{\ul{\ul{x}}}$ 
\begin{subequations}
\label{3linfdglob}
\begin{align}
&g_0{}^{-1}dg_0(\pi)-[\sigma_g,\pi]-\partial\tau_g(\pi)=0,
\vphantom{\Big]}
\label{3linfdgloba}
\\
&g_1{}^{-1}dg_1(\varPi)-[\sigma_g,\varPi]-\tau_g(\partial \varPi)=0,
\vphantom{\Big]} 
\label{3linfdglobb}
\\
&g_1{}^{-1}(dg_2(\pi,\pi)-2g_2(g_0{}^{-1}dg_0(\pi),\pi))
\vphantom{\Big]}
\label{3linfdglobc}
\\
&\qquad\qquad -[\sigma_g,\pi,\pi]-\tau_g([\pi,\pi])-2[\pi,\tau_g(\pi)]=0.
\vphantom{\Big]}
\nonumber
\end{align}
\end{subequations} %xxxxxxxxx
hold. %, for $x,y\in\mathfrak{v}_0$, $X\in \mathfrak{v}_1$.
In the following, we are going to denote a $1$--gauge transformation
such as the above as $(g,\sigma_g,\varSigma_g,\tau_g)$ or simply as $g$.
Again, in so doing, we are not implying that $\sigma_g$, $\varSigma_g$, $\tau_g$
are determined by $g$, but only that they are the partners of $g$ in the 
gauge transformation. We shall denote the set of all higher $1$--gauge 
transformations by $\Gau_1(M,\mathfrak{v})$. 

The above definition of higher gauge transformation is at first glance
a bit mysterious and needs to be justified. It is the minimal
extension of the ordinary notion to the higher setting. 
When the Lie algebra $\mathfrak{g}$ is replaced by the 
Lie $2$--algebra $\mathfrak{v}$, $g$ turns from an $\Aut(\mathfrak{g})$--valued 
map into $\Aut(\mathfrak{v})$--valued one and  
the flat connection $\sigma_g$ gets
promoted to a flat connection doublet $(\sigma_g,\varSigma_g)$, as is natural. 
This leads immediately to eqs. \eqref{1linfdglob}. 
The reason for introducing the further datum $\tau_g$ satisfying
\eqref{3linfdglob} is not as evident and must be explained. 

For an ordinary gauge transformation $(g,\sigma_g)$ the Maurer--Cartan equation
$d(g^{-1}dg)+g^{-1}dgg^{-1}dg=0$ is satisfied. For this to be
consistent with eq. \eqref{x1linfdglobb},  it is sufficient that $\sigma_g$ 
is flat. Showing this involves crucially the use of the Jacobi identity of the Lie algebra 
$\mathfrak{g}$. When we pass to a 
Lie $2$--algebra $\mathfrak{v}$, that identity is no longer available.
For this reason, we must introduce the new datum $\tau_g$ and modify the naive 
relations $g_0^{-1}dg_0(\pi)=[\sigma_g,\pi]$, $g_1^{-1}dg_1(\varPi)=[\sigma_g,\varPi]$,  
as indicated in \eqref{3linfdgloba}, \eqref{3linfdglobb}. In fact, 
if $\tau_g$ vanished, for the Maurer--Cartan equations
$d(g_0{}^{-1}dg_0)+g_0{}^{-1}dg_0g_0{}^{-1}dg_0=0$,
$d(g_1{}^{-1}dg_1)+g_1{}^{-1}dg_1g_1{}^{-1}dg_1=0$ 
to be verified, the flatness relations \eqref{1linfdglob} would
not suffice by themselves:
one would need to add an extra purely algebraic condition 
on the flat connection doublet $(\sigma_g,\varSigma_g)$, namely
$-[x,\varSigma_g]+\frac{1}{2}[\sigma_g,\sigma_g,x]=0$, which 
does not fit naturally into our higher gauge theoretic set-up. 
Once we allow for $\tau_g$, however, this condition becomes 
a differential consistency relation satisfied by $\tau_g$, viz
\eqref{2linfdglob}. This latter deserves therefore to be called ``$2$--Maurer--Cartan equation''.

In semistrict higher gauge theory, one has in addition gauge for gauge symmetry. 
For any two %Lie $2$--algebra 
$1$--gauge transformations $g,h\in\Gau_1(M,\mathfrak{v})$,
a {\it higher $2$--gauge transformation from $g$ to $h$} consists of the following data.
\begin{enumerate}

\item a map $F\in\Map(M,\Aut_2(\mathfrak{v}))(g,h)$, where $\Map(M,\Aut_2(\mathfrak{v}))(g,h)$ is the space of 
sections of the fiber bundle $\bigcup_{m\in M}\Aut_2(\mathfrak{v})(g(m),h(m))\rightarrow M$ 
(cf. app. \ref{sec:linftyauto});

\item an element $A_F\in\Omega^1(M,\mathfrak{v}_1)$.

\end{enumerate}
$F$, $A_F$ are required to satisfy the relations, 
\begin{subequations}
\label{0linfdglob}
\begin{align}
&\sigma_g-\sigma_h=\partial A_F, 
\vphantom{\Big]}
\label{0linfdgloba}
\\
&\varSigma_g-\varSigma_h=dA_F+[\sigma_h,A_F]+\frac{1}{2}[\partial A_F,A_F], \hspace{2.6cm}
\vphantom{\Big]}
\label{0linfdglobb}
\\
&\tau_g(\pi)-\tau_h(\pi)=-[\pi,A_F]+g_1{}^{-1}\big(dF(\pi)-F([\sigma_h,\pi]+\partial\tau_h(\pi))\big). %xxxxx
\vphantom{\Big]}
\label{0linfdglobc}
\end{align}
\end{subequations}
In the following, we are going to denote a $2$--gauge transformation
like the above as $(F,A_F)$, meaning that $A_F$ is the partner of $F$ in the 
transformation,  or simply as $F$. We shall also write $F:g\Rightarrow h$ to indicate its source and target. 
We shall denote the set of all $2$--gauge transformations 
$F:g\Rightarrow h$ by $\Gau_2(M,\mathfrak{v})(g,h)$ 
and that of all $2$--gauge transformations $F$ by $\Gau_2(M,\mathfrak{v})$.

The above definition of $2$--gauge transformation is again a bit
puzzling and needs to be justified. Suppose we ask what the most natural 
class of deformations of a $1$--gauge transformation
$(g,\sigma_g,\varSigma_g,\tau_g)$ which preserve 
its being such and can be  expressed in terms of elementary fields is. 
As $g,h\in\Map(M,\Aut_1(\mathfrak{v}))$, it is reasonable to demand that $g,h$ are the source and the target
of some $F\in\Map(M,\Aut_2(\mathfrak{v}))(g,h)$. Granting this, the only remaining
deformational field datum is an element $A_F\in\Omega^1(M,\mathfrak{v}_1)$ turning
$\sigma_g$ into $\sigma_h=\sigma_g-\partial A_F$. We take $A_F$
$\mathfrak{v}_1$-- rather than $\mathfrak{v}_0$--valued 
in order to be able to employ it to deform $\varSigma_g$ into 
$\varSigma_h=\varSigma_g-dA_F+\frac{1}{2}[\partial A_F,A_F]+\cdots$
and $\tau_g(x)$ into $\tau_h(x)=\tau_g(x)-[x,A_F]+\cdots$. 
Demanding that $(h,\sigma_h,\varSigma_h,\tau_h)$ is a $1$--gauge transformation 
fixes the form of the remaining terms not explicitly shown.  

{\it Higher gauge transformation $2$--group}

$\Gau(M,\mathfrak{v})$ is an infinite dimensional strict Lie $2$--group, the gauge transformation 
$2$--group of the theory. 
The composition and inversion laws and the unit $1$--gauge transformation
and the horizontal and vertical composition and inversion laws and the unit $2$--gauge
transformations of $\Gau(M,\mathfrak{v})$ are defined by %the relations 
\begin{subequations}
\label{4linfdglob}
\begin{align}
&h\diamond g=h\circ g, 
\vphantom{\Big]}
\label{4linfdglobz}
\\
&\sigma_{h\,\diamond \,g}
=\sigma_g+ g_0{}^{-1}(\sigma_h),
\vphantom{\Big]}
\label{4linfdgloba}
\\
&\varSigma_{h\,\diamond \,g}
=\varSigma_g+ g_1{}^{-1}\Big(\varSigma_h
+\frac{1}{2} g_2(g_0{}^{-1}(\sigma_h),g_0{}^{-1}(\sigma_h))\Big)-\tau_g(g_0{}^{-1}(\sigma_h)),
\vphantom{\Big]}
\label{4linfdglobb}
\\
&\tau_{h\,\diamond \,g}(\pi)
=\tau_g(\pi)+ g_1{}^{-1}\big(\tau_h(g_0(\pi))-g_2(g_0{}^{-1}(\sigma_h),\pi)\big),
\vphantom{\Big]}
\label{4linfdglobc}
\\
&g^{-1_\diamond}=g^{-1_\circ},
\vphantom{\Big]}
\label{6linfdglobz}
\\
&\sigma_{g^{-1_\diamond}}=-g_0(\sigma_g),
\vphantom{\Big]}
\label{6linfdgloba}
\\
&\varSigma_{g^{-1_\diamond}}=- g_1(\varSigma_g+\tau_g(\sigma_g))-\frac{1}{2} g_2(\sigma_g,\sigma_g),
\vphantom{\Big]}
\label{6linfdglobb}
\\
&\tau_{g^{-1_\diamond}}(\pi)=- g_1(\tau_g( g_0{}^{-1}(\pi)))- g_2(\sigma_g, g_0{}^{-1}(\pi)),
\vphantom{\Big]}
\label{6linfdglobc}
\\
&i=\id,    %_{\mathfrak{v}},
\vphantom{\Big]}
\label{5linfdglobz}
\\
&\sigma_i=0,
\vphantom{\Big]}
\label{5linfdgloba}
\\
&\varSigma_i=0,
\vphantom{\Big]}
\label{5linfdglobb}
\\
&\tau_i(\pi)=0, 
\vphantom{\Big]}
\label{5linfdglobc}
\\
&G\diamond F=G\circ F,   %\hspace{3.5cm}  %=k_1F(\pi)+Gh_0(\pi), 
\vphantom{\Big]}
\label{50linfdgloba}
\\
&A_{G\,\diamond\, F}=A_F+h^{-1}{}_1(A_G)-g_1{}^{-1}Fh_0{}^{-1}(\sigma_k),
\vphantom{\Big]}
\label{50linfdglobb}
\\
&F^{-1_\diamond}=F^{-1_\circ},%=-g_1{}^{-1}Fh_0{}^{-1},
\vphantom{\Big]}
\label{50linfdglobc}
\\
&A_{F^{-1_\diamond }}=-g_1(A_F)-F(\sigma_h),
\vphantom{\Big]}
\label{50linfdglobd}
\\
&K\bullet H=K\bfdot H,%=H(\pi)+K(\pi),
\vphantom{\Big]}
\label{50linfdglobe}
\\
&A_{K\,\bullet\, H}=A_H+A_K, \hspace{3.5cm}
\vphantom{\Big]}
\label{50linfdglobf}
\\
&H^{-1_\bullet}=H^{-1_\bfdot},%=-H(\pi)
\vphantom{\Big]}
\label{50linfdglobg}
\\
&A_{H^{-1_\bullet}}=-A_H,
\vphantom{\ul{\ul{\ul{\ul{\ul{\ul{g}}}}}}}
\vphantom{\Big]}
\label{50linfdglobh}
%\\
\end{align}
\begin{align}
&I_g=\mathrm{Id}_g, %=0,
\vphantom{\Big]}
\label{50linfdglobi}
\\
&A_{I_g}=0, \hspace{9.5cm}
\vphantom{\Big]}
\label{50linfdglobj}
\end{align}
\end{subequations}  
where $g,h,k,l\in\Gau_1(M,\mathfrak{v})$ and $F,G,H,K\in\Gau_2(M,\mathfrak{v})$, 
with $F:g\Rightarrow h$, $G:k\Rightarrow l$ and $H,K$ composable. 
In \eqref{4linfdglobz}, \eqref{6linfdglobz}, \eqref{5linfdglobz},
the composition, inversion and unit in the right hand side are those of $\Aut_1(\mathfrak{v})$
thought of as holding pointwise on $M$ (cf. eqs. \eqref{mor3tlinalga}--\eqref{mor3tlinalgc},
\eqref{mor3/2tlinalgd}--\eqref{mor3/2tlinalgf},
\eqref{mor3tlinalgg}--\eqref{mor3tlinalgi}). 
In \eqref{50linfdgloba}, \eqref{50linfdglobc}, \eqref{50linfdglobe}, \eqref{50linfdglobg},
\eqref{50linfdglobi},
the horizontal and vertical composition and inversion and the units in the right hand side 
are those of $\Aut_2(\mathfrak{v})$ thought of as holding pointwise on $M$ (cf. eqs. 
\eqref{mor4tlinalga}, \eqref{mor4/1tlinalgb},
\eqref{mor4tlinalgb}, \eqref{mor4/1tlinalgd}, \eqref{mor4tlinalgc}).

The strict $2$--group  $\Gau(M,\mathfrak{v})$ can be described also
as a crossed module, though we shall not use such description in the following. The two 
groups underlying it are $\Gau_1(M,\mathfrak{v})$ and $\Gau_2{}^*(M,\mathfrak{v})=
\bigcup_{g\in\Gau_1(M,\mathfrak{v})}\Gau_2{}(M,\mathfrak{v})(i,g)$. $\Gau_2{}^*(M,\mathfrak{v})$ 
can be characterized as the
set of pairs $(F,A_F)$ with:
\begin{enumerate}

\item $F\in\Map(M,\Aut_2{}^*(\mathfrak{v}))$ (cf. app. \ref{sec:linftyauto}); 

\item $A_F\in\Omega^1(M,\mathfrak{v}_1)$.

\end{enumerate}
The crossed module multiplications, inversions, units, target map and action are given by 
the expressions \hphantom{xxxxxxxxxxxxxx}
\begin{subequations}
\label{4croslinfdglob}
\begin{align}
&h\diamond g=h\circ g, 
\vphantom{\Big]}
\label{4croslinfdglobz}
\\
&\sigma_{h\,\diamond \,g}
=\sigma_g+ g_0{}^{-1}(\sigma_h),
\vphantom{\Big]}
\label{4croslinfdgloba}
\\
&\varSigma_{h\,\diamond \,g}
=\varSigma_g+ g_1{}^{-1}\Big(\varSigma_h
+\frac{1}{2} g_2(g_0{}^{-1}(\sigma_h),g_0{}^{-1}(\sigma_h))\Big)-\tau_g(g_0{}^{-1}(\sigma_h)),
\vphantom{\Big]}
\label{4croslinfdglobb}
\\
&\tau_{h\,\diamond \,g}(\pi)
=\tau_g(\pi)+ g_1{}^{-1}\big(\tau_h(g_0(\pi))-g_2(g_0{}^{-1}(\sigma_h),\pi)\big),
\vphantom{\Big]}
\label{4croslinfdglobc}
\\
&g^{-1_\diamond}=g^{-1_\circ},
\vphantom{\Big]}
\label{6croslinfdglobz}
\\
&\sigma_{g^{-1_\diamond}}=-g_0(\sigma_g),
\vphantom{\Big]}
\label{6croslinfdgloba}
%\\
\end{align}
\begin{align}
&\varSigma_{g^{-1_\diamond}}=- g_1(\varSigma_g+\tau_g(\sigma_g))-\frac{1}{2} g_2(\sigma_g,\sigma_g),
%\vphantom{\Big]}
\label{6croslinfdglobb}
\\
&\tau_{g^{-1_\diamond}}(\pi)=- g_1(\tau_g( g_0{}^{-1}(\pi)))- g_2(\sigma_g, g_0{}^{-1}(\pi)),
\vphantom{\Big]}
\label{6croslinfdglobc}
\\
&i=\id,      %_{\mathfrak{v}},
\vphantom{\Big]}
\label{5croslinfdglobz}
\\
&\sigma_i=0,
\vphantom{\Big]}
\label{5croslinfdgloba}
\\
&\varSigma_i=0,
\vphantom{\Big]}
\label{5croslinfdglobb}
\\
&\tau_i(\pi)=0, 
\vphantom{\Big]}
\label{5croslinfdglobc}
\\
&G\diamond F=G\circ F,   %\hspace{3.5cm}  %=k_1F(\pi)+Gh_0(\pi), 
\vphantom{\Big]}
\label{50croslinfdgloba}
\\
&A_{G\,\diamond\, F}=A_F+(1_{\mathfrak{v}_1}-F\partial)^{-1}(A_G),
\vphantom{\Big]}
\label{50croslinfdglobb}
\\
&F^{-1_\diamond}=F^{-1_\circ},
\vphantom{\Big]}
\label{50croslinfdglobc}
\\
&A_{F^{-1_\diamond }}=-(1_{\mathfrak{v}_1}-F\partial)(A_F),
\vphantom{\Big]}
\label{50croslinfdglobd}
\\
&I=\mathrm{Id}_i, %=0,
\vphantom{\Big]}
\label{50croslinfdglobi}
\\
&\mathsans{t}(F)=t(F)
\vphantom{\Big]}
\label{50croslinfdglobe}
\\
&\sigma_{\mathsans{t}(F)}=-\partial A_F,
%\vphantom{\ul{\ul{\ul{\ul{\ul{\ul{g}}}}}}}
\vphantom{\Big]}
\label{50croslinfdglobf}
\\
&\varSigma_{\mathsans{t}(F)}=-dA_F+\frac{1}{2}[\partial A_F,A_F],
\vphantom{\Big]}
\label{50croslinfdglobg}
\\
&\tau_{\mathsans{t}(F)}(\pi)=[\pi,A_F]-(1_{\mathfrak{v}_1}-F\partial)^{-1}dF(\pi)
\vphantom{\Big]}
\label{50croslinfdglobh}
\\
&A_{I_g}=0, \hspace{9.5cm}
\vphantom{\Big]}
\label{50croslinfdglobj}
\\
&\mathsans{m}(g)(F)=m(g)(F),
\vphantom{\Big]}
\label{50croslinfdglobzz}
\\
&A_{\mathsans{m}(g)(F)}=g_1(A_F-F(1_{\mathfrak{v}_0}-\partial F)^{-1}(\sigma_g)),
\vphantom{\Big]}
\label{50croslinfdglobww}
\end{align}
\end{subequations}  
where $g,h\in\Gau_1(M,\mathfrak{v})$ and $F,G\in\Gau_2{}^*(M,\mathfrak{v})$.
In \eqref{4croslinfdglobz}, \eqref{6croslinfdglobz}, \eqref{5croslinfdglobz},
the composition, inversion and unit in the right hand side are those of $\Aut_1(\mathfrak{v})$
thought of as holding pointwise on $M$ 
(cf. eqs. \eqref{cmmor3tlinalga}--\eqref{cmmor3tlinalgc}, \eqref{cmmor3/2tlinalgd}--\eqref{cmmor3/2tlinalgf},
\eqref{cmmor3tlinalgg}--\eqref{cmmor3tlinalgi}).
In \eqref{50croslinfdgloba}, \eqref{50croslinfdglobc},  \eqref{50croslinfdglobi},
the composition, inversion and unit in the right hand side are those of $\Aut_2{}^*(\mathfrak{v})$
thought of as holding pointwise on $M$ 
(cf. eqs. \eqref{cmmor4tlinalga}, \eqref{cmmor4tlinalgb}, \eqref{cmmor4tlinalgc}).
In \eqref{50croslinfdglobe}, the target map in  the right hand side is that of $\Aut_2{}^*(\mathfrak{v})$
thought of as holding pointwise on $M$ 
(cf. eqs. \eqref{cmmor4tlinalgd}--\eqref{cmmor4tlinalgf}).
Finally, in \eqref{50croslinfdglobzz}, the crossed module action in the right hand side
is that of $\Aut_1(\mathfrak{v})$ on $\Aut_2{}^*(\mathfrak{v})$ thought of as holding pointwise on $M$ 
(cf. eq. \eqref{cmmor4tlinalgg}). 

%\vfil\eject

\subsection{\normalsize \textcolor{blue}{The Lie $2$--algebra of
    infinitesimal higher gauge transformations}}
\label{subsec:inflinfgautrsf}

\hspace{.5cm} In higher gauge theory, as in ordinary gauge theory, many aspects 
of gauge symmetry are often conveniently studied by switching to the infinitesimal 
form of gauge transformation.

{\it Ordinary infinitesimal gauge transformations}

Consider again an ordinary gauge theory with symmetry Lie algebra $\mathfrak{g}$.
An {\it infinitesimal gauge transformation} is a gauge transformation in linearized form.
It consists of: % the following set of data:
%\vspace{.1cm}
\begin{enumerate}

\item a map $u\in\Map(M,\mathfrak{aut}(\mathfrak{g}))$ (cf. app. \ref{sec:linftyinfaut}),

\item a linearized flat connection $\dot\sigma_u$, \hphantom{xxxxxxxxxxxxx}
\begin{equation}
d\dot\sigma_u=0,
\vphantom{\Big]}
\label{1inflinfdgloba0}
\end{equation}

\end{enumerate} 
obeying the relation \hphantom{xxxxxxxxxxxxxxxxxx}
\begin{equation}
du(\pi)-[\dot\sigma_u,\pi]=0,
\vphantom{\Big]}
\label{3inflinfdgloba0}
\end{equation} %xxxxxxxxx
as follows from expanding \eqref{x1linfdgloba}, \eqref{x1linfdglobb} to first order around 
the unit transformation $i$. 
We shall denote the transformation as $(u,\dot\sigma_u)$, understanding as usual only that $\dot\sigma_u$ 
%is not determined by $u$, but only that it 
is the partner of $u$ in the 
gauge transformation, or simply as $u$. 
We shall denote the set of all infinitesimal gauge 
transformations by $\mathfrak{gau}(M,\mathfrak{g})$. 

{\it Ordinary infinitesimal gauge transformation Lie algebra}

$\mathfrak{gau}(M,\mathfrak{g})$ is \pagebreak an infinite dimensional Lie algebra, in fact that %the Lie algebra
of the gauge transformation Lie group $\Gau(M,\mathfrak{g})$. 
The brackets of $\mathfrak{gau}(M,\mathfrak{g})$ are defined by 
\begin{subequations}
\label{4inflinfdglob0}
\begin{align}
&[u,v]_\diamond =[u,v]_\circ,
\vphantom{\Big]}
\label{4inflinfdglobz0}
\\
&\dot\sigma_{[u,v]_\diamond }=u(\dot\sigma_v)-v(\dot\sigma_u),
\vphantom{\Big]}
\label{4inflinfdgloba0}
\end{align}
\end{subequations} %xxxxxxxxx
where $u,v\in \mathfrak{gau}(M,\mathfrak{g})$. 
In \eqref{4inflinfdglobz0}, the brackets in the right hand side are those of $\mathfrak{aut}(\mathfrak{g})$
thought of as holding pointwise on $M$ (cf. eq. \eqref{mor5tlinalgb0/1}).

{\it Adjoint type infinitesimal gauge transformations}

With any $s\in\Omega^0(M,\mathfrak{g})$, there is associated an element $\ad_M s
\in \mathfrak{gau}(M,\mathfrak{g})$ by 
\begin{subequations}
\label{4/4inflinfdglob0}
\begin{align}
&\ad_M s=\ad s,
\vphantom{\Big]}
\label{4/4inflinfdglobz0}
\\
&\dot\sigma_{\ad_M s}=ds,
\vphantom{\Big]}
\label{4/4inflinfdgloba0}
\end{align}
\end{subequations} %xxxxxxxxx
the adjoint of $s$.
In \eqref{4/4inflinfdglobz0}, the adjoint operator in the right hand side is that of $\mathfrak{g}$
holding pointwise on $M$ (cf. eq. \eqref{ad}). 

{\it Ordinary gauge transformation exponential map}

Infinitesimal gauge transformations can be exponentiated to finite ones.
The exponential map $\exp_\diamond :\mathfrak{gau}(M,\mathfrak{g})\rightarrow \Gau(M,\mathfrak{g})$
is given by 
\begin{subequations}
\label{gauexp0}
\begin{align}
&\exp_\diamond (u)=\exp_\circ (u),
\vphantom{\Big]}
\label{gauexpa0}
\\
&\sigma_{\exp_\diamond (u)}=\frac{1_{\mathfrak{g}}-\exp(-u)}{u}(\dot\sigma_u),
%\Ein'(u)(\dot\sigma_u),
\vphantom{\Big]}
\label{gauexpb0}
\end{align}
\end{subequations}
where $u\in\mathfrak{gau}(M,\mathfrak{g})$. %$\Ein'(t)$ is the entire function $(1-\exp(-t))/t$.
In \eqref{gauexpa0}, the exponentiation in the right hand side is that of
$\mathfrak{aut}(\mathfrak{g})$ thought of as holding pointwise on $M$.

{\it Higher infinitesimal gauge transformations}

Consider next a higher gauge theory with symmetry Lie $2$--algebra $\mathfrak{v}$.
A {\it infinitesimal higher $1$--gauge transformation} is a $1$--gauge transformation
in linearized form as in the ordinary case. Expanding \eqref{1linfdglob}, 
\eqref{2linfdglob} around the unit transformation $i$ to first order reveals that 
it consists of a set of data of the following form:
%\vspace{.1cm}
\begin{enumerate}

\item a map $u\in\Map(M,\mathfrak{aut}_0(\mathfrak{v}))$ (cf. app. \ref{sec:linftyinfaut});

\item a linearized flat connection doublet $(\dot\sigma_u,\dot\varSigma_u)$, 
\begin{subequations}
\label{1inflinfdglob}
\begin{align}
&d\dot\sigma_u-\partial\dot\varSigma_u=0,
\vphantom{\Big]}
\label{1inflinfdgloba}
\\
&d\dot\varSigma_u=0;
\vphantom{\Big]}
\label{1inflinfdglobb}
\end{align}
\end{subequations}

\item an element $\dot\tau_u$ of $\Omega^1(M,\mathfrak{aut}_1(\mathfrak{v}))$  
such that %the ``Maurer--Cartan equation'' 
\begin{align}
&d\dot\tau_u(\pi)-[\pi,\dot\varSigma_u]=0.
\vphantom{\Big]}
\label{2inflinfdglob}
\end{align}
\end{enumerate} %xxxxxxxxx
$u$, $\dot\sigma_u$, $\dot\varSigma_u$, $\dot\tau_u$ 
are required to satisfy the relations
stemming from \eqref{3linfdglob} by linearization. %\pagebreak 
If $u=(u_0,u_1,u_2)$ (cf. app. \ref{sec:linftyinfaut}), then these read
\begin{subequations}
\label{3inflinfdglob}
\begin{align}
&du_0(\pi)-[\dot\sigma_u,\pi]-\partial\dot\tau_u(\pi)=0,
\vphantom{\Big]}
\label{3inflinfdgloba}
\\
&du_1(\varPi)-[\dot\sigma_u,\varPi]-\dot\tau_u(\partial \varPi)=0,
\vphantom{\Big]} 
\label{3inflinfdglobb}
\\
&du_2(\pi,\pi) -[\dot\sigma_u,\pi,\pi]-\dot\tau_u([\pi,\pi])-2[\pi,\dot\tau_u(\pi)]=0.
\vphantom{\Big]}
\label{3inflinfdglobc}
\end{align}
\end{subequations}
%, for $x,y\in\mathfrak{v}_0$, $X\in \mathfrak{v}_1$
In the following, we shall denote the infinitesimal $1$--gauge transformation
as $(u,\dot\sigma_u,\dot\varSigma_u,\dot\tau_u)$, indicating as usual 
$\dot\sigma_u$, $\dot\varSigma_u$, $\dot\tau_u$ as the partners of $u$ in the 
gauge transformation data, or simply as $u$. We shall 
denote the set of all infinitesimal Lie $2$--algebra $1$--gauge 
transformations by $\mathfrak{gau}_0(M,\mathfrak{v})$. 

The gauge for gauge symmetry of semistrict higher gauge theory also has an infinitesimal version.
An {\it infinitesimal higher $2$--gauge transformation} is a 
linearized version of a $2$--gauge transformation. 
Expansion around the unit transformation $I_i$ to first order shows that 
it consists of the data
\begin{enumerate}

\item a map $P\in\Map(M,\mathfrak{aut}_1(\mathfrak{v}))$;

\item an element $\dot A_P\in\Omega^1(M,\mathfrak{v}_1)$.

\end{enumerate}
There are no further relations these objects must obey.
We shall denote the infinitesimal  $2$--gauge transformation
as $(P,\dot A_P)$, indicating $\dot A_P$ as the partner of $P$ in the 
gauge transformation, or simply as $P$. 
We shall denote the set of all infinitesimal higher $2$--gauge 
transformations by $\mathfrak{gau}_1(M,\mathfrak{v})$. 

{\it Higher infinitesimal gauge transformation Lie $2$--algebra}

$\mathfrak{gau}(M,\mathfrak{v})$ is an infinite dimensional strict Lie $2$--algebra,
in fact that of the gauge transformation Lie $2$--group $\Gau(M,\mathfrak{v})$. 
The boundary map and the brackets of $\mathfrak{gau}(M,\mathfrak{v})$ are given by the expressions
\begin{subequations}
\label{4inflinfdglob}
\begin{align}
&\partial_\diamond P=\partial_\circ P,
\vphantom{\Big]}
\label{6inflinfdglobb}
\\
&\dot\sigma_{\partial_\diamond P}=-\partial \dot A_P,
\vphantom{\Big]}
\label{6inflinfdglobc}
\\
&\dot\varSigma_{\partial_\diamond P}=-d \dot A_P,
\vphantom{\Big]}
\label{5inflinfdglobz}
\\
&\dot\tau_{\partial_\diamond P}(\pi)=[\pi,\dot A_P]-dP(\pi),
\vphantom{\Big]}
\label{5inflinfdgloba}
\\
&[u,v]_\diamond =[u,v]_\circ,
\vphantom{\Big]}
\label{4inflinfdglobz}
\\
&\dot\sigma_{[u,v]_\diamond }=u_0(\dot\sigma_v)-v_0(\dot\sigma_u),
\vphantom{\Big]}
\label{4inflinfdgloba}
\\
&\dot\varSigma_{[u,v]_\diamond }=u_1(\dot\varSigma_v)-v_1(\dot\varSigma_u)
+\dot\tau_u(\dot\sigma_v)-\dot\tau_v(\dot\sigma_u),
\vphantom{\Big]}
\label{4inflinfdglobb}
\\
&\dot\tau_{[u,v]_\diamond }(\pi)
=u_1\dot\tau_v(\pi)-v_1\dot\tau_u(\pi)+\dot\tau_uv_0(\pi)
\vphantom{\Big]}
\label{4inflinfdglobc}
\\
&\hspace{5cm}-\dot\tau_vu_0(\pi)
+u_2(\dot\sigma_v,\pi)-v_2(\dot\sigma_u,\pi),
\vphantom{\Big]}
\nonumber
\\
&[u,P]_\diamond =[u,P]_\circ ,
\vphantom{\Big]}
\label{6inflinfdglobz}
\\
&\dot A_{[u,P]_\diamond }=u_1(\dot A_P)-P(\dot\sigma_u), 
\vphantom{\Big]}
\label{6inflinfdgloba}
\\
&[u,v,w]_\diamond =[u,v,w]_\circ =0,
\vphantom{\Big]}
\label{5inflinfdglobb}
\end{align}
\end{subequations} %xxxxxxxxx
where $u,v,w\in \mathfrak{gau}_0(M,\mathfrak{v})$
and $P\in \mathfrak{gau}_1(M,\mathfrak{v})$. 
In \eqref{6inflinfdglobb}, \eqref{4inflinfdglobz}, \eqref{6inflinfdglobz}, \eqref{5inflinfdglobb},
the boundary and the brackets in the right hand side are those of $\mathfrak{aut}(\mathfrak{v})$
thought of as holding pointwise on $M$ (cf. eqs. \eqref{mor7tlinalgx}--\eqref{mor7tlinalgz},
\eqref{mor7tlinalga}--\eqref{mor7tlinalgc},
\eqref{mor7tlinalgv}, \eqref{mor7tlinalgw}). %\eqref{mor7tlinalg}).

The strict Lie $2$--algebra $\mathfrak{gau}(M,\mathfrak{v})$ can also be described
as a differential Lie crossed module. The two underlying Lie algebras are
$\mathfrak{gau}_0(M,\mathfrak{v})$ and $\mathfrak{gau}_1(M,\mathfrak{v})$. The differential Lie crossed 
module Lie brackets, target map and action are given by the expressions
\begin{subequations}
\label{4infcrlinfdglob}
\begin{align}
&[u,v]_\diamond =[u,v]_\circ ,
\vphantom{\Big]}
\label{4infcrlinfdglobz}
\\
&\dot\sigma_{[u,v]_\diamond }=u_0(\dot\sigma_v)-v_0(\dot\sigma_u),
\vphantom{\Big]}
\label{4infcrlinfdgloba}
\\
&\dot\varSigma_{[u,v]_\diamond }=u_1(\dot\varSigma_v)-v_1(\dot\varSigma_u)
+\dot\tau_u(\dot\sigma_v)-\dot\tau_v(\dot\sigma_u),
\vphantom{\Big]}
\label{4infcrlinfdglobb}
\\
&\dot\tau_{[u,v]_\diamond }(\pi)
=u_1\dot\tau_v(\pi)-v_1\dot\tau_u(\pi)+\dot\tau_uv_0(\pi)
\vphantom{\Big]}
\label{4infcrlinfdglobc}
\\
&\hspace{5cm}-\dot\tau_vu_0(\pi)
+u_2(\dot\sigma_v,\pi)-v_2(\dot\sigma_u,\pi),
\vphantom{\Big]}
\nonumber
\\
&[P,Q]_\diamond=[P,Q]_\circ
\vphantom{\Big]}
\label{4infcrlinfdglobu}
\\
&\dot A_{[P,Q]_\diamond}
=-P(\partial\dot A_Q)+Q(\partial\dot A_P)
\vphantom{\Big]}
\label{4infcrlinfdglobv}
\\
&\tau_\diamond P=\tau_\circ P,
\vphantom{\Big]}
\label{6infcrlinfdglobb}
\\
&\dot\sigma_{\tau_\diamond P}=-\partial \dot A_P,
\vphantom{\Big]}
\label{6infcrlinfdglobc}
\\
&\dot\varSigma_{\tau_\diamond P}=-d \dot A_P,
\vphantom{\Big]}
\label{5infcrlinfdglobz}
\\
&\dot\tau_{\tau_\diamond P}(\pi)=[\pi,\dot A_P]-dP(\pi),
\vphantom{\Big]}
\label{5infcrlinfdgloba}
\\
&\mu_\diamond(u)(P)=\mu_\circ(u)(P) ,
\vphantom{\Big]}
\label{6infcrlinfdglobz}
\\
&\dot A_{\mu_\diamond(u)(P)}=u_1(\dot A_P)-P(\dot\sigma_u), 
\vphantom{\Big]}
\label{6infcrlinfdgloba}
\end{align}
\end{subequations} %xxxxxxxxx \partial_\partial_
where $u,v\in \mathfrak{gau}_0(M,\mathfrak{v})$
and $P,Q\in \mathfrak{gau}_1(M,\mathfrak{v})$. 
In \eqref{4infcrlinfdglobz}, \eqref{4infcrlinfdglobu}, \eqref{6infcrlinfdglobb}, \eqref{6infcrlinfdglobz}, 
the brackets, the target map and the Lie algebra morphism in the right hand side are those of 
$\mathfrak{aut}(\mathfrak{v})$ thought of as holding pointwise on $M$ 
(cf. eqs. \eqref{dlcmmor7tlinalga}--\eqref{dlcmmor7tlinalgc}, \eqref{dlcmmor7tlinalgw},
\eqref{dlcmmor7tlinalgx}--\eqref{dlcmmor7tlinalgz}, \eqref{dlcmmor7tlinalgv}).

{\it Adjoint type higher infinitesimal gauge transformations}

For any $s\in\Omega^0(M,\mathfrak{v}_0)$, an element $\ad_M s\in \mathfrak{gau}_0(M,\mathfrak{v})$, %\pagebreak 
\begin{subequations}
\label{4/4inflinfdglob}
\begin{align}
&\ad_M s=\ad s,
\vphantom{\Big]}
\label{4/4inflinfdglobz}
%\\
\end{align}
\begin{align}
&\dot\sigma_{\ad_M s}=ds,
\vphantom{\Big]}
\label{4/4inflinfdgloba}
\\
&\dot\varSigma_{\ad_M s}=0,
\vphantom{\Big]}
\label{4/4inflinfdglobx}
\\
&\dot\tau_{\ad_M s}(\pi)=0
\vphantom{\Big]}
\label{4/4inflinfdgloby}
\end{align}
\end{subequations}
is defined, the adjoint of $s$.
In \eqref{4/4inflinfdglobz}, the adjoint operator in the right hand side is that of $\mathfrak{v}_0$
holding pointwise on $M$ (cf. eqs. \eqref{ada}--\eqref{adc}). 
Similarly, with any $s,t\in\Omega^0(M,\mathfrak{v}_0)$ and any $S\in\Omega^0(M,\mathfrak{v}_1)$, there are
associated elements $\ad_M s\wedge t, \ad_M S\in \mathfrak{gau}_1(M,\mathfrak{v})$ by 
\begin{subequations}
\label{4/5inflinfdglob}
\begin{align}
&\ad_M s\wedge t=\ad s\wedge t,
\vphantom{\Big]}
\label{4/5inflinfdglobz}
\\
&\dot A_{\ad_M s\wedge t}=0,
\vphantom{\Big]}
\label{4/5inflinfdgloba}
\\
&\ad_M S=\ad S,
\vphantom{\Big]}
\label{4/5inflinfdglobw}
\\
&\dot A_{\ad_M S}=0,
\vphantom{\Big]}
\label{4/5inflinfdglobb}
\end{align}
\end{subequations} %xxxxxxxxx
the adjoints of $s,t$ and $S$, respectively.
In \eqref{4/5inflinfdglobz}, \eqref{4/5inflinfdglobw}, the adjoint operators in the right hand side 
are those of $\mathfrak{v}_1$ holding pointwise on $M$ (cf. eqs. \eqref{ade}. \eqref{add}). 

{\it Higher gauge transformation exponential map}.

Infinitesimal Lie $2$--algebra gauge transformations can be exponentiated to finite ones.
The exponential map $\exp_\diamond :\mathfrak{gau}(M,\mathfrak{v})\rightarrow \Gau(M,\mathfrak{v})$
can be described explicitly. We have 
\begin{subequations}
\label{gauexp}
\begin{align}
&\exp_\diamond (u)=\exp_\circ (u),
\vphantom{\Big]}
\label{gauexpa}
\\
&\sigma_{\exp_\diamond (u)}=\frac{1_{\mathfrak{v}_0}-\exp(-u_0)}{u_0}(\dot\sigma_u),
%=\Ein'(u_0)(\dot\sigma_u),
\vphantom{\Big]}
\label{gauexpb}
\\
&\varSigma_{\exp_\diamond (u)}=\frac{1_{\mathfrak{v}_1}-\exp(-u_1)}{u_1}(\dot\varSigma_u)
%=\Ein'(u_1)(\dot\varSigma_u)
\vphantom{\Big]}
\label{gauexpc}
\\
&\hspace{1cm}
-\int_0^1 ds\,\exp(-su_1)\dot\tau_u\frac{1_{\mathfrak{v}_0}-\exp(-(1-s)u_0)}{u_0}(\dot\sigma_u)
\vphantom{\Big]}
\nonumber
\\&\hspace{1cm}+\int_0^1 ds\int_0^s dt\exp(-(s-t)u_1)
\vphantom{\Big]}
\nonumber
%\\
\end{align}
\begin{align}
&\hspace{2.3cm}
\times u_2\bigg(\exp(-tu_0)(\dot\sigma_u),\exp(-tu_0)\frac{1_{\mathfrak{v}_0}-\exp(-(1-s)u_0)}{u_0}(\dot\sigma_u)\bigg),
\vphantom{\Big]}
\nonumber
\\
&\tau_{\exp_\diamond (u)}(\pi)=\int_0^1 ds\,\exp(-su_1)\dot\tau_u\exp(su_0)(\pi)
\vphantom{\Big]}
\label{gauexpd}
\\
&\hspace{1.5cm}+\int_0^1 ds\,\exp(-su_1)u_2\bigg(\exp(su_0)(\pi),\frac{1_{\mathfrak{v}_0}-\exp(-(1-s)u_0)}{u_0}(\dot\sigma_u)\bigg),
\vphantom{\Big]}
\nonumber
\\
&\exp_\diamond (P)=\exp_\circ (P),
\vphantom{\Big]}
\label{gauexpe}
%\\
\end{align}
\begin{align}
&\dot A_{\exp_\diamond (P)}=\frac{\exp(P\partial)-1_{\mathfrak{v}_1}}{P\partial}(\dot A_P) \hspace{6cm}
%=\Ein'(-P\partial)(\dot A_P),
\vphantom{\Big]}
\label{gauexpf}
\end{align}
\end{subequations}
where $u\in\mathfrak{gau}_0(M,\mathfrak{v})$, $P\in\mathfrak{gau}_1(M,\mathfrak{v})$.
In \eqref{gauexpa}, the exponentiation in the right hand side is that of
$\mathfrak{aut}_0(\mathfrak{v})$ thought of as holding pointwise on $M$ (cf. eqs. 
\eqref{exp0a}--\eqref{exp0c}). Similarly, in \eqref{gauexpe}, 
the exponentiation in the right hand side is that of
$\mathfrak{aut}_1(\mathfrak{v})$ pointwise on $M$ (cf. eq. \eqref{exp0d}).

%\vfil\eject

\subsection{\normalsize \textcolor{blue}{Orthogonal gauge transformations}}\label{subsec:orthogauge}

\hspace{.5cm}
In the higher Chern--Simons theory that we are going to construct later, one of the basic datum
is an invariant form on the relevant algebra.

{\it Ordinary orthogonal gauge transformations}

We consider an ordinary gauge theory with symmetry Lie algebra $\mathfrak{g}$ equipped with 
an invariant bilinear symmetric form $(\cdot,\cdot)$ (cf. app. \ref{sec:linftyform}).
A gauge transformation $(g,\sigma_g)$ of $\Gau(M,\mathfrak{g})$ is said orthogonal if
$g$ is pointwise orthogonal,
\begin{enumerate}

\item $g\in\Map(M,\OAut(\mathfrak{g}))$ (cf. eq. \eqref{linftyform40}).

\end{enumerate}
We shall denote by $\OGau(M,\mathfrak{g})$ the set of all orthogonal elements $g\in \Gau(M,\mathfrak{g})$. 
$\OGau(M,\mathfrak{g})$ is an infinite dimensional 
Lie subgroup of the gauge Lie group $\Gau(M,\mathfrak{g})$. 

{\it Ordinary infinitesimal orthogonal gauge transformations}

An infinitesimal gauge transformation $(u,\dot\sigma_u)$ of 
$\mathfrak{gau}(M,\mathfrak{g})$ is accordingly orthogonal if $u$ is pointwise orthogonal, 
\begin{enumerate}

\item $u\in \Map(M,\mathfrak{oaut}(\mathfrak{g}))$.

\end{enumerate}
We let $\mathfrak{ogau}(M,\mathfrak{g})$ be the set of all orthogonal 
elements $u\in\mathfrak{gau}(M,\mathfrak{g})$.
$\mathfrak{ogau}(M,\mathfrak{g})$
is an infinite dimensional Lie subalgebra of the gauge Lie algebra $\mathfrak{gau}(M,\mathfrak{g})$. 
$\mathfrak{ogau}(M,\mathfrak{g})$ is also the Lie algebra of 
the orthogonal gauge Lie group $\OGau(M,\mathfrak{g})$. 

{\it Adjoint type ordinary orthogonal infinitesimal gauge transformations}

For $s\in\Omega^0(M,\mathfrak{g})$, the adjoint type infinitesimal gauge transformation
$\ad_M s\in\mathfrak{gau}(M,\mathfrak{g})$
is orthogonal, $\ad_M s\in\mathfrak{ogau}(M,\mathfrak{g})$ (cf. eqs. \eqref{4/4inflinfdglob0}).
%\eqref{4/4inflinfdglobz0}, \eqref{4/4inflinfdgloba0}). 

{\it Ordinary gauge transformation exponential and orthogonality}

The exponential map $\exp_\diamond:\mathfrak{ogau}(M,\mathfrak{g})\rightarrow\OGau(M,\mathfrak{g})$ 
of $\mathfrak{ogau}(M,\mathfrak{g})$ is simply the restriction of the exponential map $\exp_\diamond 
:\mathfrak{gau}(M,\mathfrak{g})\rightarrow \Gau(M,\mathfrak{g})$ of $\mathfrak{gau}(M,\mathfrak{g})$ 
to $\mathfrak{ogau}(M,\mathfrak{g})$. In particular, the orthogonal exponential 
is still computed by the expressions \eqref{4/4inflinfdglob0}. 

{\it Higher orthogonal gauge transformations}

We consider now a semistrict higher gauge theory having as symmetry algebra
a balanced Lie $2$--algebra $\mathfrak{v}$ equipped with an invariant bilinear form 
$(\cdot,\cdot)$ (cf. apps. \ref{sec:linftybal}, \ref{sec:linftyform}).
A $1$--gauge transformation $(g,\sigma_g,\varSigma_g,\tau_g)$ of $\Gau_1(M,\mathfrak{v})$ is said orthogonal if:
\begin{enumerate}

\item $g\in \Map(M,\OAut_1(\mathfrak{v}))$ (cf. eqs. \eqref{linftyform4}, \eqref{linftyform5});

\item For $x,y\in\mathfrak{v}_0$, one has 
\begin{equation}
(x,\tau_g(y))+(y,\tau_g(x))=0.
\label{orthogauge1}
\end{equation}

\end{enumerate}
We shall denote by $\OGau_1(M,\mathfrak{v})$ the set of all orthogonal elements $g\in\Gau_1(M,\mathfrak{v})$. 

An invariant form $(\cdot,\cdot)$ can be seen as a special invariant symmetric bilinear form on the direct sum
$\mathfrak{v}_0\oplus\mathfrak{v}_1$ with non vanishing off--diagonal elements only. From this perspective, 
the relationship to the ordinary case is more evident. 
Condition \eqref{orthogauge1} is at first glance a bit mysterious, but it emerges naturally in many contexts and is 
a necessary condition for orthogonal symmetry invariance in higher Chern--Simons theory. 

A $2$--gauge transformation $(F,A_F)$ of $\Gau_2(M,\mathfrak{v})(g,h)$, $g,h\in\Gau_1(M,\mathfrak{v})$ 
being two $1$--gauge transformations, is said orthogonal if both $g$, $h$ are orthogonal. For
$g,h\in \OGau_1(M,\mathfrak{v})$, we shall set $\OGau_2(M,\mathfrak{v})(g,h)=\Gau_2(M,\mathfrak{v})(g,h)$. 
We further set $\OGau_2(M,\mathfrak{v})=\bigcup_{g,h\in\OGau_1(M,\mathfrak{v})}\Gau_2(M,\mathfrak{v})(g,h)$. 

Remarkably, $\OGau(M,\mathfrak{v})=(\OGau_1(M,\mathfrak{v}),\OGau_2(M,\mathfrak{v}))$ 
is a Lie $2$--subgroup of the strict Lie $2$--group 
$\Gau(M,\mathfrak{v})=(\Gau_1(M,\mathfrak{v}),\Gau_2(M,\mathfrak{v}))$, 
meaning that $\OGau(M,\mathfrak{v})$ is closed under all $2$--group operations of 
$\Gau(M,\mathfrak{v})$ (cf. subsect. \ref{subsec:linfgautrsf}).

$\OGau(M,\mathfrak{v})$ can be described as a crossed module. The two groups underlying it
are $\OGau_1(M,\mathfrak{v})$ and 
$\OGau_2{}^*(M,\mathfrak{v})=\bigcup_{g\in\OGau_1(M,\mathfrak{v})}\Gau_2(M,\mathfrak{v})(i,g)$.
$\OGau_2{}^*(M,\mathfrak{v})$ can be characterized as the set of pairs $(F,A_F)$ with:
\begin{enumerate}

\item $F\in\Map(M,\OAut_2{}^*(\mathfrak{v}))$ (cf. app. \ref{sec:linftyform},
eqs. \eqref{linftyform6}, \eqref{linftyform7}) and 
\begin{equation}
(x,dF(y))+(y,dF(x))-d(\partial F(x),F(y))=0,
\label{orthogauge1/1}
\end{equation}
for $x,y\in\mathfrak{v}_0$. % (cf. subsect. \ref{subsec:linfgautrsf}). 

\item $A_F\in\Omega^1(M,\mathfrak{v}_1)$.

\end{enumerate}
Condition \eqref{orthogauge1/1} is required by compatibility with \eqref{orthogauge1}.
In this description, as expected, $\OGau(M,\mathfrak{v})$ is a Lie crossed submodule
of the Lie crossed module $\Gau(M,\mathfrak{v})$ (cf. subsect. \ref{subsec:linfgautrsf}).   

{\it Higher infinitesimal orthogonal gauge transformations}

An infinitesimal higher $1$--gauge transformation $(u,\dot\sigma_u,\dot\varSigma_u,\dot\tau_u)$ of 
$\mathfrak{gau}_0(M,\mathfrak{v})$ is othogonal if:
\begin{enumerate}

\item $u\in \Map(M,\mathfrak{oaut}_0(\mathfrak{v}))$;

\item For $x,y\in\mathfrak{v}_0$, one has 
\begin{equation}
(x,\dot\tau_u(y))+(y,\dot\tau_u(x))=0.
\label{orthogauge2}
\end{equation}

\end{enumerate}
\eqref{orthogauge2} arises from \eqref{orthogauge1} by linearization around $i$. 
We shall denote by $\mathfrak{ogau}_0$ $(M,\mathfrak{v})$ the set of all orthogonal 
elements $u\in\mathfrak{gau}_0(M,\mathfrak{v})$.

An infinitesimal $2$--gauge transformation $(P,\dot A_P)$ of 
$\mathfrak{gau}_1(M,\mathfrak{v})$ is said orthogonal if;
\begin{enumerate}

\item $P\in \Map(M,\mathfrak{oaut}_1(\mathfrak{v}))$ and 
\begin{equation}
(x,dP(y))+(y,dP(x))=0,
\label{orthogauge3}
\end{equation}
for $x,y\in\mathfrak{v}_0$. 

\end{enumerate}
\eqref{orthogauge3} stems from \eqref{orthogauge1/1} through linearization around $I_i$.  
We shall denote by $\mathfrak{ogau}_1(M,\mathfrak{v})$ the set of all orthogonal 
elements $P\in\mathfrak{gau}_1(M,\mathfrak{v})$.

$\mathfrak{ogau}(M,\mathfrak{v})=(\mathfrak{ogau}_0(M,\mathfrak{v}),\mathfrak{ogau}_1(M,\mathfrak{v}))$ 
is an infinite dimensional strict Lie $2$--subalgebra of the gauge algebra $\mathfrak{gau}(M,\mathfrak{v})
=(\mathfrak{gau}_0(M,\mathfrak{v}),\mathfrak{gau}_1(M,\mathfrak{v}))$, 
meaning that $\mathfrak{ogau}(M,\mathfrak{v})$ is closed under all $2$--algebra operations of 
$\mathfrak{gau}(M,\mathfrak{v})$. 
Furthermore, $\mathfrak{ogau}(M,\mathfrak{v})$ is the strict Lie $2$--algebra 
of the orthogonal gauge Lie $2$--group $\OGau(M,\mathfrak{v})$.  

{\it Adjoint type higher orthogonal infinitesimal gauge transformations}

For $s\in\Omega^0(M,\mathfrak{v}_0)$, the infinitesimal $1$--gauge transformation
$\ad_M s\in\mathfrak{gau}_0(M,$ $\mathfrak{v})$ is orthogonal, 
$\ad_M s\in\mathfrak{ogau}_0(M,\mathfrak{v})$ (cf. eqs. \eqref{4/4inflinfdglob}).
Likewise, for and $s,t\in\Omega^0(M,\mathfrak{v}_0)$
and any $S\in\Omega^0(M,\mathfrak{v}_1)$, the infinitesimal $2$--gauge transformations 
$\ad_M s\wedge t,\ad_M S\in\mathfrak{gau}_1(M,\mathfrak{v})$
are orthogonal, $\ad_M s\wedge t,\ad_M S\in\mathfrak{oaut}_1(M,\mathfrak{v})$
(cf. eqs. \eqref{4/5inflinfdglob}).

{\it Higher gauge transformation exponential and orthogonality}

The exponential map $\exp_\diamond:\mathfrak{ogau}(M,\mathfrak{v})\rightarrow\OGau(M,\mathfrak{v})$ 
of $\mathfrak{ogau}(M,\mathfrak{v})$ is simply the restriction of the exponential map $\exp_\diamond 
:\mathfrak{gau}(M,\mathfrak{v})\rightarrow \Gau(M,\mathfrak{v})$ of $\mathfrak{gau}(M,\mathfrak{v})$ 
to $\mathfrak{ogau}(M,\mathfrak{v})$. In particular, the orthogonal exponential 
is still computed by the expressions \eqref{gauexp}.

\vfil\eject

\section{\normalsize \textcolor{blue}{$4$--d higher gauge theoretic Chern--Simons theory}}\label{sec:4dchern}

\hspace{.5cm}  In this section, we shall construct and analyse a
$4$--dimensional semistrict analog of the standard Chern--Simons theory \cite{Witten:1988hf}.
.Beside providing a potentially interesting example of higher gauge
theory, our construction, if it turns out successful, may furnish a basic
field theoretic framework for the study of $4$--dimensional topology. 

Our model was already introduced in lesser generality in
ref. \cite{Zucchini:2011aa}, where it was analyzed mainly 
employing the Batalin--Vilkovisky quantization algorithm
\cite{BV1,BV2} in the geometric AKSZ formulation \cite{AKSZ}. 
Generalized Chern-Simons theory were studied in 
\cite{Sati:2008eg} and in \cite{Kotov:2007nr,Fiorenza:2011jr} in an AKSZ framework.
See also \cite{Schreiber2011}.

Below, we assume tacitly that manifold on which 
fields are defined is oriented and that the fields satisfy 
asymptotic or boundary conditions allowing for the convergence of the 
integration and integration by parts. 

%\vfil\eject 

\subsection{\normalsize \textcolor{blue}{The gauge transformation action}}
\label{subsec:gauact}

\hspace{.5cm} 
In ordinary gauge theory the construction of gauge invariant action functionals 
requires a prior definition of a gauge transformation action on gauge connections.
In the same way, in semistrict higher gauge theory the construction of higher gauge invariant 
action functionals is possible upon defining a higher gauge transformation action on
connection doublets. This is the topic of this subsection. We follow here the 
formulation of ref. \cite{Zucchini:2011aa}. 

In the familiar geometrical formulation of ordinary gauge theory, 
the basic geometrical datum is a principal $G$--bundle $P$ on a
manifold $N$. Connections are $\mathfrak{g}$--valued $1$--forms on $P$
satisfying the so called Ehrensmann conditions. 
Fields are horizontal and equivariant $\mathfrak{g}$--valued forms on $P$.
Gauge transformations are automorphisms of $P$
projecting to the identity $\id_N$ on $N$. The gauge transformation
action is then defined in terms of the pull-back action of automorphisms on
connections and fields. 
Because of the way we have have formulated the theory of gauge
transformation in subsect. \ref{subsec:linfgautrsf}, this type of approach is not immediately
extendable to higher gauge theory. We proceed therefore in an
alternative way closer in spirit to the physical approach to gauge symmetry.

{\it Gauge transformation action in ordinary gauge theory}

In ordinary gauge theory with symmetry Lie algebra $\mathfrak{g}$,
gauge transformation action is a left action of the gauge
transformation group $\Gau(N,\mathfrak{g})$ on 
connections $\omega$ and fields $\phi$
compatible with covariant differentiation (cf. eq. \eqref{xfcovder}), in the sense that
for any gauge transformation $g\in\Gau(N,\mathfrak{g})$
\begin{equation}
{}^gD{}^g\phi={}^g(D\phi).
\label{xgauact0}
\end{equation}
This requirement essentially determines the gauge transformation action.
The gauge transform ${}^g\omega$ of the connection
$\omega$ is
\begin{equation}
{}^g\omega=g(\omega-\sigma_g). 
\label{xgauact1}
\end{equation}
Further, the gauge transform ${}^g\phi$ of the field $\phi$ reads as 
\begin{equation}
{}^g\phi=g(\phi).
\label{xgauact3}
\end{equation}
In virtue \eqref{xgauact1}, \eqref{xgauact3}, one has as required that 
\begin{equation}
{}^gD^g\phi=g(D\phi).
\label{xgauact4}
\end{equation}
For a gauge transformation of the familiar form $(g,\sigma_g)=(\Ad \gamma$, $\gamma^{-1}d\gamma)$ 
with $\gamma\in\Map(M,G)$, \eqref{xgauact1}--\eqref{xgauact4} reduce to the usual expressions. 

The gauge transform ${}^gf$ of the curvature $f$ of $\omega$ (cf. eq. \eqref{xfcurv}) is
\begin{equation}
{}^gf=g(f).
\label{xgauact2}
\end{equation}
in compliance with \eqref{xgauact3}. 
%The Bianchi identity \eqref{xcpfBianchi} is gauge invariant as required.

Turning to the Lie algebra $\mathfrak{gau}(M,\mathfrak{g})$ of $\Gau(M,\mathfrak{g})$, we can write 
\eqref{xgauact1} in infinitesimal form (cf. subsect. \ref{subsec:inflinfgautrsf}). For an infinitesimal gauge transformation
$u\in\mathfrak{gau}(M,\mathfrak{g})$, the gauge variation $\delta_u\omega$  
of $\omega$ is 
\begin{equation}
\delta_u\omega=u(\omega)-\dot\sigma_u. 
\label{xinfgauact1}
\end{equation}
The gauge variation $\delta_uf$ of $f$ reads similarly as 
\begin{equation}
\delta_uf=u(f).
\label{xinfgauact2}
\end{equation}
For the infinitesimal gauge transformation $(u,\dot\sigma_u)=(\ad s, ds)$, \eqref{xinfgauact1}, \eqref{xinfgauact2}
take the well--known adjoint form. 

{\it BRST cohomology in ordinary gauge theory}

In standard gauge theory, gauge symmetry is most efficiently analyzed concentrating  
on infinitesimal gauge transformation of the adjoint type. 
This is codified by a bidegree $(0,1)$ ghost field 
$c$ %$c\in\Omega^0(M,\mathfrak{g}[1])$ 
through the ghost degree $1$ infinitesimal gauge transformation 
$w\in\mathfrak{gau}(M,\mathfrak{g})[1]$  %$(w,\dot\sigma_w)$, 
given by $w=-\ad_M c$ and $\dot\sigma_w=dc$ (cf. eqs. \eqref{4/4inflinfdglob0})
and is implemented by the odd BRST operator $s=\delta_w$. By \eqref{xinfgauact1}, then, 
\begin{equation}
s\omega=-Dc %=-dc-[\omega,c]
\label{cinfgauact1}
\end{equation}
(cf. eq. \eqref{xfcovder}).
We can make $s$ nilpotent by suitably defining the variation $sc$ of $c$. 
As by \eqref{cinfgauact1} by a simple computation 
\begin{equation}
s{}^2\omega=D\Big(sc+\frac{1}{2}[c,c]\Big),
\label{cinfgauact3}
\end{equation}
we can enforce $s{}^2\omega=0$ by setting
\begin{equation}
sc=-\frac{1}{2}[c,c].
\label{cinfgauact4}
\end{equation}
$s{}^2c=0$, as is readily verified, and so $s$ is nilpotent as required,
\begin{equation}
s{}^2=0.
\label{cinfgauact4/1}
\end{equation} 
For completeness, we report also the BRST variation of the curvature $f$ of $\omega$
which, by \eqref{xinfgauact2}, reads as \hphantom{xxxxxxxxxxxxxxxx}
\begin{equation}
sf=-[c,f].
\label{cinfgauact2}
\end{equation}

BRST cohomology 
plays an important role in gauge theory, ranging from the classification of observables 
to that of anomalies. 

{\it The ordinary orthogonal case} 

The results of above analysis continue to  hold with no modifications in the case where the Lie algebra
$\mathfrak{g}$ is equipped with an invariant bilinear form, % simply by restricting
 the gauge group $\Gau(M,\mathfrak{g})$ and the gauge Lie algebra $\mathfrak{gau}(M,\mathfrak{g})$ 
being replaced by their orthogonal counterparts $\OGau(M,\mathfrak{g})$ and $\mathfrak{ogau}(M,\mathfrak{g})$,
respectively (cf. subsect. \ref{subsec:orthogauge}). 
In particular, no additional restriction  on the ghost field $c$ is required
by orthogonality.

{\it Gauge transformation action in semistrict higher gauge theory}

In semistrict higher gauge theory with symmetry Lie $2$--algebra $\mathfrak{v}$,
we may define by analogy with the ordinary case the gauge transformation action as 
a left action of the $1$--gauge transformation group $\Gau_1(N,\mathfrak{v})$ on
connection doublets $(\omega,\varOmega_\omega)$
and field doublets $(\phi,\varPhi_\phi)$ compatible %\pagebreak 
with covariant differentiation (cf. eqs. \eqref{fFcovder}).
The straightforward generalization of \eqref{xgauact0} 
to the higher setting, %that is 
\begin{subequations}
\label{07linfdglob}
\begin{align}
&{}^gD{}^g\phi={}^g(D\phi),
\vphantom{\Big]}
\label{07linfdgloba}
\\
&{}^gD{}^g\varPhi_\phi={}^g(D\varPhi_\phi)
\vphantom{\Big]}
\label{07linfdglobb}
\end{align}
\end{subequations}
however cannot be made to hold unless a natural restriction on the curvature of the connection doublet 
is imposed. Through selection by way of selfconsistency, a coherent definition of 
the gauge transformation action can be worked out \cite{Zucchini:2011aa}.
The gauge transform $({}^g\omega,{}^g\varOmega_\omega)$ of $(\omega,\varOmega_\omega)$ is found to be 
\begin{subequations}
\label{7linfdglob}
\begin{align}
&{}^g\omega=g_0(\omega-\sigma_g), %\hspace{6cm}
\vphantom{\Big]}
\label{7linfdgloba}
\\
&{}^g\varOmega_\omega=g_1(\varOmega_\omega-\varSigma_g+\tau_g(\omega-\sigma_g))
-\frac{1}{2}g_2(\omega-\sigma_g,\omega-\sigma_g).
\vphantom{\Big]}
\label{7linfdglobb}
\end{align}
\end{subequations}
Further, the gauge transform $({}^g\phi,{}^g\varPhi_\phi)$ of $(\phi,\varPhi_\phi)$ reads as
\begin{subequations}
\label{9linfdglob}
\begin{align}
&{}^g\phi=g_0(\phi),
\vphantom{\Big]}
\label{9linfdgloba}
\\
&{}^g\varPhi_\phi=g_1(\varPhi_\phi-(-1)^{p+q}\tau_g(\phi))+(-1)^{p+q}g_2(\omega-\sigma_g,\phi),
\vphantom{\Big]}
\label{9linfdglobb}
\end{align}
\end{subequations}
$(p,q)$ being the bidegree of $(\phi,\varPhi_\phi)$.
We observe that the action \eqref{9linfdglob} explicitly depends on 
and cannot be defined without the prior assignment of a connection doublet.
Under the action \eqref{7linfdglob}, \eqref{9linfdglob}, one has now
\begin{subequations}
\label{10linfdglob}
\begin{align}
&{}^gD{}^g\phi=g_0(D\phi), 
\vphantom{\Big]}
\label{10linfdgloba}
\\
&{}^gD^g\varPhi_\phi=g_1(D\varPhi_\phi+(-1)^{p+q}\tau_g(D\phi))
%-(-1)^{p+q}g_2(\omega-\sigma_g,D\phi)+(-1)^{p+q}g_2(f,\phi).
\vphantom{\Big]}
\label{10linfdglobb}
\\
&\qquad\qquad\qquad-(-1)^{p+q}g_2(\omega-\sigma_g,D\phi)
+(-1)^{p+q}g_2(f,\phi), 
\vphantom{\Big]}
\nonumber
\end{align}
\end{subequations}
from which it emerges that \eqref{7linfdglob} holds provided the restriction $f=0$ 
on the curvature of the connection doublet, known as vanishing fake curvature condition
in the literature, is imposed.

The gauge transform of the curvature doublet $f=(f,F_f)$ of $\omega$ is 
\begin{subequations}
\label{8linfdglob}
\begin{align}
&{}^gf=g_0(f), 
\vphantom{\Big]}
\label{8linfdgloba}
\\
&{}^gF_f=g_1(F_f-\tau_g(f))+g_2(\omega-\sigma_g,f),
\vphantom{\Big]}
\label{8linfdglobb}
\end{align}
\end{subequations}  
in agreement with \eqref{9linfdglob}.

Turning to the Lie $2$--algebra $\mathfrak{gau}(M,\mathfrak{v})$ of $\Gau(M,\mathfrak{v})$, we can write 
\eqref{7linfdglob} in infinitesimal form (cf. subsect. \ref{subsec:inflinfgautrsf}). For an infinitesimal $1$--gauge transformation
$u\in\mathfrak{gau}_0(M,\mathfrak{v})$, the gauge variation $(\delta_u\omega,\delta_u\varOmega_\omega)$  
of $(\omega,\varOmega_\omega)$ reads
\begin{subequations}
\label{7inflinfdglob}
\begin{align}
&\delta_u\omega=u_0(\omega)-\dot\sigma_u, %\hspace{6cm}
\vphantom{\Big]}
\label{7inflinfdgloba}
\\
&\delta_u\varOmega_\omega=u_1(\varOmega_\omega)
-\dot\varSigma_u+\dot\tau_u(\omega)-\frac{1}{2}u_2(\omega,\omega).
\vphantom{\Big]}
\label{7inflinfdglobb}
\end{align}
\end{subequations}
The gauge variation $(\delta_uf,\delta_uF_f)$ of $(f,F_f)$ reads similarly as 
\begin{subequations}
\label{8inflinfdglob}
\begin{align}
&\delta_uf=u_0(f), \hspace{4.3cm}
\vphantom{\Big]}
\label{8inflinfdgloba}
\\
&\delta_uF_f=u_1(F_f)-\dot\tau_u(f)+u_2(\omega,f).
\vphantom{\Big]}
\label{8inflinfdglobb}
\end{align}
\end{subequations}  

A $2$--gauge transformation $G\in \Gau_2{}^*(M,\mathfrak{v})$ 
acts on a $1$--gauge transformation $g\in \Gau_1(M,\mathfrak{v})$ as 
\begin{subequations}
\label{Fgauact1,2,3,,4}
\begin{align}
&{}^Gg=t(G)g,
\vphantom{\Big]}
\label{Fgauact1}
\\
&\sigma_{{}^Gg}=\sigma_g-\partial g_1{}^{-1}(A_G),
\vphantom{\Big]}
\label{Fgauact2}
\\
&\varSigma_{{}^Gg}=\varSigma_g-d(g_1{}^{-1}(A_G))-[\sigma_g,g_1{}^{-1}(A_G)]
+\frac{1}{2}[\partial g_1{}^{-1}(A_G),g_1{}^{-1}(A_G)],
\vphantom{\Big]}
\label{Fgauact3}
\\
&\tau_{{}^Gg}(\pi)=\tau_g(\pi)+[\pi,g_1{}^{-1}(A_G)]-g_1{}^{-1}(1_{\mathfrak{v}_1}-G\partial)^{-1}dGg_0(\pi)
\vphantom{\Big]}
\label{Fgauact4}
\end{align}
\end{subequations}  
(cf. subsect. \ref{subsec:linfgautrsf}). The action of an infinitesimal $2$--gauge transformation 
$P\in \mathfrak{gau}_1(M,\mathfrak{v})$ on a $1$--gauge transformation $g\in \Gau_1(M,\mathfrak{v})$ 
correspondingly is
\begin{subequations}
\label{Uinfgauact1,2,3,4}
\begin{align}
&g^{-1}\delta_P g=\tau_\circ P,
\vphantom{\Big]}
\label{Uinfgauact1}
\\
&\delta_P\sigma_g=-\partial g_1{}^{-1}(\dot A_P),
\vphantom{\Big]}
\label{Uinfgauact2}
\\
&\delta_P\varSigma_g=-d(g_1{}^{-1}(\dot A_P))-[\sigma_g,g_1{}^{-1}(\dot A_P)],
\vphantom{\Big]}
\label{Uinfgauact3}
\\
&\delta_P\tau_g(\pi)=[\pi,g_1{}^{-1}(\dot A_P)]-g_1{}^{-1}dPg_0(\pi).
\vphantom{\Big]}
\label{Uinfgauact4}
\end{align}
\end{subequations}
This in turn induces an action of $P$ on an infinitesimal $1$--gauge transformation 
$u\in\mathfrak{gau}_0(M,\mathfrak{v})$ given by  
\begin{subequations}
\label{Uuinfgauact1,2,3,4}
\begin{align}%
&\delta_P u=\tau_\circ P,
\vphantom{\Big]}
\label{Uuinfgauact1}
\\
&\delta_P\dot\sigma_u=-\partial\dot A_P,
\vphantom{\Big]}
\label{Uuinfgauact2}
\\
&\delta_P\dot\varSigma_u=-d\dot A_P, %-[\sigma_g,g_1{}^{-1}(\dot A_P)],
\vphantom{\Big]}
\label{Uuinfgauact3}
\\
&\delta_P\dot\tau_u(\pi)=[\pi,\dot A_P]-dP(\pi).
\vphantom{\Big]}
\label{Uuinfgauact4}
\end{align}
\end{subequations}
$2$--gauge symmetry represents gauge for gauge symmetry, that is gauge symmetry of $1$--gauge transformation. 
Note that eqs. \eqref{Uuinfgauact1,2,3,4}
can be concisely written as $\delta_Pu=\tau_\diamond u$ by \eqref{6infcrlinfdglobb}--\eqref{5infcrlinfdgloba}.

{\it BRST cohomology in semistrict higher gauge theory} 

In semistrict higher gauge theory, analogously to ordinary gauge theory, 
higher gauge symmetry is most efficiently analyzed concentrating  
on higher infinitesimal gauge transformation of the adjoint type. 
Infinitesimal higher $1$--gauge transformation is codified 
by a bidegree $(0,1)$ ghost field doublet of 
$(c,C_c)$ %$\in\Omega^0(M,\mathfrak{v}_0[1])\times\Omega^1(M,\mathfrak{v}_1[1])$
through the ghost degree $1$ infinitesimal $1$--gauge transformation 
$w\in \mathfrak{gau}_0(M,\mathfrak{v})[1]$ given by 
%$(w,\dot\sigma_w,\dot\varSigma_w,\dot\tau_w)$, where 
$w=-\ad_M c$ and $\dot\sigma_w=dc-\partial C_c$,  
$\dot\varSigma_w=dC_c$ and $\dot\tau_w(\pi)=-[\pi,C_c]$
(cf. eqs. \eqref{4/4inflinfdglob} for a special case) and is implemented by the odd 
BRST operator $s_1=\delta_w$.
Infinitesimal $2$--gauge transformation turns out to be field dependent necessitating the
specification of a connection doublet $(\omega,\varOmega_\omega)$ by the requirement of 
BRST nilpotence. It is is codified by a bidegree 
$(-1, 2)$ ghost field doublet $(0, \varGamma)$ 
%with $\varGamma\in\Omega^0(M,\mathfrak{v}_1[2])$
through the ghost degree $2$ infinitesimal $2$--gauge transformation 
$W\in \mathfrak{gau}_1(M,\mathfrak{v})[2]$ given by 
%$(w,\dot\sigma_w,\dot\varSigma_w,\dot\tau_w)$, where 
$W=-\ad_M \varGamma$ and $\dot A_W=-[\omega,\varGamma]$
(cf. eqs. \eqref{4/5inflinfdglobz}, \eqref{4/5inflinfdgloba}
for a special case) and is implemented by the odd BRST operator $s_2=\delta_W$.
The total BRST operator is therefore given by \hphantom{xxxxxxxxxxxxxxx}
\begin{equation}
s=s_1+s_2.
\label{7inflinfdgloba/1}
\end{equation}

By \eqref{7inflinfdgloba}, \eqref{7inflinfdglobb}, then, 
\begin{subequations}
\label{7cinflinfdglob}
\begin{align}
&s_1\omega=-Dc, %=-dc-[\omega,c]+\partial C_c
\vphantom{\Big]}
\label{7cinflinfdgloba}
\\
&s_1\varOmega_\omega=-DC_c  %=-dC_c-[\omega,C_c]-[c,\varOmega]+\frac{1}{2}[\omega,\omega,c]
\vphantom{\Big]}
\label{7cinflinfdglobb}
\end{align}
\end{subequations}
(cf. eqs. \eqref{fcovder}, \eqref{Fcovder}).
As $2$--gauge transformations are inert on 
$\omega$, $\varOmega_\omega$, %we have \hphantom{xxxxxxxxxxxx}
\begin{subequations}
\label{7cinflinfdglob/1}
\begin{align}
&s_2\omega=0, 
\vphantom{\Big]}
\label{7cinflinfdgloba/1}
\\
&s_2\varOmega_\omega=0, 
\vphantom{\Big]}
\label{7cinflinfdglobb/1}
\end{align}
\end{subequations} 
trivially. In conclusion, we have 
\begin{subequations}
\label{7cinflinfdglob/2}
\begin{align}
&s\omega=-Dc, \hspace{1cm} %=-dc-[\omega,c]+\partial C_c
\vphantom{\Big]}
\label{7cinflinfdgloba/2}
\\
&s\varOmega_\omega=-DC_c.  %=-dC_c-[\omega,C_c]-[c,\varOmega]+\frac{1}{2}[\omega,\omega,c]
\vphantom{\Big]}
\label{7cinflinfdglobb/2}
\end{align}
\end{subequations}

We can try to make $s$ nilpotent by suitably defining the variations $sc$, $sC_c$ of $c$, $C_c$. 
From \eqref{7cinflinfdgloba}, \eqref{7cinflinfdglobb}, we find the relations
\begin{subequations}
\label{9cinflinfdglob}
\begin{align}
&s_1{}^2\omega=D\Big(s_1c+\frac{1}{2}[c,c]\Big), %-\partial\varGamma\Big),
\vphantom{\Big]}
\label{9cinflinfdgloba}
\\
&s_1{}^2\varOmega=D\Big(s_1C_c+[c,C_c]-\frac{1}{2}[\omega,c,c]\Big) %+D\varGamma\Big)-[f,\varGamma],
+\frac{1}{2}[f,c,c],
\vphantom{\Big]}
\label{9cinflinfdglobb}
\end{align}
\end{subequations}  
where above the covariant differentiation is applied to the field doublet
defined by the expressions within brackets  acording to eqs. \eqref{fFcovder}.
This suggests to set 
\begin{subequations}
\label{15cinflinfdglob}
\begin{align}
&s_1c=-\frac{1}{2}[c,c],
\vphantom{\Big]}
\label{15cinflinfdgloba}
\\
&s_1C_c=-[c,C_c]+\frac{1}{2}[\omega,c,c].
\vphantom{\Big]}
\label{15cinflinfdglobb}
\end{align}
\end{subequations}  
Of course, this is not enough to eventually make $s^2\varOmega$ vanish unless $f=0$, but 
it is the best we can do. % in the present situation.
From \eqref{Uuinfgauact1}--\eqref{Uuinfgauact4}, we find the relations 
\begin{subequations}
\label{13cinflinfdglob}
\begin{align}
&[s_2c-\partial\varGamma,\pi]=0,
\vphantom{\Big]}
\label{13cinflinfdgloba}
\\
&d(s_2c-\partial\varGamma)+\partial(s_2C+D\varGamma)=0,
\vphantom{\Big]}
\label{13cinflinfdglobb}
\\
&d(s_2C+D\varGamma)=0,
\vphantom{\Big]}
\label{13cinflinfdglobc}
\\
&[\pi,s_2C+D\varGamma]=0
\vphantom{\Big]}
\label{13cinflinfdglobd}
\end{align}
\end{subequations}  
which reveal that 
\begin{subequations}
\label{16cinflinfdglob}
\begin{align}
&s_2c=\partial\varGamma
\vphantom{\Big]}
\label{16cinflinfdgloba}
\\
&s_2C_c=-D\varGamma.
\vphantom{\Big]}
\label{16cinflinfdglobb}
\end{align}
\end{subequations}  
From \eqref{15cinflinfdglob}, \eqref{16cinflinfdglob}, we conclude that
\begin{subequations}
\label{10cinflinfdglob}
\begin{align}
&sc=-\frac{1}{2}[c,c]+\partial\varGamma
\vphantom{\Big]}
\label{10cinflinfdgloba}
\\
&sC_c=-[c,C_c]+\frac{1}{2}[\omega,c,c]-D\varGamma.
\vphantom{\Big]}
\label{10cinflinfdglobb}
\end{align}
\end{subequations}
We can now check that, with above definition of $sc$, $sC_c$, one has 
$s{}^2\omega=0$ and $s{}^2\varOmega=0$ for connection doublets 
$(\omega,\varOmega_\omega)$ satisfying the condition $f=0$, called vanishing fake curvature condition in
the literature. 
To make $s$ nilpotent, we have to suitably define also the variation $s\varGamma$ of $\varGamma$. 
To this end, we note that 
\begin{subequations}
\label{11cinflinfdglob}
\begin{align}
&s{}^2c=\partial\Big(s\varGamma+[c,\varGamma]-\frac{1}{6}[c,c,c]\Big),
\vphantom{\Big]}
\label{11cinflinfdgloba}
\\
&s{}^2C_c=D\Big(s\varGamma+[c,\varGamma]-\frac{1}{6}[c,c,c]\Big).
\vphantom{\Big]}
\label{11cinflinfdglobb}
\end{align}
\end{subequations}
Thus, we succeed to enforce $s{}^2c=0$ and $s{}^2C_c=0$ by requiring that 
\begin{equation}
s\varGamma=-[c,\varGamma]+\frac{1}{6}[c,c,c].
\label{12cinflinfdglob}
\end{equation}
$s{}^2\varGamma=0$ as wished.  

In conclusion $s$ is nilpotent as desired 
\begin{equation}
s{}^2=0,
\label{12cinflinfdglob/1}
\end{equation}
provided  we restrict to
connection doublets $(\omega,\varOmega_\omega)$ such that $f=0$.
We note here that the ghost sector here is not pure, as the BRST variation 
$sC_c$ explicitly depends on the connection component $\omega$. 

For completeness, we report the BRST variation of curvature doublet $(f,F_f)$ of
$(\omega,\varOmega_\omega)$, which by \eqref{8inflinfdglob}, \eqref{8inflinfdglobb} read 
\begin{subequations}
\label{8cinflinfdglob}
\begin{align}
&sf=-[c,f], 
\vphantom{\Big]}
\label{8cinflinfdgloba}
\\
&sF_f=-[c,F_f]+[f,C_c]-[c,\omega,f].
\vphantom{\Big]}
\label{8cinflinfdglobb}
\end{align}
\end{subequations} 

We expect BRST cohomology to play the same basic role in semistrict higher gauge theory, 
which it does in ordinary gauge theory.  

{\it The higher orthogonal case} 

The results of above analysis keep holding with no modifications in the case where the Lie $2$--algebra 
$\mathfrak{v}$ is balanced and equipped with an invariant bilinear form, % simply by restricting
the gauge $2$--group $\Gau(M,\mathfrak{v})$ and the gauge Lie $2$--algebra 
$\mathfrak{gau}(M,\mathfrak{v})$ being replaced by their orthogonal counterparts 
$\OGau(M,\mathfrak{v})$ and $\mathfrak{ogau}(M,\mathfrak{v})$,
respectively (cf. subsect. \ref{subsec:orthogauge}). 
In particular, no additional restriction  on the ghost fields $c$ $C_c$ and $\varGamma$ 
is required.

%\vfil\eject

\subsection{\normalsize \textcolor{blue}{Semistrict higher Chern--Simons theory}}
\label{subsec:2tchern}
 
\hspace{.5cm} 
In this section, we shall describe in detail Lie $2$--algebra 
Chern--Simons theory. To highlight the way in which the model 
generalizes ordinary Chern--Simons theory \cite{Witten:1988hf},  
we first review this latter using the gauge theoretic framework 
developed in the previous section.

{\it Ordinary Chern--Simons Theory} 

The basic algebraic datum of ordinary Chern--Simons theory
is a Lie algebra $\mathfrak{g}$ equipped with an
invariant symmetric form $(\cdot,\cdot)$ (cf. app. \ref{sec:linftyform}).
The topological background is a compact oriented $3$--fold $N$. 
The field content consists in a $\mathfrak{g}$--connection 
$\omega$ on $N$. The classical action functional reads
\begin{equation}
\CS_1(\omega)=\kappa_1\int_N\bigg[(\omega,f)-\frac{1}{6}(\omega,[\omega,\omega])\bigg],
\label{1tchern1}
\end{equation}
where the curvature $f$ is given by \eqref{xfcurv}. The classical field equations are
\begin{equation}
f=0,
\label{1tchern2}
\end{equation}
(cf. eq. \eqref{xfcurv}) and entail that the connection $\omega$ is flat.
We shall denote this classical field theory by $\CS_1(N,\mathfrak{g})$ or simply $\CS_1$. 

Let $X$ be any manifold. In gauge theory, the de Rham complex 
$\Omega^*(X)$ contains the special subcomplex $\Omega_{\mathfrak{g}}{}^*(X)$
formed by those forms that are polynomials in one or more connections $\omega_a$ and 
their differentials $d\omega_a$. In turn, $\Omega_{\mathfrak{g}}{}^*(X)$
includes the subcomplex $\Omega_{\mathfrak{g}\mathrm{inv}}{}^*(X)$ 
of the elements invariant under the action \eqref{xgauact1} of the orthogonal gauge transformation 
group $\OGau(X,\mathfrak{g})$. For any $\mathfrak{g}$--connection $\omega$ on $X$, 
a form $\mathcal{L}_1\in\Omega^3(X)$, 
\begin{equation}
\mathcal{L}_1=(\omega,f)-\frac{1}{6}(\omega,[\omega,\omega]),
\label{h1tchern1}
\end{equation}
formally identical to the Lagrangian density of the $\CS_1$ action is defined. 
While $\mathcal{L}_1\in\Omega_{\mathfrak{g}}{}^3(X)$, one has 
$\mathcal{L}_1\not\in\Omega_{\mathfrak{g}\mathrm{inv}}{}^3(X)$, since, as is well--known, 
\begin{equation}
{}^g\mathcal{L}_1=\mathcal{L}_1-\frac{1}{3}(\sigma_g,d\sigma_g)+d(\sigma_g,\omega)
\label{h1tchern2}
\end{equation}
for $g\in\OGau(X,\mathfrak{g})$. It is a standard result of gauge theory that 
\begin{equation}
d\mathcal{L}_1=\mathcal{C}_1,
\label{h1tchern3}
\end{equation}
where $\mathcal{C}_1\in\Omega^4(X)$ is the curvature bilinear 
\begin{equation}
\mathcal{C}_1=(f,f).
\label{h1tchern4}
\end{equation}
Clearly, $\mathcal{C}_1\in\Omega_{\mathfrak{g}}{}^4(X)$. 
Unlike $\mathcal{L}_1$, however, $\mathcal{C}_1$ is invariant under %any orthogonal $1$--gauge transformation 
$\OGau(X,\mathfrak{g})$, 
\begin{align}
{}^g\mathcal{C}_1=\mathcal{C}_1.
\label{h1tchern6}
\end{align}
Thus, $\mathcal{C}_1\in\Omega_{\mathfrak{g}\mathrm{inv}}{}^4(X)$ as well. 
By \eqref{h1tchern2} and \eqref{h1tchern3}, $\mathcal{C}_1$, while exact in the
complex $\Omega_{\mathfrak{g}}{}^*(X)$, is generally only closed in the 
$\OGau(X,\mathfrak{g})$--invariant complex $\Omega_{\mathfrak{g}\mathrm{inv}}{}^*(X)$.
It thus defines a class $[\mathcal{C}_1]_{\mathrm{inv}}\in H_{\mathfrak{g}\mathrm{inv}}{}^4(X)$. 
More can be said. The variation $\delta\mathcal{C}_1$ of $\mathcal{C}_1$ under 
arbitrary variations of $\delta\omega$ of $\omega$ is given by \hphantom{xxxxxxxxxxx}
\begin{equation}
\delta\mathcal{C}_1=2d(\delta\omega,f).
\label{h1tchern7}
\end{equation}
where the $3$--form in the right hand side is $\OGau(X,\mathfrak{g})$ invariant 
\begin{equation}
({}^g\delta\omega,{}^gf)=(\delta\omega,f).
\label{h1tchern8}
\end{equation}
It follows that, albeit $\mathcal{C}_1$ is not necessarily exact in $\Omega_{\mathfrak{g}\mathrm{inv}}{}^*(X)$, 
its variation $\delta\mathcal{C}_1$ always is. 
This property characterizes $\mathcal{L}_1$ as the Chern--Simons form of a
characteristic class $[\mathcal{C}_1]_{\mathrm{inv}}$, in fact the $2$nd Chern class. 

The $\CS_1$ action is not invariant under the $\OGau(N,\mathfrak{g})$ 
action \eqref{xgauact1}. In fact, from \eqref{h1tchern2}, one has \hphantom{xxxxxxxxx}
\begin{equation}
\CS_1({}^g\omega)=\CS_1(\omega)-\kappa_1 Q_1(g)
\label{1tchern3}
\end{equation}
for $g\in\OGau(N,\mathfrak{g})$, where the anomaly $Q_1(g)$ is given by 
\begin{equation}
Q_1(g)=\frac{1}{3}\int_N(\sigma_g,d\sigma_g).
%=\frac{1}{4}\int_N\bigg[([\sigma_g,\sigma_g],\varSigma_g)-\frac{1}{6}(\sigma_g,[\sigma_g,\sigma_g,\sigma_g])\bigg].
\label{1tchern4}
\end{equation}
$Q_1(g)$ is in fact simply related to the $\CS_1$ functional itself, 
\begin{equation}
Q_1(g)=\kappa_1{}^{-1}\CS_1(\sigma_g).
\label{1tchern5}
\end{equation}
The independence of $Q_1(g)$ from the connection $\omega$ %\eqref{1tchern3} 
implies so that the field equations \eqref{1tchern2} are gauge 
invariant. Indeed this follows directly and independently from eq. \eqref{xgauact2}.

From \eqref{1tchern4}, the anomaly density is the form $q_1\in\Omega^3(N)$
\begin{equation}
q_1=\frac{1}{3}(\sigma_g,d\sigma_g).
\label{d1tchern1}
\end{equation}
Note that, since $\sigma_g$ is a connection, $q_1\in\Omega_{\mathfrak{g}}{}^3(N)$. 
From \eqref{h1tchern2}, \eqref{h1tchern3} and \eqref{h1tchern6}, it is readily seen that 
$q_1$ is closed. The variation of $q_1$ under 
continuous deformations of the gauge transformation $g$ is instead exact
\begin{equation}
\delta q_1=d(\delta\sigma_g,\sigma_g).
\label{d1tchern2}
\end{equation}
$Q_1(g)$ is so a topological invariant of $g$. 
Another way of showing this is by using relation \eqref{1tchern5}:
since flat connections $\omega$ are 
the ones solving the classical field equations \eqref{1tchern2}, 
and $\sigma_g$ is a flat connection for any $g$ (cf. eq. \eqref{x1linfdgloba}), 
the variation of $Q_1(g)=\kappa_1{}^{-1}\CS_1(\sigma_g)$ 
under an infinitesimal variation of $g$ necessarily vanishes. 
$Q_1(g)$ reduces in fact up to a factor to the customary winding number of the gauge transformation $g$
when $g=\Ad\gamma$, $\sigma_g=\gamma^{-1}d\gamma$ for a map $\gamma\in\Map(N,G)$, $G$ being 
a Lie group integrating $\mathfrak{g}$. 

By \eqref{x1linfdgloba}, the anomaly density $q_1$ can be cast as 
\begin{equation}
q_1=-\frac{1}{6}(\sigma_g,[\sigma_g,\sigma_g]).
\label{e1tchern1}
\end{equation}
This relation indicates that with $q_1$ there is associated a special Chevalley--Eilenberg cochain 
$\chi_1\in\CE^3(\mathfrak{g})$, 
\begin{equation}
\chi_1=-\frac{1}{6}(\pi,[\pi,\pi]), 
\label{e1tchern2}
\end{equation}
which is in fact a cocycle (cf. app. \ref{sec:linfty}). By \eqref{x1linfdgloba}
and \eqref{x2tlinalgQ}, if $\chi_1$ is exact in $\CE(\mathfrak{g})$, then 
$q_1$ is exact in $\Omega_{\mathfrak{g}}{}^*(N)$. 
In order the anomaly $Q_1(g)$ to be non vanishing, so, 
it is necessary that $H_{CE}{}^3(\mathfrak{g})\not=0$. This is the case if $\mathfrak{g}$ is semisimple. 

Since $Q_1(g)$ vanishes for any gauge transformation $g$ continuously connected with the identity $i$, 
$\CS_1$ is annihilated by the BRST operator $s$ (cf. eq. \eqref{cinfgauact1}), 
\begin{equation}
s\!\CS_1(\omega)=0,
\label{1tchern/15}
\end{equation}
as can be directly verified from \eqref{1tchern1}. This property opens the way to
the gauge invariant perturbative quantization of the model.

Due to the $\OGau(N,\mathfrak{g})$ gauge non invariance of the $\CS_1$ action functional, 
the gauge invariant path integral quantization of the $\CS_1$ field theory is possible
only if the value of $\kappa_1$ is such that $\kappa_1 Q_1(g)\in 2\pi\mathbb{Z}$ for all 
$g\in\OGau(N,\mathfrak{g})$. For $\mathfrak{g}=\mathfrak{u}(n)$ and 
$(\cdot,\cdot)=-\tr_{\mathrm{fund}}(\,\cdot\,\cdot\,)$ 
this is achieved if 
\begin{equation}
\kappa_1=-\frac{k}{4\pi},
\label{1tchern7}
\end{equation}
where $k\in\mathbb{Z}$ is an integer called level. 

{\it Semistrict higher Chern--Simons theory}
 
After reviewing ordinary Chern--Simons theory, we introduce the 
semistrict higher Chern--Simons theory, which is the main topic of this paper. 
The basic algebraic datum of the model 
is a balanced Lie $2$--algebra $\mathfrak{v}$ equipped with an 
invariant form $(\cdot,\cdot)$ (cf. apps. \ref{sec:linftybal}, \ref{sec:linftyform}).
The topological background is a compact oriented $4$--fold $N$. 
The field content consists in a $\mathfrak{v}$--connection doublet 
$(\omega,\varOmega_\omega)$ on $N$. The classical action functional is
\begin{equation}
\CS_2(\omega,\varOmega_\omega)=\kappa_2\int_N\bigg[\frac{1}{2}(2f+\partial\varOmega_\omega,\varOmega_\omega)
-\frac{1}{24}(\omega, [\omega,\omega,\omega])\bigg],
\label{2tchern1}
\end{equation}
where $f$ is given by \eqref{fcurv}.
The classical field equations of $\CS_2(N,\mathfrak{v})$ are
\begin{subequations}
\label{2tchern2}
\begin{align}
&f=0,
\vphantom{\Big]}
\label{2tchern2a}
%\\
\end{align}
\begin{align}
&F_f=0
\vphantom{\Big]}
\label{2tchern2b}
\end{align}
\end{subequations}
(cf. eqs. \eqref{fcurv}, \eqref{Fcurv}).
They imply that the connection doublet $(\omega,\varOmega_\omega)$ is flat, analogously to standard CS theory. 
We shall denote this classical field theory  by $\CS_2(N,\mathfrak{v})$ or simply $\CS_2$. 

Let $X$ be any manifold. In semistrict gauge theory, in analogy to ordinary gauge theory, 
the de Rham complex $\Omega^*(X)$ contains the special subcomplex $\Omega_{\mathfrak{v}}{}^*(X)$
formed by those forms that are polynomials in the components of one or more connection doublets 
$(\omega_a,\varOmega_a)$ and their differentials $(d\omega_a,d\varOmega_a)$. In turn, 
$\Omega_{\mathfrak{v}}{}^*(X)$ includes the subcomplex $\Omega_{\mathfrak{v}\mathrm{inv}}{}^*(X)$ 
of the elements invariant under the action \eqref{7linfdglob} of the orthogonal $1$--gauge 
transformation group $\OGau_1(X,\mathfrak{v})$.
For any $\mathfrak{v}$--connection doublet $(\omega,\varOmega_\omega)$ 
on $X$, a form $\mathcal{L}_2\in\Omega^4(X)$ 
\begin{equation}
\mathcal{L}_2=\frac{1}{2}(2f+\partial\varOmega_\omega,\varOmega_\omega)
-\frac{1}{24}(\omega, [\omega,\omega,\omega]).
\label{h2tchern1}
\end{equation}
formally identical to the Lagrangian density of the $\CS_2$ action is defined. 
While $\mathcal{L}_2\in\Omega_{\mathfrak{v}}{}^4(X)$, one has 
$\mathcal{L}_2\not\in\Omega_{\mathfrak{v}\mathrm{inv}}{}^4(X)$, since
\begin{align}
&{}^g\mathcal{L}_2=\mathcal{L}_2-\frac{1}{4}(\sigma_g,d\varSigma_g)
-d\bigg[\frac{1}{2}(\sigma_g,\varSigma_g)
\vphantom{\Big]}
\label{h2tchern2}
\\
&\hspace{2cm}
+\frac{1}{6}(\omega-\sigma_g, g_1{}^{-1}g_2(\omega-\sigma_g,\omega-\sigma_g)
+6\varSigma_g-3\tau_g(\omega-\sigma_g))\bigg].
\vphantom{\Big]}
\nonumber
\end{align}
for $g\in\OGau_1(X,\mathfrak{v})$. Similarly to standard gauge theory, one has
\begin{equation}
d\mathcal{L}_2=\mathcal{C}_2,
\label{h2tchern3}
\end{equation}
where $\mathcal{C}_2\in\Omega^5(X)$ is the curvature bilinear 
\begin{equation}
\mathcal{C}_2=(f,F_f).
\label{h2tchern4}
\end{equation}
Clearly, $\mathcal{C}_2\in\Omega_{\mathfrak{v}}{}^5(X)$. 
Unlike $\mathcal{L}_2$, however, $\mathcal{C}_2$ is invariant under %any orthogonal $1$--gauge transformation 
$\OGau_1(X,\mathfrak{v})$, 
\begin{align}
{}^g\mathcal{C}_2=\mathcal{C}_2,
\label{h2tchern6}
\end{align}
implying that $\mathcal{C}_2\in\Omega_{\mathfrak{v}\mathrm{inv}}{}^5(X)$. 
By \eqref{h2tchern2} and \eqref{h2tchern3}, $\mathcal{C}_2$, 
while exact in the complex $\Omega_{\mathfrak{v}}{}^*(X)$, 
is generally only closed in the $\OGau_1(X,\mathfrak{v})$--invariant complex 
$\Omega_{\mathfrak{v}\mathrm{inv}}{}^*(X)$. 
It thus defines a class $[\mathcal{C}_2]_{\mathrm{inv}}\in H_{\mathfrak{v}\mathrm{inv}}{}^5(X)$. 
Further, the variation $\delta\mathcal{C}_2$ of $\mathcal{C}_2$ under 
arbitrary variations  variations  $\delta\omega$, $\delta\varOmega_\omega$ of $\omega$, $\varOmega_\omega$
is given by 
\begin{equation}
\delta\mathcal{C}_2=d\big[(\delta\omega,F_f)+(f,\delta\varOmega_\omega)\big].
\label{h2tchern7}
\end{equation}
where the $5$--form in the right hand side is $\OGau_1(X,\mathfrak{v})$ invariant 
\begin{equation}
({}^g\delta\omega,{}^gF_f)+({}^gf,{}^g\delta\varOmega_\omega)=(\delta\omega,F_f)+(f,\delta\varOmega_\omega).
\label{h2tchern8}
\end{equation}
It follows that, although $\mathcal{C}_2$ is not necessarily exact in 
$\Omega_{\mathfrak{v}\mathrm{inv}}{}^*(X)$, its variation $\delta\mathcal{C}_2$ always is. 
This property characterizes then $\mathcal{L}_2$ as the Chern--Simons form of a higher
characteristic class $[\mathcal{C}_2]_{\mathrm{inv}}$. 

The $\CS_2$ action is not invariant under the $\OGau_1(N,\mathfrak{v})$ action \eqref{7linfdglob}. 
In fact, from \eqref{h2tchern2},  analogously to ordinary Chern--Simons theory, one has
\begin{equation}
\CS_2({}^g\omega,{}^g\varOmega_\omega)=\CS_2(\omega,\varOmega_\omega)-\kappa_2 Q_2(g)
\label{2tchern3}
\end{equation}
for $g\in\OGau_1(N,\mathfrak{v})$, where the anomaly $Q_2(g)$ is given by 
\begin{equation}
Q_2(g)=\frac{1}{4}\int_N\big[2(d\sigma_g,\varSigma_g)-(\sigma_g,d\varSigma_g)\big].
%=\frac{1}{4}\int_N\bigg[([\sigma_g,\sigma_g],\varSigma_g)-\frac{1}{6}(\sigma_g,[\sigma_g,\sigma_g,\sigma_g])\bigg].
\label{2tchern4}
\end{equation}
$Q_2(g)$ is in fact simply related to the $\CS_2$ action itself, 
\begin{equation}
Q_2(g)=\kappa_2{}^{-1}\CS_2(\sigma_g,\varSigma_g).
\label{2tchern5}
\end{equation}
Again, the independence of $Q_2(g)$ from the connection doublet $(\omega,\varOmega_\omega)$
implies that the field equations \eqref{2tchern2} are gauge 
invariant, a property that follows also directly and independently from eqs. \eqref{8linfdglob}. 

From \eqref{2tchern4}, the anomaly density is the form $q_2\in\Omega^4(N)$
\begin{equation}
q_2=\frac{1}{4}\big[2(d\sigma_g,\varSigma_g)-(\sigma_g,d\varSigma_g)\big].
\label{d2tchern1}
\end{equation}
Note that, since $(\sigma_g,\varSigma_g)$ is a connection doublet, $q_2\in\Omega_{\mathfrak{v}}{}^4(N)$. 
From \eqref{h2tchern2}, \eqref{h2tchern3} and \eqref{h2tchern6}, it is readily seen that 
$q_2$ is closed. The variation of $q_2$ under 
continuous deformations of the gauge transformation $g$ is instead exact
\begin{equation}
\delta q_2=d(\delta\sigma_g,\varSigma_g).
\label{d2tchern2}
\end{equation}
In $\CS_2$ too, $Q_2(g)$ is so a topological invariant of $g$. Another way of showing this is by 
using relation \eqref{2tchern5}: since flat connections $(\omega,\varOmega_\omega)$ are 
the ones solving the classical field equations \eqref{2tchern2} 
and $(\sigma_g,\varSigma_g)$ is a flat connection doublet for any $g$ (cf. eqs. \eqref{1linfdglob}), 
the variation of $Q_2(g)=\kappa_2{}^{-1}\CS_2(\sigma_g,\varSigma_g)$ 
under an infinitesimal variation of $g$ necessarily vanishes. 
In analogy to ordinary Chern--Simons theory, $Q_2(g)$ 
represents a higher winding number of the higher gauge transformation $g$. 

By using \eqref{1linfdglobb}, the anomaly density $q_2$ can be cast as 
\begin{equation}
q_2=-\frac{1}{24}(\sigma_g,[\sigma_g,\sigma_g,\sigma_g])+\frac{1}{2}(\partial\varSigma_g,\varSigma_g).
\label{e2tchern1}
\end{equation}
With $q_2$ there is therefore associated a special higher Chevalley--Eilenberg cochain 
$\chi_2\in\CE^4(\mathfrak{v})$, \hphantom{xxxxxxxxxxxx}
\begin{equation}
\chi_2=-\frac{1}{24}(\pi,[\pi,\pi,\pi])+\frac{1}{2}(\partial\varPi,\varPi),
\label{e2tchern2}
\end{equation}
which is in fact a cocycle (cf. app. \ref{sec:linfty}). By \eqref{1linfdglob}
and \eqref{2tlinalgQ}, if $\chi_2$ is exact in $\CE(\mathfrak{v})$, then 
$q_2$ is exact in $\Omega_{\mathfrak{v}}{}^*(N)$. In this way, in order the anomaly $Q_2(g)$ to be non trivial, 
it is necessary that $H_{CE}{}^4(\mathfrak{v})\not=0$. %qqqqqqqqq

Since $Q_2(g)$ vanishes for any $1$--gauge transformation $g$ continuously connected with the identity $i$, 
$\CS_2$ is invariant under the BRST operator %$s$ acting on $(\omega,\varOmega_\omega)$ by 
\eqref{7cinflinfdglob/2}, %\eqref{7cinflinfdgloba/2}, \eqref{7cinflinfdglobb/2}, \hphantom{xxxxxxxxxxx}
\begin{equation}
s\!\CS_2(\omega,\varOmega_\omega)=0,
\label{2tchern/15}
\end{equation}
a property that can be directly verified from \eqref{2tchern1}.  
As shown in subsect. \ref{subsec:gauact}, defining the BRST variations of the ghost fields $c$, $C_c$, $\varGamma$ 
according to \eqref{10cinflinfdgloba}, \eqref{10cinflinfdglobb}
\eqref{12cinflinfdglob}, the BRST operator $s$ turns out to be nilpotent provided the vanishing 
fake curvature condition $f=0$ is satisfied, since $s{}^2\mathcal{F}=0$
for all fields and ghost fields $\mathcal{F}$ except for $\varOmega_\omega$, in which case one has
\begin{equation}
s{}^2\varOmega_\omega=-[f,\varGamma]+\frac{1}{2}[f,c,c].
\label{2tchern6}
\end{equation}
Being $f=0$ one of the field equations, $s$ is nilpotent on shell. 
Perturbative quantization of the model is still possible, but it requires the 
Batalin--Vilkovisky quantization algorithm \cite{Zucchini:2011aa}. 

As in ordinary Chern--Simons theory, 
the fact that the $\CS_2$ action is not $\OGau_1(N,\mathfrak{v})$ invariant makes 
the gauge invariant path integral quantization of the $\CS$ field theory impossible
unless certain conditions are met. 
The pair of the $4$--fold $N$ and the balanced Lie $2$--algebra $\mathfrak{v}$ with invariant form is said 
{\it admissible} if there exists a
positive value of $\kappa_2$ such that $\kappa_2 Q_2(g)\in 2\pi\mathbb{Z}$ for all $g\in\OGau_1(N,\mathfrak{v})$.
Letting $\kappa_{2N\mathfrak{v}}$ be the smallest value of $\kappa_2$ with such property,
the gauge invariant path integral quantization of the $\CS_2(N,\mathfrak{v})$ theory is possible, 
at least in principle, provided that 
\begin{equation}
\kappa_2=k\kappa_{2N\mathfrak{v}},
\label{2tchern7}
\end{equation}
where $k\in\mathbb{Z}$ is an integer, which we shall call level as in the ordinary theory. 

An important issue of the theory is the classification of
the admissible pairs $(N,\mathfrak{v})$. We cannot provide any
solution of it presently. This is also related to the fact that
the integrability of a semistrict Lie $2$--algebra $\mathfrak{v}$ to a semistrict 
Lie $2$--group $V$ is not guaranteed in general. 
In the canonical quantization of semistrict higher Chern--Simons theory carried out in the next 
subsections, we assume as a working hypothesis that $\mathfrak{v}$ is a balanced Lie $2$--algebra 
with invariant form such that $(N,\mathfrak{v})$ is admissible for a sufficiently ample class of 
closed $4$--folds $N$.

\vfil\eject

\subsection{\normalsize \textcolor{blue}{Canonical quantization}}  %of semistrict higher Chern--Simons theory}}
\label{subsec:2tcano}

\hspace{.5cm} 
In this section, we shall briefly review the canonical quantization of ordinary Chern--Simons theory 
and then pass to that of the semistrict higher Chern--Simons theory. 

To carry out the canonical quantization of a field theory, we restrict to the case where 
the base manifold $N$ is of the form $N=\mathbb{R}\times M$ with $M$ a compact oriented manifold. 
Let $t$ denote the standard coordinate of $\mathbb{R}$. Then, the derivation operator $d_t$ is a globally defined 
nowhere vanishing vector field on $\mathbb{R}\times M$. We denote by $\Omega_{\mathrm{h}}{}^p(\mathbb{R}\times M)$ the subspace
of $\Omega^p(\mathbb{R}\times M)$ consisting of those $p$--forms $\alpha$ such that $i_{d_t}\alpha=0$. 
Every $p$--form $\alpha\in\Omega^p(\mathbb{R}\times M)$ decomposes uniquely as  
$\alpha=dt\alpha_t+\alpha_s$, where $\alpha_t\in\Omega_{\mathrm{h}}{}^{p-1}(\mathbb{R}\times M)$, 
$\alpha_s\in\Omega_{\mathrm{h}}{}^p(\mathbb{R}\times M)$.
Analogously, the differential $d$ of $\mathbb{R}\times M$ decomposes as
$d=dtd_t+d_s$, $d_s$ being the differential along $M$ in $\mathbb{R}\times M$. 

{\it Ordinary Chern--Simons theory}

In the $\CS_1(\mathbb{R}\times M,\mathfrak{g})$ theory, the $\mathfrak{g}$--connection $\omega$ decomposes as
\begin{equation}
\omega=dt\omega_t+\omega_s,
\vphantom{\Big]}
\label{1tcano1}
\end{equation}
where $\omega_t\in\Omega_{\mathrm{h}}{}^0(\mathbb{R}\times M,\mathfrak{g})$, 
$\omega_s\in\Omega_{\mathrm{h}}{}^1(\mathbb{R}\times M,\mathfrak{g})$.
The curvature $f$ of $\omega$ splits as
\begin{equation}
f=dtf_t+f_s,
\vphantom{\Big]}
\label{1tcano2}
\end{equation}
where $f_t\in\Omega_{\mathrm{h}}{}^1(\mathbb{R}\times M,\mathfrak{g})$, 
$f_s\in\Omega_{\mathrm{h}}{}^2(\mathbb{R}\times M,\mathfrak{g})$, in similar fashion (cf. eqs. \eqref{xfcurv}).
$\omega_s$ is itself a $\mathfrak{g}$--connection and $f_s$ is the associated curvature. 
The $\CS_1$ action \eqref{1tchern1} reads then as
\begin{equation}
\CS_1(\omega)=\kappa_1\int_{\mathbb{R}\times M}dt \bigg[-(\omega_s,d_t\omega_s)+2(\omega_t,f_s)\bigg].
\label{1tcano3}
\end{equation}
The field equations read then as
\begin{subequations}
\label{1tcano4}
\begin{align}
&f_s=0,
\vphantom{\Big]}
\label{1tcano4a}
\\
&d_t\omega_s-D_s\omega_t=0,
%d_t\varOmega_{\omega s}+[\omega_t,\varOmega_{\omega s}]
%-\frac{1}{2}[\omega_t,\omega_s,\omega_s]-d_s\varOmega_{\omega t}-[\omega_s,\varOmega_{\omega t}]=0,
\vphantom{\Big]}
\label{1tcano4b}
\end{align}
\end{subequations}
where $D_s$ denotes the covariant differentiation operator associated with the connection $\omega_s$ 
defined according to  \eqref{xfcovder} and $\omega_t$ is treated as a bidegree $(0,0)$ field. 

The momenta $\xi_t$, $\xi_s$ canonically conjugate to $\omega_t$, $\omega_s$ can easily 
be read off from \eqref{1tcano3}.
In virtue of the linear isomorphisms $\mathfrak{g}{}^\vee\simeq \mathfrak{g}$
induced by the bilinear form $(\cdot,\cdot)$, we have 
$\xi_t\in\Omega_{\mathrm{h}}{}^2(\mathbb{R}\times M,\mathfrak{g})$,
$\xi_s\in\Omega_{\mathrm{h}}{}^1(\mathbb{R}\times M,\mathfrak{g})$,
\begin{subequations}
\label{1tcano5}
\begin{align}
&\xi_t=0,
\vphantom{\Big]}
\label{1tcano5a}
\\
&\xi_s=-\kappa_1\omega_s. 
\vphantom{\Big]}
\label{1tcano5b}
\end{align}
\end{subequations}
Ordinary Chern--Simons theory is therefore constrained. 
This requires the application of Dirac's quantization algorithm. 

To this end, we set below 
\begin{equation}
\langle g,g' \rangle=\int_M(g,g') 
\label{1tcano6}
\end{equation}
for $g\in \Omega^p(M,\mathfrak{g})$, $g'\in\Omega^{2-p}(M,\mathfrak{g})$, for notational convenience.
Further, for any $\Omega^p(M,\mathfrak{g})$--valued phase function $\psi$, we denote by $g_\psi$ 
a $\Omega^{2-p}(M,\mathfrak{g})$--valued phase constant.

In the Hamiltonian formulation of $\CS_1(\mathbb{R}\times M,\mathfrak{g})$, the canonical field coordinates are
$\omega_t\in\Omega^0(M,\mathfrak{g})$, $\omega_s\in\Omega^1(M,\mathfrak{g})$ 
and their canonically con\-jugate momenta are respectively
$\xi_t\in\Omega^2(M,\mathfrak{g})$, $\xi_s\in\Omega^1(M,\mathfrak{g})$.
The basic Poisson brackets are 
\begin{subequations}
\label{1tcano7}
\begin{align}
&\{\langle g_{\omega_t}, \omega_t\rangle,\langle \xi_t,g_{\xi_t}\rangle\}_P=\langle g_{\omega_t},g_{\xi_t}\rangle,
\vphantom{\Big]}
\label{1tcano7a}
\\
&\{\langle g_{\omega_s},\omega_s\rangle,\langle \xi_s,g_{\xi_s}\rangle\}_P=\langle g_{\omega_s},g_{\xi_s}\rangle, 
\vphantom{\Big]}
\label{1tcano7b}
\end{align}
\end{subequations}
The canonical Hamiltonian drawn from \eqref{1tcano3} is 
\begin{equation}
H=-2\kappa_1\langle\omega_t,f_s\rangle.
\label{1tcano7d}
\end{equation}
The primary constraints corresponding to the relations \eqref{1tcano5a}, \eqref{1tcano5b} are  
\begin{subequations}
\label{1tcano8}
\begin{align}
%a_t:=
&\xi_t\approx 0, 
\vphantom{\Big]}
\label{1tcano8a}
\\%a_s:=
&\kappa_1\omega_s+\xi_s\approx 0. 
\vphantom{\Big]}
\label{1tcano8b}
\end{align}
\end{subequations}
Implementation  of the Dirac's algorithm leads to the secondary constraints
\begin{equation}
f_s\approx 0,
\vphantom{\Big]}
\label{1tcano9}
\end{equation}
and no higher order constraints. Further, the phase functions $\xi_t$ and $f_s$ are 
identified as generators of gauge symmetries. Gauge fixing is thus required.
A complete fixing of the symmetry, however, leads to unwanted non locality in the resulting gauge fixed theory. 
To remain in the framework of local field theory, we fix only the gauge symmetry associated with $\xi_t$
leaving that corresponding to $f_s$ unfixed. The gauge fixing condition we choose to impose is 
\begin{equation}
\omega_t\approx 0,
\vphantom{\Big]}
\label{1tcano10}
\end{equation}
The constraints \eqref{1tcano8a}, \eqref{1tcano8b}, \eqref{1tcano10}
form a second class set and, so, they can be used to construct the Dirac brackets
on the associated constrained phase space. The only independent phase variable remaining after the
constraints are taken into  account is $\omega_s$, whose Dirac brackets are 
\begin{equation}
\{\langle g_{\omega_s},\omega_s \rangle,\langle\omega_s,g_{\omega_s}{}'\rangle\}_D
=-\frac{1}{2\kappa_1}\langle g_{\omega_s},g_{\omega_s}{}'\rangle.
\label{1tcano11}
\end{equation}
The constraint \eqref{1tcano9} remains pending. $f_s$ generates now the constrained 
phase space BRST transformations. Introducing a ghost field $c_s\in\Omega^0(M,\mathfrak{g}[1])$,
we have 
\begin{equation}
\{\langle f_s, c_s\rangle,\langle \omega_s,g_{\omega_s} \rangle\}_D
=\frac{1}{2\kappa_1}\langle s_s\omega_s,g_{\omega_s} \rangle, 
\vphantom{\Big]}
\label{1tcano12}
\end{equation}
where $s_s\omega_s$ is given by \hphantom{xxxxxxxxxx}
\begin{equation}
s_s\omega_s=-D_sc_s,
\vphantom{\Big]}
\label{1tcano13}
\end{equation}
in agreement with \eqref{cinfgauact1}. 

We quantize $\CS_1(\mathbb{R}\times M,\mathfrak{g})$ by replacing the classical field $\omega_s$
satisfying the Dirac brackets \eqref{1tcano11} with a corresponding quantum field $\widehat{\omega}_s$
satisfying the commutation relations
\begin{equation}
[\langle  g_{\omega_s},\widehat{\omega}_s\rangle,\langle\widehat{\omega}_s,g_{\omega_s}{}'\rangle]
=-\frac{i}{2\kappa_1}\langle g_{\omega_s},g_{\omega_s}{}' \rangle.
\label{1tcano14}
\end{equation}
The constraint \eqref{1tcano9}, which we left pending in the classical theory, becomes
a condition obeyed by the state vectors $\varPsi$ of the theory,
\begin{equation}
\langle \widehat{f}_s,g_{f_s}\rangle\varPsi=0.
\vphantom{\Big]}
\label{1tcano15}
\end{equation}

{\it Semistrict higher Chern--Simons theory} 

The canonical quantization of semistrict higher Chern--Simons theory proceeds on the same lines
as the ordinary case. The structural similarities and differences of the two models should be 
evident to the reader. 

In the $\CS_2(\mathbb{R}\times M,\mathfrak{v})$ theory, the 
$\mathfrak{v}$--connection doublet $(\omega,\varOmega_\omega)$ splits as
\begin{subequations}
\label{2tcano1}
\begin{align}
&\omega=dt\omega_t+\omega_s,
\vphantom{\Big]}
\label{2tcano1a}
\\
&\varOmega_\omega=dt\varOmega_{\omega t}+\varOmega_{\omega s},
\vphantom{\Big]}
\label{2tcano1b}
\end{align}
\end{subequations}
where $\omega_t\in\Omega_{\mathrm{h}}{}^0(\mathbb{R}\times M,\mathfrak{v}_0)$, 
$\omega_s\in\Omega_{\mathrm{h}}{}^1(\mathbb{R}\times M,\mathfrak{v}_0)$, 
$\varOmega_{\omega t}\in\Omega_{\mathrm{h}}{}^1(\mathbb{R}\times M,\mathfrak{v}_1)$, 
$\varOmega_{\omega s}\in\Omega_{\mathrm{h}}{}^2(\mathbb{R}\times M,\mathfrak{v}_1)$.
Similarly, the curvature doublet $(f,F_f)$ of $(\omega,\varOmega_\omega)$ splits as
\begin{subequations}
\label{2tcano2}
\begin{align}
&f=dtf_t+f_s,
\vphantom{\Big]}
\label{2tcano2a}
\\
&F_f=dtF_{ft}+F_{fs}
\vphantom{\Big]}
\label{2tcano2b}
\end{align}
\end{subequations}
(cf. eqs. \eqref{fcurv}, \eqref{Fcurv}), where $f_t\in\Omega_{\mathrm{h}}{}^1(\mathbb{R}\times M,\mathfrak{v}_0)$, 
$f_s\in\Omega_{\mathrm{h}}{}^2(\mathbb{R}\times M,\mathfrak{v}_0)$, 
$F_{ft}\in\Omega_{\mathrm{h}}{}^2(\mathbb{R}\times M,\mathfrak{v}_1)$, 
$F_{f s}\in\Omega_{\mathrm{h}}{}^3(\mathbb{R}\times M,\mathfrak{v}_1)$. Here, %Notice that 
$(\omega_s,\Omega_{\omega s})$ is itself a $\mathfrak{v}$--connection doublet and $(f_s,F_{fs})$ is 
the associated curvature doublet. 
%$\omega_t\in\Gamma(\mathbb{R}\times M, \mathfrak{v}_0)$, 
%$\omega_s\in\Gamma(\mathbb{R}\times M,\wedge^1\pi^*T^*M\otimes\mathfrak{v}_0)$,
%$\varOmega_{\omega t}\in\Gamma(\mathbb{R}\times M,\wedge^1\pi^*T^*M$ $\otimes\,\mathfrak{v}_1)$, 
%$\varOmega_{\omega s}\in\Gamma(\mathbb{R}\times M,\wedge^2\pi^*T^*M\otimes\mathfrak{v}_0)$.
The $\CS_2$ action \eqref{2tchern1} reads then as
\begin{align}
&\CS_2(\omega,\varOmega_\omega)
=\kappa_2\int_{\mathbb{R}\times M}dt \bigg[\frac{1}{2}(d_t\omega_s,\varOmega_{\omega s})
\vphantom{\Big]}
\label{2tcano3}
\\
&\hspace{4.8cm}-\frac{1}{2}(\omega_s,d_t\varOmega_{\omega s})
+(\omega_t,F_{fs})+(f_s,\varOmega_t)\bigg].
\vphantom{\Big]}
\nonumber
\end{align}
The field equations read then as
\begin{subequations}
\label{2tcano4}
\begin{align}
&f_s=0,
\vphantom{\Big]}
\label{2tcano4a}
\\
&F_{fs}=0,
\vphantom{\Big]}
\label{2tcano4b}
%\\
\end{align}
\begin{align}
&d_t\omega_s-D_s\omega_t=0, 
%d_t\omega_s+[\omega_t,\omega_s]-d_s\omega_t-\partial\varOmega_{\omega t}=0,
\vphantom{\Big]}
\label{2tcano4d}
\\
&d_t\varOmega_{\omega s}-D_s\varOmega_{\omega t}=0,
%d_t\varOmega_{\omega s}+[\omega_t,\varOmega_{\omega s}]
%-\frac{1}{2}[\omega_t,\omega_s,\omega_s]-d_s\varOmega_{\omega t}-[\omega_s,\varOmega_{\omega t}]=0,
\vphantom{\Big]}
\label{2tcano4c}
\end{align}
\end{subequations}
where $D_s$ denotes the covariant differentiation
operator associated with the connection doublet $(\omega_s,\Omega_{\omega s})$ 
defined according to \eqref{fcovder}, \eqref{Fcovder} and 
$(\omega_t,\Omega_{\omega t})$ is treated as a bidegree $(0,0)$ field doublet. 

The expressions of momenta $\varXi_{\xi t}$, $\varXi_{\xi s}$, $\xi_t$, $\xi_s$
canonically conjugate to $\omega_t$, $\omega_s$, $\varOmega_{\omega t}$, $\varOmega_{\omega s}$ can easily 
be read off from \eqref{2tcano3}.
In virtue of the linear isomorphisms $\mathfrak{v}_0{}^\vee\simeq \mathfrak{v}_1$, 
$\mathfrak{v}_1{}^\vee\simeq \mathfrak{v}_0$ induced by the non singular 
bilinear pairing  $(\cdot,\cdot)$ of $\mathfrak{v}_0$ and $\mathfrak{v}_1$, we have 
$\varXi_{\xi t}\in\Omega_{\mathrm{h}}{}^3(\mathbb{R}\times M,\mathfrak{v}_1)$, 
$\varXi_{\xi s}\in\Omega_{\mathrm{h}}{}^2(\mathbb{R}\times M,\mathfrak{v}_1)$, 
$\xi_t\in\Omega_{\mathrm{h}}{}^2(\mathbb{R}\times M,\mathfrak{v}_0)$,
$\xi_s\in\Omega_{\mathrm{h}}{}^1(\mathbb{R}\times M,\mathfrak{v}_0)$ and \hphantom{xxxxxxxxxxxx}
\begin{subequations}
\label{2tcano5}
\begin{align}
&\varXi_{\xi t}=0,
\vphantom{\Big]}
\label{2tcano5a}
\\
&\varXi_{\xi s}=\frac{\kappa_2}{2}\varOmega_{\omega s},
\vphantom{\Big]}
\label{2tcano5b}
\\
&\xi_t=0,
\vphantom{\Big]}
\label{2tcano5c}
\\
&\xi_s=-\frac{\kappa_2}{2}\omega_s. 
\vphantom{\Big]}
\label{2tcano5d}
\end{align}
\end{subequations}
Higher semistrict Chern--Simons theory, as ordinary one, is therefore constrained. 
This requires once more the application of Dirac's quantization algorithm. Its implementation 
turns out to be straightforward. %, in spite of increased algebraic complexity. 

For notational convenience, below we set  
\begin{equation}
\langle g,G \rangle=\int_M(g,G) 
\label{2tcano6}
\end{equation}
for $g\in \Omega^p(M,\mathfrak{v}_0)$,
$G\in\Omega^{3-p}(M,\mathfrak{v}_1)$. 
Further, for any $\Omega^p(M,\mathfrak{v}_0)$--valued phase function $\psi$, we denote by $G_\psi$ 
a $\Omega^{3-p}(M,\mathfrak{v}_1)$--valued phase constant and, for any 
$\Omega^p(M,\mathfrak{v}_1)$--valued phase function $\varPsi$,
we denote by $g_\varPsi$ a $\Omega^{3-p}(M,\mathfrak{v}_0)$--valued phase constant.

In the Hamiltonian formulation of $\CS_2(\mathbb{R}\times M,\mathfrak{v})$, the canonical field coordinates are
$\omega_t\in\Omega^0(M,\mathfrak{v}_0)$, $\omega_s\in\Omega^1(M,\mathfrak{v}_0)$, 
$\varOmega_{\omega t}\in\Omega^1(M,\mathfrak{v}_1)$, $\varOmega_{\omega s}\in\Omega^2(M,\mathfrak{v}_1)$
and their canonically con\-jugate momenta are respectively
$\varXi_{\xi t}\in\Omega^3(M,\mathfrak{v}_1)$, $\varXi_{\xi s}\in\Omega^2(M,\mathfrak{v}_1)$, 
$\xi_t\in\Omega^2(M,\mathfrak{v}_0)$, $\xi_s\in\Omega^1(M,\mathfrak{v}_0)$.
The basic Poisson brackets are 
\begin{subequations}
\label{2tcano7}
\begin{align}
&\{\langle \omega_t,G_{\omega_t} \rangle,\langle g_{\varXi_{\xi t}}, \varXi_{\xi t}\rangle\}_P=\langle g_{\varXi_{\xi t}},G_{\omega_t} \rangle,
\vphantom{\Big]}
\label{2tcano}
\\
&\{\langle \omega_s,G_{\omega_s} \rangle,\langle g_{\varXi_{\xi s}}, \varXi_{\xi s}\rangle\}_P=\langle g_{\varXi_{\xi s}},G_{\omega_s} \rangle, 
\vphantom{\Big]}
\label{2tcano7a}
\\
&\{\langle g_{\varOmega_{\omega t}},\varOmega_{\omega t}\rangle,\langle \xi_t,G_{\xi_t}\rangle\}_P=\langle g_{\varOmega_{\omega t}},G_{\xi_t} \rangle, 
\vphantom{\Big]}
\label{2tcano7b}
\\
&\{\langle g_{\varOmega_{\omega s}},\varOmega_{\omega s}\rangle,\langle \xi_s,G_{\xi_s}\rangle\}_P=\langle g_{\varOmega_{\omega s}},G_{\xi_s} \rangle.
\vphantom{\Big]}
\label{2tcano7c}
\end{align}
\end{subequations}
The canonical Hamiltonian implied by \eqref{2tcano3} is 
\begin{equation}
H=-\kappa_2[\langle\omega_t,F_{fs}\rangle+\langle f_s,\varOmega_{\omega t}\rangle].
\label{2tcano7d}
\end{equation}

The primary constraints stemming from relations \eqref{2tcano5a}--\eqref{2tcano5d} are  
\begin{subequations}
\label{2tcano8}
\begin{align}
&%A_t:=
\varXi_{\xi t}\approx 0, 
\vphantom{\Big]}
\label{2tcano8a}
\\
&%A_s:=
\frac{\kappa_2}{2}\varOmega_{\omega s}-\varXi_{\xi s}\approx 0, 
\vphantom{\Big]}
\label{2tcano8b}
\\
&%a_t:=
\xi_t\approx 0, 
\vphantom{\Big]}
\label{2tcano8c}
\\
&%a_s:=
\frac{\kappa_2}{2}\omega_s+\xi_s\approx 0. 
\vphantom{\Big]}
\label{2tcano8d}
\end{align}
\end{subequations}
Implementation  of the Dirac's algorithm leads to the secondary constraints
\begin{subequations}
\label{2tcano9}
\begin{align}
&f_s\approx 0,
\vphantom{\Big]}
\label{2tcano9a}
\\
&F_{f s}\approx 0
\vphantom{\Big]}
\label{2tcano9b}
\end{align}
\end{subequations}
and no higher order constraints. Further, the phase functions $\xi_t$, $\varXi_{\xi t}$, $f_s$ and $F_{f s}$ are 
identified as generators of gauge symmetries. Gauge fixing is thus required.
A complete fixing of the symmetry, however, leads to a problematic non local gauge fixed theory
as in the ordinary case. 
To remain in the framework of local field theory, we fix only the gauge symmetry associated with $\xi_t$, $\varXi_{\xi t}$
leaving that corresponding to $f_s$ and $F_{f s}$ unfixed. The gauge fixing conditions we %choose to 
impose are 
\begin{subequations}
\label{2tcano10}
\begin{align}
&\omega_t\approx 0,
\vphantom{\Big]}
\label{2tcano10a}
%\\
\end{align}
\begin{align}
&\varOmega_{\omega t}\approx 0.
\vphantom{\Big]}
\label{2tcano10b}
\end{align}
\end{subequations}
The constraints \eqref{2tcano8a}--\eqref{2tcano8d}, \eqref{2tcano10a}, \eqref{2tcano10b}
form a second class set and, so they can be used to construct the Dirac brackets
on the associated constrained phase space. The only independent phase variables remaining after the
constraints are taken into  account are $\omega_s$, $\varOmega_{\omega s}$ and their Dirac brackets are 
\begin{equation}
\{\langle \omega_s,G_{\omega_s} \rangle,\langle g_{\varOmega_{\omega s}}, \varOmega_{\omega s}\rangle\}_D
=\frac{1}{\kappa_2}\langle g_{\varOmega_{\omega s}},G_{\omega_s} \rangle.
\label{2tcano11}
\end{equation}
The constraints \eqref{2tcano9a}, \eqref{2tcano9b}
are left pending. As it is immediate to see, $f_s$, $F_{fs}$ generate constrained 
phase space BRST transformations. Introducing ghost fields $c_s\in\Omega^0(M,\mathfrak{v}_0[1])$ 
and $C_{cs}\in\Omega^1(M,\mathfrak{v}_1[1])$, we have 
\begin{subequations}
\label{2tcano12}
\begin{align}
&\{\langle f_s, C_{cs}\rangle+\langle c_s,F_{fs}\rangle, \langle \omega_s,G_{\omega_s} \rangle\}_D
=\frac{1}{\kappa_2}\langle s_s\omega_s,G_{\omega_s} \rangle, 
\vphantom{\Big]}
\label{2tcano12a}
\\
&\{\langle f_s, C_{cs}\rangle+\langle c_s,F_{fs}\rangle, \langle g_{\varOmega_{\omega s}}, \varOmega_{\omega s}\}_D
=-\frac{1}{\kappa_2}\langle g_{\varOmega_{\omega s}},s_s\varOmega_{\omega_s} \rangle.
\vphantom{\Big]}
\label{2tcano12b}
\end{align}
\end{subequations}
where $s_s\omega_s$, $s_s\varOmega_{\omega s}$ are given by 
\begin{subequations}
\label{2tcano13}
\begin{align}
&s_s\omega_s=-D_sc_s,
\vphantom{\Big]}
\label{2tcano13a}
\\
&s_s\varOmega_{\omega s}=-D_sC_{cs},
\vphantom{\Big]}
\label{2tcano13b}
\end{align}
\end{subequations}
in agreement with  \eqref{7cinflinfdgloba/2}, \eqref{7cinflinfdglobb/2}. 
%with $\omega$, $\varOmega_\omega$, $c$, $C_c$ replaced by $\omega_s$, $\varOmega_{\omega s}$, $c_s$, $C_{cs}$, 

We quantize $\CS_2(\mathbb{R}\times M,\mathfrak{v})$ by replacing the classical fields $\omega_s$, $\Omega_{\omega s}$ 
satisfying the Dirac brackets \eqref{2tcano11} with corresponding quantum fields $\widehat{\omega}_s$, $\widehat{\Omega}_{\omega s}$ 
satisfying the commutation relations
\begin{equation}
[\langle \widehat{\omega}_s,G_{\omega_s} \rangle,\langle g_{\varOmega_{\omega s}}, \widehat{\varOmega}_{\omega s}\rangle]
=\frac{i}{\kappa_2}\langle g_{\varOmega_{\omega s}},G_{\omega_s} \rangle.
\label{2tcano14}
\end{equation}
The constraints \eqref{2tcano9a}, \eqref{2tcano9b}, which we left pending in the classical theory, translate into 
conditions obeyed by the state vectors $\varPsi$ of the theory
\begin{subequations}
\label{2tcano15a,b}
\begin{align}
&\langle \widehat{f}_s,G_{f_s}\rangle\varPsi=0,
\vphantom{\Big]}
\label{2tcano15a}
\\
&\langle g_{F_{fs}},\widehat{F}_{fs}\rangle\varPsi=0.
\vphantom{\Big]}
\label{2tcano15b}
\end{align}
\end{subequations}

%\vfil\eject 

\subsection{\normalsize \textcolor{blue}{Choice of polarization  and Ward identities }}
\label{subsec:2tpolar}

\hspace{.5cm} 
To build a representation of the operator algebra yielded by canonical quantization, 
we must choose a polarization, a maximal integrable 
distribution on the classical phase space, the restriction of the Dirac symplectic form 
to which vanishes. The polarization must be gauge invariant by consistency.

Henceforth, we shall make reference exclusively to the space manifold $M$. 
We shall thus suppress the index $s$ throughout as it is no longer necessary
lightening in this way the notation.

{\it Ordinary Chern--Simons theory}

In the canonically quantized $\CS_1(\mathbb{R}\times M,\mathfrak{g})$ theory
reviewed in subsect. \ref{subsec:2tcano}, the space manifold $M$ is a $2$--dimensional 
surface. The conventionally normalized Dirac symplectic form is in this case 
\begin{equation}
\langle\delta\omega,\delta\omega\rangle
=-2\kappa_1\int_M(\delta\omega,\delta\omega).
\label{1tpolar0}
\end{equation}
This can be checked to be invariant under any gauge transformation 
$g\in\OGau$ $(M,\mathfrak{g})$ acting by \eqref{xgauact1}.

A generic phase space vector field is of the form
\begin{equation}
\Big\langle g_{\frac{\delta}{\delta \omega}},\frac{\delta}{\delta\omega}\Big\rangle F
=\int_M\Big(g_{\frac{\delta}{\delta \omega}},\frac{\delta F}{\delta\omega}\Big)
\label{1tpolar1}
\end{equation}
where $\delta/\delta\omega$ is a $\Omega^1(M,\mathfrak{g})$--valued vector field.
A standard polarization of the phase space $\omega$ is built as follows.
One picks a complex structure on the surface $M$ 
and uses the marks $10$, $01$ to denote the holomorphic 
and antiholomorphic components of a $1$--form. Setting 
$\delta/\delta\omega^{10}=-i(\delta/\delta\omega)^{01}$, 
$\delta/\delta\omega^{01}=i(\delta/\delta\omega)^{10}$, % and c. c., 
the polarization is defined by the integrable distribution
of the vector fields \hphantom{xxxxxxxxxxxxxxxxx}
\begin{equation}
\Big\langle v_{\frac{\delta}{\delta\omega}}{}^{10},\frac{\delta}{\delta\omega^{10}}\Big\rangle,
\label{1tpolar2}
\end{equation}
%where the marks $10$, $01$ denote the holomorphic and antiholomorphic components of a $1$--form
%and $\delta/\delta\omega^{10}=-i(\delta/\delta\omega)^{01}$, $\delta/\delta\omega^{01}=i(\delta/\delta\omega)^{10}$. 
where $v_{\delta/\delta\omega}{}^{10}(\omega)$ is a phase function. 
The distribution is gauge invariant, since one has  ${}^g\delta/\delta\omega^{10}=g(\delta/\delta\omega^{10})$ for 
$g\in\OGau(M,\mathfrak{g})$. 

With the above choice of polarization, the quantum Hilbert space $\mathcal{H}$ of the $\CS_1$ theory 
consists of phase space functionals $\varPsi(\omega)$ satisfying 
\begin{equation}
\Big\langle v_{\frac{\delta}{\delta\omega}}{}^{10},\frac{\delta\varPsi}{\delta\omega^{10}}\Big\rangle=0,
\label{1tpolar5}
\end{equation}
that is of holomorphic wave functionals $\varPsi(\omega^{01})$.
% since $\delta/\delta\omega^{01}=id\overline{z}\delta/\delta\omega_z$ in local holomorphic coordinates. 
The Hilbert structure appropriate for $\mathcal{H}$, as realized in \cite{Axelrod:1989xt}, is thus 
of the Bargmann type. The $\varPsi$ belonging to $\mathcal{H}$
must satisfy the formal square integrability condition
%of the form \hphantom{xxxxxxxxxxxxxxxxx}
\begin{equation}
\int \mathcal{D}\omega^{01}\mathcal{D}\omega^{10}\,\exp\big(2i\kappa_1\langle \omega^{10},\omega^{01}\rangle\big)
|\varPsi(\omega^{01})|^2<\infty,
\label{1tpolar6}
\end{equation}
where $\mathcal{D}\omega^{01}\mathcal{D}\omega^{10}$ is a formal functional measure. Note that a restriction on 
the sign of $\kappa_1$ is implied by the convergence of \eqref{1tpolar6}. 
The Hilbert inner product is correspondingly given by Bargmann expression 
\begin{equation}
\langle\varPsi_1,\varPsi_2\rangle=\int \mathcal{D}\omega^{01}\mathcal{D}\omega^{10}\,
\exp\big(2i\kappa_1\langle \omega^{10},\omega^{01}\rangle\big)
\varPsi_1(\omega^{01})^*\,\varPsi_2(\omega^{01}).
\label{1tpolar7}
\end{equation}
The field operators $\widehat{\omega}^{01}$, $\widehat{\omega}^{10}$ satisfying \eqref{2tcano14} are 
represented by 
\begin{subequations}
\label{1tpolar8,9}
\begin{align}
&\langle g_{\omega}{}^{10},\widehat{\omega}^{01}\rangle=\langle g_{\omega}^{10},\omega^{01}\,\cdot\,\rangle,
\vphantom{\Big]}
\label{1tpolar8}
\\
&\langle \widehat{\omega}^{10},g_\omega{}^{01}\rangle
=\Big\langle -\frac{1}{2\kappa_1}\frac{\delta}{\delta\omega^{01}},g_\omega{}^{01}\Big\rangle.
\vphantom{\Big]}
\label{1tpolar9}
\end{align}
\end{subequations}
In virtue of the exponential factor in the inner product, one has 
$\widehat{\omega}^{01+}=\widehat{\omega}^{10}$ as required. 

In the representation \eqref{1tpolar8,9}, the vanishing curvature constraint \eqref{1tcano15} takes the form
\begin{equation}
\Big\langle d^{10}\omega^{01}-\frac{1}{2\kappa_1}\Big(d^{01}\frac{\delta}{\delta\omega^{01}}
+\Big[\omega^{01},\frac{\delta}{\delta\omega^{01}}\Big]\Big), g_f\Big\rangle\varPsi(\omega^{01})=0,
\vphantom{\Big]}
\label{1tpolar10}
\end{equation}
This is a WZW type Ward identity determining the variation of $\varPsi(\omega^{01})$ 
under an infinitesimal gauge transformation $u\in\mathfrak{oaut}(M,\mathfrak{g})$ 
with $u=\ad \theta$, $\dot\sigma_u=d\theta$ with 
$\theta$ being a bidegree $(0,0)$ field. Noting that the resulting variation of $\omega$ is 
\begin{equation}
\delta_u\omega^{01}=D^{01}\theta
\label{1tpolar12}
\end{equation}
by \eqref{1tcano13}, we have 
\begin{equation}
\delta_u\varPsi(\omega^{01})=2i\kappa_1\langle d^{10}\omega^{01},\theta\rangle\varPsi(\omega^{01}).
\label{1tpolar13}
\end{equation}
Therefore, the gauge variation of $\varPsi(\omega^{01})$ under a finite gauge transformation
$g\in\OGau(M,\mathfrak{g})$ is given by a universal multiplicative factor
\begin{equation}
\varPsi({}^g\omega^{01})=\exp(iS_{WZW1}(g,\omega^{01}))\varPsi(\omega^{01}),
\label{1tpolar14}
\end{equation}
where $S_{WZW1}(g,\omega^{01})$ is the gauged WZW action. 
By consistency with the group action property of gauge transformation on connections, 
$S_{WZW1}(g,\omega^{01})$ obeys the Polyakov-Wiegmann identity
\begin{equation}
S_{WZW1}(h\diamond g,\omega^{01})=S_{WZW1}(h,{}^g\omega^{01})+S_{WZW1}(g,\omega^{01})\quad \text{mod}~~2\pi.
\label{1tpolar15}
\end{equation}
To reproduce the infinitesimal variation \eqref{2tpolar13}, $S_{WZW1}(g,\omega)$ must satisfy
the normalization condition
\begin{equation}
\delta_uS_{WZW1}(g,\widetilde{\omega}^{01})|_{g=i}=2\kappa_1\langle d^{10}\omega^{01},\theta\rangle, 
\label{1tpolar16}
\end{equation}
where the tilde notation indicates that $\delta_u$ is inert on $\omega^{01}$. 
%This relation can be used in principle to determine the expressiono $S_{WZW1}(g,\omega)$. 
%Instead of proceeding in this way, w
\eqref{1tpolar15}, \eqref{1tpolar16} essentially determine 
the expression of $S_{WZW1}(g,\omega)$. When $M$ is the boundary of a $3$--fold $B$
and $g$ can be extended to an element of $\OGau(B,\mathfrak{g})$, we have 
\begin{align}
S_{WZW1}(g,\omega^{01})&=\kappa_1\int_M\big[(\sigma_g{}^{10},\sigma_g{}^{01})-2(\sigma_g{}^{10},\omega^{01})\big]
\label{1tpolar17}
\\
&\hspace{4.5cm}%+2(g^{-1}(\omega^{10}),\sigma_g{}^{01})-2(g^{-1}(\omega^{10})-\omega^{10},\omega^{01})\Big]
+\frac{\kappa_1}{3}\int_B(\sigma_g,d\sigma_g)\quad \text{mod}~~2\pi,
\nonumber
\end{align}
a classic result \cite{Witten:1983ar}. 
The independence of $\exp(iS_{WZW1}(g,\omega^{01}))$ from the choice of $B$ requires that the 
$\CS_1$ anomaly density $3$--form $\kappa_1q_1$ (cf. eq. \eqref{d1tchern1}) integrates to an 
integer multiple of $2\pi$ on any closed $3$--fold of the form $N=B\cup-B'$ with 
$\partial B=\partial B'=M$. This is how the quantization condition of $\kappa_1$ emerges 
in the canonical quantization of the $\CS_1$ theory. $\vphantom{\ul{\ul{\ul{x}}}}$

{\it Semistrict Chern--Simons theory}

In the canonically quantized $\CS_2(\mathbb{R}\times M,\mathfrak{v})$ theory
worked out in subsect. \ref{subsec:2tcano}, the space manifold $M$ is a $3$--dimensional 
space. The associated normalized Dirac symplectic form is in this case 
%The Dirac symplectic form of $\CS_2$ theory is
\begin{equation}
\langle\delta\omega,\delta\varOmega_\omega\rangle
=\kappa_2\int_M(\delta\omega,\delta\varOmega_\omega).
\label{2tpolar0}
\end{equation}
The form is invariant under any $1$--gauge transformation $g\in\OGau_1(M,\mathfrak{v})$
acting via \eqref{7linfdglob}. In $3$ dimensions, $1$-- and $2$--forms have the same number 
of functional degrees of freedom. The phase space has thus the usual  Hamiltonian form. 

The vector fields $\delta/\delta\omega$, $\delta/\delta\varOmega_\omega$ are specified by the relation  
\begin{align}
&\bigg[\Big\langle g_{\frac{\delta}{\delta \omega}},\frac{\delta}{\delta\omega}\Big\rangle
+\Big\langle \frac{\delta}{\delta\varOmega_\omega}, G_{\frac{\delta}{\delta \varOmega_\omega}}\Big\rangle\bigg] F
\vphantom{\Big]}
\label{2tpolar1}
\\
&\hspace{3.5cm}=\int_M\bigg[\Big(g_{\frac{\delta}{\delta \omega}},\frac{\delta F}{\delta\omega}\Big)
+\Big(\frac{\delta F}{\delta\varOmega_\omega}, G_{\frac{\delta}{\delta \varOmega_\omega}}\Big)
\bigg],
\vphantom{\Big]}
\nonumber
\end{align}
for any phase function $F(\omega,\varOmega_\omega)$. 
A canonical polarization in the phase space $(\omega,\varOmega_\omega)$ is defined as follows. 
It is spanned by the vector fields of the form 
\begin{equation}
\Big\langle \frac{\delta}{\delta\varOmega_\omega}, V_{\frac{\delta}{\delta \varOmega_\omega}}\Big\rangle,
\label{2tpolar2}
\end{equation}
where $V_{\delta/\delta\varOmega_\omega}(\omega,\varOmega_\omega)$ is a phase function and it is understood that 
$\delta/\delta\varOmega_\omega$ does not act on $V_{\delta/\delta\varOmega_\omega}$. The distribution
\eqref{2tpolar2} is clearly integrable. It is also checked that it is gauge invariant by noting that
${}^g\delta/\delta\omega=g_1(\delta/\delta\omega)\,+$ terms linear in $\delta/\delta\varOmega_\omega$,
${}^g\delta/\delta\varOmega_\omega=g_0(\delta/\delta\varOmega_\omega)$ 
under a gauge transformation $g\in\OGau_1(M,\mathfrak{v})$.

With the above choice of polarization, the quantum Hilbert space $\mathcal{H}$ consists of phase space functionals 
$\varPsi(\omega,\varOmega_\omega)$ satisfying 
\begin{equation}
\Big\langle \frac{\delta\varPsi}{\delta\varOmega_\omega}, V_{\frac{\delta}{\delta \varOmega_\omega}}\Big\rangle=0,
\label{2tpolar5}
\end{equation}
that is of wave functionals $\varPsi(\omega)$ depending on $\omega$ only. 
The $\varPsi$ belonging to $\mathcal{H}$ must satisfy a square integrability condition
of the form %\hphantom{xxxxxxxxxxxxxxxxx}
\begin{equation}
\int \mathcal{D}\omega\,|\varPsi(\omega)|^2<\infty.
\label{2tpolar6}
\end{equation}
where $\mathcal{D}\omega$ is a suitable formal functional measure. % to be specified more precisely. 
The Hilbert inner product has then the familiar form \hphantom{xxxxxxxxxxxxxxxxx}
\begin{equation}
\langle\varPsi_1,\varPsi_2\rangle=\int \mathcal{D}\omega\,\varPsi_1(\omega)^*\,\varPsi_2(\omega).
\label{2tpolar7}
\end{equation}
The field operators $\widehat{\omega}$, $\widehat{\varOmega}_\omega$ satisfying \eqref{2tcano14} 
are represented by
\begin{subequations}
\label{2tpolar8,9}
\begin{align}
&\langle\widehat{\omega},G_{\omega}\rangle=\langle\omega\,\cdot\,,G_{\omega}\rangle,
\vphantom{\Big]}
\label{2tpolar8}
\\
&\langle g_{\varOmega_\omega},\widehat{\varOmega}_\omega\rangle
=\Big\langle g_{\varOmega_\omega},-\frac{i}{\kappa_2}\frac{\delta}{\delta\omega}\Big\rangle.
\vphantom{\Big]}
\label{2tpolar9}
\end{align}
\end{subequations}
They are manifestly formally selfadjoint with respect to the Hilbert product \eqref{2tpolar7}: 
$\widehat{\omega}^+=\widehat{\omega}$ and $\widehat{\varOmega}_\omega{}^+=\widehat{\varOmega}_\omega$.

By \eqref{2tpolar8,9}, the constraints \eqref{2tcano15a,b} take the form
\begin{subequations}
\label{2tpolar10,11}
\begin{align}
&\Big\langle d\omega+\frac{1}{2}[\omega,\omega]+\frac{i}{\kappa_2}\partial\frac{\delta}{\delta \omega}, 
G_f\Big\rangle\varPsi(\omega)=0,
\vphantom{\Big]}
\label{2tpolar10}
\\
&\Big\langle g_F, -\frac{i}{\kappa_2}\Big(d\frac{\delta}{\delta \omega}+\Big[\omega,\frac{\delta}{\delta \omega}\Big]\Big)
-\frac{1}{6}[\omega,\omega,\omega]\Big\rangle\varPsi(\omega)=0.
\vphantom{\Big]}
\label{2tpolar11}
\end{align}
\end{subequations}
These are the Ward identities obeyed by $\varPsi$. They determine the variation of $\varPsi(\omega)$ 
under an infinitesimal gauge transformation $u\in\mathfrak{oaut}_0(M,\mathfrak{v})$ 
with $u=\ad \theta$, $\dot\sigma_u$ $=d\theta+\partial \varTheta_\theta$, 
$\dot \varSigma_u=d\varTheta_\theta$, $\dot \tau_u(\pi)=-[\pi,\varTheta_\theta]$,
$(\theta,\varTheta_\theta)$ being a bidegree $(0,0)$ field doublet. Noting that the resulting variation of $\omega$ is 
\begin{equation}
\delta_u\omega=D\theta
\label{2tpolar12}
\end{equation}
(cf. eq. \eqref{2tcano13a}), we have 
\begin{equation}
\delta_u\varPsi(\omega)=i\kappa_2\bigg[\Big\langle d\omega+\frac{1}{2}[\omega,\omega],\varTheta_\theta\Big\rangle
-\frac{1}{6}\langle \theta,[\omega,\omega,\omega]\rangle\bigg]\varPsi(\omega).
\label{2tpolar13}
\end{equation}
Therefore, the gauge variation of $\varPsi(\omega)$ under a finite gauge transformation
$g\in\OGau_1(M,\mathfrak{v})$ is given by a universal multiplicative factor
\begin{equation}
\varPsi({}^g\omega)=\exp(iS_{WZW2}(g,\omega))\varPsi(\omega),
\label{2tpolar14}
\end{equation}
where $S_{WZW2}(g,\omega)$ is a higher analog of the gauged WZW action. In analogy to its ordinary counterpart, 
$S_{WZW2}(g,\omega)$ obeys a higher version of the Polyakov-Wiegmann identity
\begin{equation}
S_{WZW2}(h\diamond g,\omega)=S_{WZW2}(h,{}^g\omega)+S_{WZW2}(g,\omega)\quad \text{mod}~~2\pi.
\label{2tpolar15}
\end{equation}
To reproduce the infinitesimal variation \eqref{2tpolar13}, $S_{WZW2}(g,\omega)$ must satisfy
further the normalization condition 
\begin{equation}
\delta_uS_{WZW2}(g,\widetilde{\omega})|_{g=i}=
\kappa_2\bigg[\Big\langle d\omega+\frac{1}{2}[\omega,\omega],\varTheta_\theta\Big\rangle
-\frac{1}{6}\langle \theta,[\omega,\omega,\omega]\rangle\bigg],
\label{2tpolar16}
\end{equation}
where the tilde indicates that $\delta_u$ is inert on $\omega$. 
%This relation can be used in principle to determine the expressiono $S_{WZW2}(g,\omega)$. 
%Instead of proceeding in this way, w
An expression of $S_{WZW2}(g,\omega)$
fulfilling relations \eqref{2tpolar15}, \eqref{2tpolar16} holding when $M$ is the boundary of a $4$--fold $B$
and $g$ can be extended to and element of $\OGau_1(B,\mathfrak{v})$ is 
\begin{align}
S_{WZW2}(g,\omega)&=-\frac{\kappa_2}{2}\int_M\Big[(\sigma_g-\omega,\tau_g(\sigma_g-\omega))-2(\omega-\sigma_g,\varSigma_g)
\vphantom{\Big]}
\label{2tpolar17}
\\
&\hspace{3.5cm}+\frac{1}{3}(\sigma_g-\omega,g_1{}^{-1}g_2(\sigma_g-\omega,\sigma_g-\omega))\Big]
\nonumber
\\
&\hphantom{=}+\frac{\kappa_2}{4}\int_B\big[2(d\sigma_g,\varSigma_g)-(\sigma_g,d\varSigma_g)\bigg]\quad \text{mod}~~2\pi.
\nonumber
\end{align}
As in the ordinary case, 
the independence of $\exp(iS_{WZW2}(g,\omega))$ from the choice of $B$ requires that the 
$\CS_2$ anomaly density $4$--form $\kappa_2q_2$ (cf. eq. \eqref{d2tchern1}) integrates to an 
integer multiple of $2\pi$ on any closed $4$--fold of the form $N=B\cup-B'$ with 
$\partial B=\partial B'=M$. This will be the case if the pair $(N,\mathfrak{v})$ 
is admissible for a sufficiently broad class of closed $4$-folds $N$, as we assumed earlier
at the end of subsect. \ref{subsec:2tchern}

%implying again the quantization condition of $\kappa_2$.

The polarization we have constructed above is fully topological in the sense that 
its definition does not require 
the choice of any auxiliary structure on the threefold $M$. In this respect, the associated semistrict Chern--Simons 
theory is manifestly topological in a way ordinary Chern--Simons theory is not. There is however another choice 
of polarization more similar in flavour to standard Chern--Simons' in that it assumes the assignment of a 
strictly pseudoconvex CR structure on $M$. 

We review briefly a few basic facts about CR structures to the reader's benefit.
(See refs. \cite{Chirka:1991,Dragomir:2006}for background material.)
In a CR $3$--fold $M$, the complexified cotangent bundle $T^*M\otimes \mathbb{C}$ 
has a direct sum decomposition $T^{*100}M\oplus T^{*010}M\oplus T^{*001}M$, where $T^{*100}M$, $T^{*010}M$, 
$T^{*001}M$ are line subbundles of $T^*M\otimes \mathbb{C}$, 
$T^{*001}M=\overline{T^{*100}M}$ and $T^{*010}M$ is the complexification of a trivial line subbundle $E$ 
of $T^*M$, the one fiberwise generated by the underlying %global nowhere vanishing
contact form. 
%A choice of trivialization of $E_r$ is a nowhere vanishing $1$--form $\lambda\in\Omega^1(M)$ 
%called a pseudo Hermitian structure.
Forms of $M$ are graded accordingly. 
%A $0$--form $\phi\in\Omega^0(M)$ has a unique component, $\phi=\phi^{000}$. 
For instance, a $1$--form $\alpha\in\Omega^1(M)$ has three components,  
$\alpha=\alpha^{100}+\alpha^{010}+\alpha^{001}$. A $2$--form $\beta\in\Omega^2(M)$ has also three components,
$\beta=\beta^{110}+\beta^{101}+\beta^{011}$. 
%A $3$--form $\tau\in\Omega^3(M)$ has again a unique component, $\tau=\tau^{111}$. 
Strictly pseudoconvex CR spaces are the closest $3$--dimensional analog 
of Riemann surfaces. In particular, with the strictly pseudoconvex CR structure of a space 
there is associated a class of metrics, called Webster metrics, 
related to each other by a change of the normalization of the contact form, 
much as with a conformal structure of a surface there is associated a conformal class
of metrics.   

A second polarization of the phase space $(\omega,\varOmega_\omega)$ is built as follows.
One picks a strictly pseudoconvex CR structure on $M$. Setting
$\delta/\delta\omega^{100}=-i(\delta/\delta\omega)^{011}$,
$\delta/\delta\omega^{010}=-i(\delta/\delta\omega)^{101}$,
$\delta/\delta\omega^{001}=-i(\delta/\delta\omega)^{110}$
and $\delta/\delta\varOmega_\omega{}^{011}=-i(\delta/\delta\varOmega_\omega)^{100}$,
$\delta/\delta\varOmega_\omega{}^{101}=-i(\delta/\delta\varOmega_\omega)^{010}$,
$\delta/\delta\varOmega_\omega{}^{110}=-i(\delta/\delta\varOmega_\omega)^{001}$, the polarization is spanned 
by the vector fields of the form 
\begin{equation}
\Big\langle \frac{\delta}{\delta\varOmega_\omega{}^{110}}, V_{\frac{\delta}{\delta \varOmega_\omega}}{}^{110}\Big\rangle
+\Big\langle \frac{\delta}{\delta\varOmega_\omega{}^{011}}, V_{\frac{\delta}{\delta \varOmega_\omega}}{}^{011}\Big\rangle
+\Big\langle v_{\frac{\delta}{\delta\omega}}{}^{010}, \frac{\delta}{\delta\omega^{010}}\Big\rangle, 
\label{3tpolar2}
\end{equation}
where $V_{\delta/\delta\varOmega_\omega}(\omega,\varOmega_\omega){}^{110}$,
$V_{\delta/\delta\varOmega_\omega}(\omega,\varOmega_\omega){}^{011}$, $v_{\delta/\delta\omega}(\omega,\varOmega_\omega){}^{010}$
are phase functions and again it is understood  that 
$\delta/\delta\varOmega_\omega{}^{110}$, $\delta/\delta\varOmega_\omega{}^{011}$  
does not act on $V_{\delta/\delta\varOmega_\omega}{}^{110}$,
$V_{\delta/\delta\varOmega_\omega}{}^{110}$. It is easily checked that the distribution
\eqref{3tpolar2} is integrable. It is also checked that it is gauge invariant by noting that
${}^g\delta/\delta\omega^{010}=g_1(\delta/\delta\omega^{010})\,+$ terms linear in $\delta/\delta\varOmega_\omega{}^{110}$,
$\delta/\delta\varOmega_\omega{}^{011}$ and ${}^g\delta/\delta\varOmega_\omega{}^{110}=g_0(\delta/\delta\varOmega_\omega{}^{110})$,
${}^g\delta/\delta\varOmega_\omega{}^{011}=g_0(\delta/\delta\varOmega_\omega{}^{011})$ 
under a gauge transformation $g\in\OGau_1(M,\mathfrak{v})$.

With the above choice of polarization, the quantum Hilbert space $\mathcal{H}$ consists of phase space functionals 
$\varPsi(\omega,\varOmega_\omega)$ satisfying 
\begin{equation}
\Big\langle \frac{\delta\varPsi}{\delta\varOmega_\omega{}^{110}}, V_{\frac{\delta}{\delta \varOmega_\omega}}{}^{110}\Big\rangle
+\Big\langle \frac{\delta\varPsi}{\delta\varOmega_\omega{}^{011}}, V_{\frac{\delta}{\delta \varOmega_\omega}}{}^{011}\Big\rangle
+\Big\langle v_{\frac{\delta}{\delta\omega}}{}^{010}, \frac{\delta\varPsi}{\delta\omega^{010}}\Big\rangle=0 
\label{3tpolar5}
\end{equation}
that is of wave functionals $\varPsi(\omega^{100},\omega^{001},\varOmega_\omega{}^{101})$. 
The $\varPsi$ must satisfy a square integrability condition
of the form \hphantom{xxxxxxxxxxxxxxxxx}
\begin{equation}
\int \mathcal{D}\omega^{100}\mathcal{D}\omega^{001}\mathcal{D}\varOmega_\omega{}^{101}\,
|\varPsi(\omega^{100},\omega^{001},\varOmega_\omega{}^{101})|^2<\infty.
\label{3tpolar6}
\end{equation}
where $\mathcal{D}\omega^{100}\mathcal{D}\omega^{001}\mathcal{D}\varOmega_\omega{}^{101}$ 
is a suitable functional measure. %to be specified more precisely. 
The Hilbert inner product is then 
\begin{align}
\langle\varPsi_1,\varPsi_2\rangle&=\int \mathcal{D}\omega^{100}\mathcal{D}\omega^{001}\mathcal{D}\varOmega_\omega{}^{101}\,
\vphantom{\Big]}
\label{3tpolar7}
\\
&\hphantom{4.5cm}\times 
\varPsi_1(\omega^{100},\omega^{001},\varOmega_\omega{}^{101})^*\,\varPsi_2(\omega^{100},\omega^{001},\varOmega_\omega{}^{101}).
\vphantom{\Big]}
\nonumber
\end{align}
The field operators $\widehat{\omega}$, $\widehat{\varOmega}_\omega$ satisfying \eqref{2tcano14} are realized as
\begin{subequations}
\label{3tpolar8,9}
\begin{align}
&\langle\widehat{\omega}^{100},G_{\omega}{}^{011}\rangle=\langle\omega^{100}\,\cdot\,,G_{\omega}{}^{011}\rangle,
\vphantom{\Big]}
\label{3tpolar8}
\\
&\langle\widehat{\omega}^{010},G_{\omega}{}^{101}\rangle
=\Big\langle-\frac{1}{\kappa_2}\frac{\delta}{\delta\varOmega_\omega{}^{101}},G_{\omega}{}^{101}\Big\rangle,
\vphantom{\Big]}
\nonumber
\\
&\langle\widehat{\omega}^{001},G_{\omega}{}^{110}\rangle=\langle\omega^{001}\,\cdot\,,G_{\omega}{}^{110}\rangle,
\vphantom{\Big]}
\nonumber
\\
&\langle g_{\varOmega_\omega}{}^{100},\widehat{\varOmega}_\omega{}^{011}\rangle
=\Big\langle g_{\varOmega_\omega}{}^{100},\frac{1}{\kappa_2}\frac{\delta}{\delta\omega^{100}}\Big\rangle,
\vphantom{\Big]}
\label{3tpolar9}
\\
&
\langle g_{\varOmega_\omega}{}^{010},\widehat{\varOmega}_\omega{}^{101}\rangle
=\langle g_{\varOmega_\omega}{}^{010},\varOmega_\omega{}^{101}\,\cdot\,\rangle,
\vphantom{\Big]}
\nonumber
\\
&\langle g_{\varOmega_\omega}{}^{001},\widehat{\varOmega}_\omega{}^{110}\rangle
=\Big\langle g_{\varOmega_\omega}{}^{001},\frac{1}{\kappa_2}\frac{\delta}{\delta\omega^{001}}\Big\rangle.
\vphantom{\Big]}
\nonumber
\end{align}
\end{subequations}
They satisfy the natural adjunction relations $\widehat{\omega}^{100+}=\widehat{\omega}^{001}$,
$\widehat{\omega}^{010+}=\widehat{\omega}^{010}$ and 
$\widehat{\varOmega}_\omega{}^{011+}=\widehat{\varOmega}_\omega{}^{110}$,
$\widehat{\varOmega}_\omega{}^{101+}=\widehat{\varOmega}_\omega{}^{101}$.

By \eqref{3tpolar8,9}, the constraints \eqref{2tcano15a,b} presently read 
\begin{subequations}
\label{3tpolar10,11}
\begin{align}
&\Big\langle \frac{1}{\kappa_2}\Big(d^{100}\frac{\delta}{\delta\varOmega_\omega{}^{101}}
+\Big[\omega^{100},\frac{\delta}{\delta\varOmega_\omega{}^{101}}\Big]
\hspace{4cm}
\vphantom{\Big]}
\label{3tpolar10}
%\\
\end{align}
\begin{align}
&\hspace{3.5cm}
+\partial\frac{\delta}{\delta\omega^{001}}\Big)-d^{010}\omega^{100}, G_f{}^{001}\Big\rangle
\varPsi(\omega^{100},\omega^{001},\varOmega_\omega{}^{101})=0,
\vphantom{\Big]}
\nonumber
\\
&\langle d^{100}\omega^{001}+d^{001}\omega^{100}
+[\omega^{100},\omega^{001}]-\partial\varOmega_\omega{}^{101}, 
G_f{}^{010}\rangle\varPsi(\omega^{100},\omega^{001},\varOmega_\omega{}^{101})=0,
%\vphantom{\dot{\dot{\dot{\dot{\dot{f}}}}}}
\vphantom{\frac{1}{\kappa_2}}
\nonumber
\\
&\Big\langle \frac{1}{\kappa_2}\Big(d^{001}\frac{\delta}{\delta\varOmega_\omega{}^{101}}
+\Big[\omega^{001},\frac{\delta}{\delta\varOmega_\omega{}^{101}}\Big]
%+\partial\frac{\delta}{\delta\omega^{100}}\Big)-d^{010}\omega^{001}, G_f{}^{100}\Big\rangle\varPsi(\omega^{100},\omega^{001},\varOmega_\omega{}^{101})=0,
\vphantom{\Big]}
\nonumber
\\
&%\Big\langle \frac{1}{\kappa_2}\Big(d^{001}\frac{\delta}{\delta\varOmega_\omega{}^{101}}
%+\Big[\omega^{001},\frac{\delta}{\delta\varOmega_\omega{}^{101}}\Big]
\hspace{3.5cm}+\partial\frac{\delta}{\delta\omega^{100}}\Big)-d^{010}\omega^{001}, 
G_f{}^{100}\Big\rangle\varPsi(\omega^{100},\omega^{001},\varOmega_\omega{}^{101})=0,
\vphantom{\Big]}
\nonumber
\\
&\Big\langle g_F, \frac{1}{\kappa_2}\Big(d^{100}\frac{\delta}{\delta \omega^{100}}+d^{001}\frac{\delta}{\delta \omega^{001}}
+\Big[\omega^{100},\frac{\delta}{\delta \omega^{100}}\Big]+\Big[\omega^{001},\frac{\delta}{\delta \omega^{001}}\Big]
\vphantom{\Big]}
\label{3tpolar11}
\\
%\hspace{.5cm}
\!-&\Big[\frac{\delta}{\delta\varOmega_\omega{}^{101}}, \varOmega_\omega{}^{101}\Big]
+\Big[\omega^{100},\frac{\delta}{\delta\varOmega_\omega{}^{101}},\omega^{001}\Big]\Big)
+d^{010}\varOmega_\omega{}^{101}\Big\rangle\varPsi(\omega^{100},\omega^{001},\varOmega_\omega{}^{101})=0.
\vphantom{\Big]}
\nonumber
\end{align}
\end{subequations}
In the fifth term of \eqref{3tpolar11}, it is understood that $\delta/\delta\varOmega_\omega{}^{101}$ is
inert on $\varOmega_\omega{}^{101}$.
These are the Ward identities obeyed by $\varPsi$ in this CR canonical quantization scheme. 
They determine the variation of a $\varPsi(\omega^{100},\omega^{001},\varOmega_\omega{}^{101})$ 
under an infinitesimal gauge transformation $u\in\mathfrak{oaut}_0(M,\mathfrak{v})$ 
of the form $u=\ad \theta$, $\dot\sigma_u$ $=d\theta+\partial \varTheta_\theta$, 
$\dot \varSigma_u=d\varTheta_\theta$, $\dot \tau_u(\pi)=-[\pi,\varTheta_\theta]$,
$(\theta,\varTheta_\theta)$ as earlier. The resulting variations of 
$\omega^{100}$, $\omega^{001}$, $\varOmega_\omega{}^{101}$ are given by 
\begin{subequations}
\label{3tpolar12a,b}
\begin{align}
&\delta_u\omega^{100}=(D\theta)^{100}=d^{100}\theta+[\omega^{100},\theta]+\partial\varTheta_\theta{}^{100},
\vphantom{\Big]}
\label{3tpolar12a}
\\
&\delta_u\omega^{001}=(D\theta)^{001}=d^{001}\theta+[\omega^{001},\theta]+\partial\varTheta_\theta{}^{001},
\vphantom{\Big]}
\nonumber
\\
&\delta_u\varOmega_\omega{}^{101}=(D\varTheta_\theta)^{101}=d^{100}\varTheta_\theta{}^{001}+[\omega^{100},\varTheta_\theta{}^{001}]
\vphantom{\Big]}
\label{3tpolar12b}
\\
&\hspace{1.5cm}+d^{001}\varTheta_\theta{}^{100}+[\omega^{001},\varTheta_\theta{}^{100}]
-[z,\varOmega_\omega{}^{101}]+[\omega^{100},\omega^{001},z]
\vphantom{\Big]}
\nonumber
\end{align}
\end{subequations}
(cf. eq. \eqref{2tcano13a}). On account of \eqref{3tpolar12a,b}, we have 
\begin{align}
&\delta_u\varPsi(\omega^{100},\omega^{001},\varOmega_\omega{}^{101})=i\kappa_2
\big[\langle \theta,d^{010}\varOmega_\omega{}^{101}\rangle
\vphantom{\Big]}
\label{3tpolar13}
\\
&\hspace{.4cm}+\langle d^{010}\omega^{100},\varTheta_\theta{}^{001}\rangle
+\langle d^{010}\omega^{001},\varTheta_\theta{}^{100}\rangle\big]\varPsi(\omega^{100},\omega^{001},\varOmega_\omega{}^{101}).
\vphantom{\Big]}
\nonumber
\end{align}
Therefore, the gauge variation of $\varPsi(\omega)$ under a finite gauge transformation
$g\in\OGau_1(M,\mathfrak{v})$ is given by a universal multiplicative factor
\begin{align}
&\varPsi({}^g\omega^{100},{}^g\omega^{001},{}^g\varOmega_\omega{}^{101})
\hspace{5.8cm}
\vphantom{\Big]}
\label{3tpolar14}
%\\
\end{align}
\begin{align}
&\hspace{3cm}=\exp(iS_{WZW2}(g,\omega^{100},\omega^{001},\varOmega_\omega{}^{101}))
\varPsi(\omega^{100},\omega^{001},\varOmega_\omega{}^{101}),
\vphantom{\Big]}
\nonumber
\end{align}
where $S_{WZW2}(g,\omega^{100},\omega^{001},\varOmega_\omega{}^{101})$ is another
higher analog of the gauged WZW action. Again, as its ordinary counterpart, 
it obeys a higher Polyakov-Wiegmann identity
\begin{align}
&S_{WZW2}(h\diamond g,\omega^{100},\omega^{001},\varOmega_\omega{}^{101})
\vphantom{\Big]}
\label{3tpolar15}
\\
&=S_{WZW2}(h,{}^g\omega^{100},{}^g\omega^{001},{}^g\varOmega_\omega{}^{101})
+S_{WZW2}(g,\omega^{100},\omega^{001},\varOmega_\omega{}^{101})\quad \text{mod}~~2\pi
\vphantom{\Big]}
\nonumber
\end{align}
To reproduce the infinitesimal variation \eqref{3tpolar13}, $S_{WZW2}(g,\omega^{100},\omega^{001},\varOmega_\omega{}^{101})$ must satisfy
the normalization condition 
\begin{align}
&\delta_uS_{WZW2}(g,\widetilde{\omega}^{100},\widetilde{\omega}^{001},\widetilde{\varOmega}_\omega{}^{101})|_{g=i}
\vphantom{\Big]}
\label{3tpolar16}
\\
&\hspace{1cm}=\kappa_2\big[\langle \theta,d^{010}\varOmega_\omega{}^{101}\rangle
+\langle d^{010}\omega^{100},\varTheta_\theta{}^{001}\rangle
+\langle d^{010}\omega^{001},\varTheta_\theta{}^{100}\rangle\big]
\vphantom{\Big]}
\nonumber
\end{align}
where again the tilde notation indicates that $\delta_u$ 
is inert on $\omega^{100}$, $\omega^{001}$, $\varOmega_\omega{}^{101}$. 
%This relation can be used in principle to determine the expressiono 
%$S_{WZW2}(g,\omega^{100},\omega^{001},\varOmega_\omega{}^{101})$. 
%Instead of proceeding in this way, %wwwwwwwww
An expression of $S_{WZW2}(g,\omega^{100},\omega^{001},\varOmega_\omega{}^{101})$ fulfilling 
relation \eqref{3tpolar15} holding when $M$ is the boundary of a $4$--fold $B$ and $g$ 
can be extended to an element of $\Gau_1(B,\mathfrak{v})$ is 
\begin{align}
&S_{WZW2}(g,\omega^{100},\omega^{001},\varOmega_\omega{}^{101})
\vphantom{\Big]}
\label{3tpolar17}
\\
&\hspace{.5cm}=-\frac{\kappa_2}{2}\int_M\Big[2(\sigma_g{}^{100}-\omega{}^{100},\tau_g{}^{010}(\sigma_g{}^{001}-\omega{}^{001}))
\vphantom{\Big]}
\nonumber
\\
&\hspace{2cm}%\hspace{3.5cm}\vphantom{\int_M}
-2(\omega{}^{100}-\sigma_g{}^{100},\varSigma_g{}^{011})
-2(\omega{}^{001}-\sigma_g{}^{001},\varSigma_g{}^{110})+2(\sigma_g{}^{010},\varOmega_\omega{}^{101})\Big]
\nonumber
\\
&\hspace{.8cm}%\hspace{4.5cm}
+\frac{\kappa_2}{4}\int_B\big[2(d\sigma_g,\varSigma_g)-(\sigma_g,d\varSigma_g)\big] \quad \text{mod}~~2\pi,
\nonumber
\end{align}
where for the last term the same considerations as before hold. 
This action does not fulfill \eqref{3tpolar16} however, but a weaker version of it,
\begin{align}
&\delta_uS_{WZW2}(g,\widetilde{\omega}^{100},\widetilde{\omega}^{001},\widetilde{\varOmega}_\omega{}^{101})|_{g=i}
\vphantom{\Big]}
\label{3tpolar18}
\\
&\hspace{1cm}=\kappa_2\big[\langle \theta,d^{010}\varOmega_\omega{}^{101}\rangle
+\langle d^{010}\omega^{100},\varTheta_\theta{}^{001}\rangle
+\langle d^{010}\omega^{001},\varTheta_\theta{}^{100}\rangle
\vphantom{\Big]}
\nonumber
\\
&\hspace{2cm}
+\langle d^{100}\omega^{001}+d^{001}\omega^{100}
+[\omega^{100},\omega^{001}]-\partial\varOmega_\omega{}^{101}, \varTheta_\theta{}^{010}\rangle
\big].
\vphantom{\Big]}
\nonumber
\end{align}
This however poses no problem. By the second Ward identity \eqref{3tpolar10}, 
the field functionals $\varPsi(\omega^{001},\varOmega_\omega{}^{101})$ are supported precisely 
on the functional hypersurface $d^{100}\omega^{001}+d^{001}\omega^{100}
+[\omega^{100},\omega^{001}]-\partial\varOmega_\omega{}^{101}=0$. Thus the last offending term
in \eqref{3tpolar18} vanishes identically upon insertion in \eqref{3tpolar14}. 

To summarize, we have found that, when certain conditions are met, 
semistrict higher Chern--Simons theory admits two distinct canonical 
quantizations and correspondingly two sets of higher WZW Ward identities 
each characterized by a gauged WZW action.

The first canonical quantization is manifestly topological, as it does not necessitate a choice of any
additional structure on the spacial $3$--fold. The second one requires instead a choice of a 
CR structure on the latter. The unitary equivalence of the quantizations associated with distinct CR structures 
is an open problem. A solution of it on the same lines as that presented in ref. \cite{Axelrod:1989xt}
for the ordinary case requires a full fledged deformation theory of CR structure, which to the best of our 
knowledge is missing presently. 
Furthermore, the relationship between the the topological and CR quantizations 
remains mysterious.

It would be interesting to investigate the properties of the solutions of the Ward identities
for both canonical quantizations. Here, we limit ourselves to observe that the solutions
are generically functional distributions. For instance, the second Ward identity \eqref{3tpolar10} entails 
that the wave functional is supported on connections with vanishing $101$ curvature component
and thus exhibits a corresponding functional Dirac delta singularity.

%\vfil\eject

\subsection{\normalsize \textcolor{blue}{Examples}}
\label{subsec:exa}

\hspace{.5cm} 
We present a few examples to illustrate the higher Chern--Simons theory
developed in subsect. \ref{subsec:2tchern}.

{\it Balanced differential Lie crossed modules}

A differential Lie crossed module $(\mathfrak{g},\mathfrak{h})$ is balanced if it is so
when viewed as a strict Lie $2$--algebra (cf. apps. \ref{sec:difliecr}, \ref{sec:linftybal}). 
Thus, $(\mathfrak{g},\mathfrak{h})$ is balanced 
if it is equipped with a non singular bilinear pairing $(\cdot,\cdot):\mathfrak{g}\times\mathfrak{h}
\rightarrow \mathbb{R}$ such that 
\begin{subequations}
\label{crmcs1}
\begin{align}
&(\tau(X),Y)-(\tau(Y),X)=0,
\vphantom{\Big]}
\label{crmcs1a}
\\
&([\pi,x],X)+(x,\mu(\pi)(X))=0
\vphantom{\Big]}
\label{crmcs1b}
\end{align}
\end{subequations}
(cf. eqs. \eqref{linftyform1}, \eqref{linftyform2}). 
Below, we assume  that $(\mathfrak{g},\mathfrak{h})$ is the differential Lie crossed module of a Lie crossed module
$(G,H)$ (cf. app. \ref{sec:crossed}). 

By \eqref{2tchern1}, since the three argument bracket vanishes in the present case, the higher Chern--Simons theory
$\CS_2(N,\mathfrak{g},\mathfrak{h})$ is formally a BF theory, with the $2$ form connection component playing the role
of the $B$ field. This conclusion is however unwarranted, because the symmetry structure 
of $\CS_2(N,\mathfrak{g},\mathfrak{h})$ is basically different from that of an ordinary BF model. 

There exists a distinguished $2$--subgroup $\overline{\Gau}(N,G,H)$
of the gauge transformation strict $2$--group $\Gau(N,\mathfrak{g},\mathfrak{h})$ \cite{Zucchini:2011aa}. 
The $1$--gauge transformations belonging to  $\overline{\Gau}(N,G,H)$ are of the form 
\begin{subequations}
\label{crmdgautrsf5}
\begin{align}
&g_\gamma=\phi_\gamma, %\hspace{6.5cm}
\vphantom{\Big]}
\label{crmdgautrsf5a}
\\
&\sigma_{g_\gamma}=\gamma^{-1}d\gamma+\Ad\gamma^{-1}(\tau(\chi_\gamma)),
\vphantom{\Big]}
\label{crmdgautrsf5b}
\\
&\varSigma_{g_\gamma}=\dot m(\gamma^{-1})\Big(d\chi_\gamma+\frac{1}{2}[\chi_\gamma,\chi_\gamma]\Big),
\vphantom{\Big]}
\label{crmdgautrsf5c}
\\
&\tau_{g_\gamma}(x)=\mu(x)(\dot m(\gamma^{-1})(\chi_\gamma)),
\vphantom{\Big]}
\label{crmdgautrsf5d}
\end{align}
\end{subequations}
where $\gamma\in\Map(N,G)$, $\chi_\gamma\in\Omega^1(N,\mathfrak{h})$. Here, for $a\in G$, $\phi_a\in \Aut_1(\mathfrak{v})$
is defined by $\phi_{a 0}(\pi)=\Ad a(\pi)$, $\phi_{a 1}(\varPi)=\dot m(a)(\varPi)$ and $\phi_{a 2}(\pi,\pi)=0$
and \eqref{crmdgautrsf5a} is understood to hold pointwise on $N$.
$\tau$, $\mu$, $t$ and $m$ are related by \eqref{st2grst2lie1}, \eqref{st2grst2lie3} and $\dot m$ is given by \eqref{crexp1a}. 
For two $1$--gauge transformations $g_\zeta$, $g_\eta$ associated with the data $\zeta,\eta\in \Map(N,G)$ and 
$\chi_\zeta,\chi_\eta\in\Omega^1(N,\mathfrak{h})$, the $2$--gauge transformations of $\overline{\Gau}(N,G,H)$ with source 
$g_\zeta$ and target $g_\eta$ are those of the form \pagebreak
\begin{subequations}
\label{2crmdgautrsf5}
\begin{align}
&F_\varLambda(x)=\varPhi_{\zeta,\varLambda}(x), %=Q(\zeta x\zeta^{-1},\varLambda),
\vphantom{\Big]}
\label{2crmdgautrsf5e}
\\
&A_{F_\varLambda}=\dot m(\zeta^{-1})(-\varLambda^{-1}d\varLambda+\chi_\zeta+\Ad\varLambda^{-1}(B_\varLambda-\chi_\zeta)),
\vphantom{\Big]}
\label{2crmdgautrsf5f}
\end{align}
\end{subequations}
where $\varLambda\in\Map(N,H)$ and $B_\varLambda\in \Omega^1(N,\mathfrak{h})$ with 
\begin{subequations}
\label{crmdgautrsf1}
\begin{align}
&\eta=t(\varTheta)\zeta,
\vphantom{\Big]}
\label{crmdgautrsf1a}
\\
&\chi_\zeta-\chi_\eta=B_\varLambda.
\vphantom{\Big]}
\label{crmdgautrsf1b}
\end{align}
\end{subequations}
Here, for $a\in G$ and $A\in H$, $\varPhi_{a,A}$
is defined by $\varPhi_{a,A}(\pi)=Q(\Ad a(\pi),A)$
and \eqref{2crmdgautrsf5e} is understood to hold pointwise on $N$. 
$Q$ is given by \eqref{crexp1b}. 

Let $(\omega,\varOmega_\omega)$ be a connection doublet and $(f,F_f)$
be its curvature doublet.
Inserting eqs. \eqref{crmdgautrsf5b}--\eqref{crmdgautrsf5d}
into the relations \eqref{7linfdglob}, we obtain
\begin{subequations}%
\label{gliecross6}
\begin{align}
&{}^{g_\gamma}\omega=\Ad\gamma(\omega)-d\gamma\gamma^{-1}-\tau(\chi_\gamma), 
\vphantom{\Big]}
\label{gliecross6a}
\\
&{}^{g_\gamma}\varOmega_\omega=\dot m(\gamma)(\varOmega_\omega)
-d\chi_\gamma-\frac{1}{2}[\chi_\gamma,\chi_\gamma].
\vphantom{\Big]}
\label{gliecross6b}
\\
&\hspace{4.9cm}
-\mu(\Ad\gamma(\omega)-d\gamma\gamma^{-1}-\tau(\chi_\gamma))(\chi_\gamma)
\vphantom{\Big]}
\nonumber
\end{align}
\end{subequations} 
Inserting eqs. \eqref{crmdgautrsf5b}--\eqref{crmdgautrsf5d} into \eqref{8linfdglob},
we find further
\begin{subequations}%
\label{gliecross7}
\begin{align}
&{}^{g_\gamma}f=\Ad \gamma(f), 
\vphantom{\Big]}
\label{gliecross7a}
\\
&{}^{g_\gamma}F_f=\dot m(\gamma)(F_f)-\mu(\Ad \gamma(f))(\chi_\gamma).
\vphantom{\Big]}
\label{gliecross7b}
\end{align}
\end{subequations} 
These expressions are identical to those obtained originally in refs.
\cite{Baez:2004in,Baez:2005qu}. 

The anomaly $Q_2(g_\gamma)$ turns out to vanish for all $1$--gauge transformations $g_\gamma$ 
of $\overline{\Gau}(N,G,H)$. Indeed, the anomaly density $q_2$ is exact
\begin{equation}
q_2=\frac{1}{2}(\tau(\varSigma_{g_\gamma}), \varSigma_{g_\gamma})
=\frac{1}{2}d\Big(\tau(\chi_\gamma),d\chi_\gamma+\frac{1}{3}[\chi_\gamma,\chi_\gamma]\Big).
\label{crmcsq2}
\end{equation}
Therefore the higher Chern--Simons theory $\CS_2(N,\mathfrak{g},\mathfrak{h})$ is non 
anomalous, at least when restricting to the $1$--gauge transformations drawn from
$\overline{\Gau}(N,G,H)$, and there is no level quantization. 

{\it Balanced Lie $2$--algebra $\mathfrak{v}$ with invertible $\partial$}

Let $\mathfrak{v}$ be a balanced Lie $2$--algebra  with invariant
form such that $\partial$ is invertible.
Then, the  gauge anomaly $Q_2(g)$ of the classical action of the Chern--Simons theory
$\CS_2(N,\mathfrak{v})$ vanishes identically.  Indeed, the Chevalley--Eilenberg 
cocycle $\chi_2\in\CE^4(\mathfrak{v})$ of eq. \eqref{e2tchern2} turns out to be exact in this case, being
%for $g\in\OGau_1(N,\mathfrak{v})$, the closed anomaly density 4--form $(\sigma_g,d\varSigma_g)/4$ is exact,
\begin{equation}
\chi_2=\mathcal{Q}_{\CE(\mathfrak{v})}\frac{1}{2}\Big(\pi,\varPi-\frac{1}{6}\partial^{-1}[\pi,\pi]\Big)
\label{exa0}
\end{equation}
and, as we have shown in sect \ref{subsec:2tchern}, 
this implies that $Q_2(g)=0$. Consequently, in this case too
the higher Chern--Simons theory $\CS_2(N,\mathfrak{v})$ is non 
anomalous and there is no level quantization.

\vspace{1mm} 

{\it Balanced Lie $2$--algebra $\mathfrak{v}$ with vanishing $\partial$}

In the category of Lie $2$--algebras, seen as $2$--term $L_\infty$ algebras,
every Lie $2$--algebra $\mathfrak{v}$ is equivalent to one with vanishing 
boundary map $\partial$. We are thus led to consider a balanced Lie $2$--algebra $\mathfrak{v}$  with invariant
form such that $\partial=0$. By \eqref{2tlinalgc},
$\mathfrak{v}_0=\mathfrak{g}$ is a Lie algebra with brackets
$[\cdot,\cdot]$. Since the invariant form $(\cdot,\cdot)$ is non singular, 
$\mathfrak{v}_1=\mathfrak{g}^*$ with duality pairing
$\langle\cdot,\cdot\rangle=(\cdot,\cdot)$. By the invariance of the pairing $(\cdot,\cdot)$,
eq. \eqref{linftyform2}, $\mathfrak{v}_1$ is just the coadjoint
$\mathfrak{g}$--module.  The property \eqref{2tlinalge} is equivalent
to the three argument bracket $[\cdot,\cdot,\cdot]$ defining a $\mathfrak{g}^*$--valued Chevalley--Eilenberg
cocycle $\phi\in \CE^3(\mathfrak{g},\mathfrak{g}^*)$ 
\footnote{$\vphantom{\dot{\dot{\dot{\dot{x}}}}}$
Recall that the Chevalley--Eilenberg complex $\CE^*(\mathfrak{g},\mathfrak{g}^*)$ of $\mathfrak{g}$ with values in 
$\mathfrak{g}^*$ 
is the graded vector space $\Fun(\mathfrak{g}[1],\mathfrak{g}^*)$ equipped with the 
coboundary operator $\mathcal{Q}_{\CE(\mathfrak{g},\mathfrak{g}^*)}$ defined by
\begin{equation}
\mathcal{Q}_{\CE(\mathfrak{g},\mathfrak{g}^*)}\phi(\pi,\ldots,\pi)
=[\pi,\phi(\pi,\ldots,\pi)]-\frac{p}{2}\phi([\pi,\pi],\pi,\ldots,\pi),
\nonumber
\end{equation}
for a $p$--cochain $\phi\in \CE^p(\mathfrak{g},\mathfrak{g}^*)$ seen as as a linear map
$\phi\in\Hom(\wedge^p\mathfrak{g},\mathfrak{g}^*)$. The associated cohomology is 
$H_{\CE}{}^*(\mathfrak{g},\mathfrak{g}^*)$. 
A  $p$--cochain $\phi\in \CE^p(\mathfrak{g},\mathfrak{g}^*)$ is  cyclic if 
\begin{equation}
\langle x,\phi(y,\pi,\ldots,\pi)\rangle+\langle y,\phi(x,\pi,\ldots,\pi)\rangle=0,
\nonumber
\end{equation}
where $\langle\cdot,\cdot\rangle$ is the duality pairing of $\mathfrak{g}$. 
The cyclic cochain form a subcomplex $\CCE^*(\mathfrak{g},\mathfrak{g}^*)$ of $\CE^*(\mathfrak{g},\mathfrak{g}^*)$ 
with cohomology $H_{\CCE}{}^*(\mathfrak{g},\mathfrak{g}^*)$ isomorphic to 
$H_{\CE}{}^*(\mathfrak{g})[-1]$, the $-1$ degree shifted 
real valued cohomology of $\mathfrak{g}$ \cite{Penkava1995}. The correspondence is defined by
\begin{equation}
\hat\phi(\pi,\ldots,\pi)=\frac{1}{p+1}\langle\pi,\phi(\pi,\ldots,\pi)\rangle
\nonumber
\end{equation}
at the level of representatives. 
(See also
\cite{deAzcarraga:1995jw} for reference.)} $\vphantom{\ul{\ul{x}}}$.
On account of the cyclicity property \eqref{linftyform3}, $\phi$ is cyclic
and, so, 
\begin{equation}
\hat\phi=\frac{1}{4}\langle\pi,[\pi,\pi,\pi]\rangle,
\label{exa1}
\end{equation}
is a Chevalley--Eilenberg cocycle $\phi\in\CE^4(\mathfrak{g})$. $\hat\phi$ is in fact simply related to 
the Chevalley--Eilenberg 
cocycle $\chi_2\in\CE^4(\mathfrak{v})$ of eq. \eqref{e2tchern2}.
\begin{equation}
\chi_2=-\hat\phi/6
\label{exa2}
\end{equation}
Since $\CE^*(\mathfrak{g})$ is a subcomplex of $\CE^*(\mathfrak{v})$ when $\partial=0$
by \eqref{x2tlinalgQ} and \eqref{2tlinalgQa}, $\chi_2$ is exact in $\CE^*(\mathfrak{v})$
if $\hat\phi$ is in $\CE^*(\mathfrak{g})$. In that case, we have $Q_2(g)=0$ and 
there is no level quantization in the associated $\CS_2(N,\mathfrak{v})$ 
Chern--Simons model. If the $4$--cocycle $\hat\phi$ is not a
coboundary, then $Q_2(g)$ may be non trivial and level quantization may
obtain. Now $H_{\CE}{}^4(\mathfrak{g})=0$ for all simple Lie algebras $\mathfrak{g}$.
$H_{\CE}{}^4(\mathfrak{g})\not=0$, e. g. $\mathfrak{g}=\mathfrak{u}(n)$
with $n\geq 2$.

\vfil\eject

\appendix

\section{\normalsize \textcolor{blue}{Lie $2$--group and $2$--algebra theory}}\label{sec:2tLinftytheor}

\hspace{.5cm} 
In the following appendices, we collect various results on $2$--groups
and Lie $2$--algebras and their automorphisms disseminated 
 in the mathematical literature in order to define our terminology and notation and for 
reference throughout in the text. A good introduction to these matters
tailored for higher gauge theoretic applications  
is provided in \cite{Baez:2010ya}.

%\vfil\eject 

\subsection{\normalsize \textcolor{blue}{Strict $2$--groups}}\label{sec:twogr}

\hspace{.5cm} 
The theory of strict $2$--groups is formulated most elegantly in the 
language of higher category theory \cite{Baez5}. Here, we shall limit ourselves to providing
the basic definitions and properties.

{\it Ordinary groups}. 

We recall first the familiar definition of group. 

A group (in delooped form) consists of the following set of data: 
\begin{enumerate}

\item a set of $1$-cells $G$;

\item a composition law of $1$--cells $\circ: G\times G\rightarrow G$;

\item a inversion law of $1$--cells ${}^{-1_\circ}: G\rightarrow G$;

\item a distinguished unit $1$--cell $1\in G$

\end{enumerate}
 These are required to satisfy the following axioms. %\pagebreak 
\begin{subequations}
\label{twogr10}
\begin{align}
&(c\circ b)\circ a=c\circ(b\circ a),
\vphantom{\Big]}
\label{twogr1a0}
\\
&a^{-1_\circ}\circ a=a\circ a^{-1_\circ}=1,
\vphantom{\Big]}
\label{twogr1b0}
\\
&a\circ 1=1\circ a=a,
\vphantom{\Big]}
\label{twogr1c0}
\end{align}
\end{subequations}
where $a,b,c,\dots\in G$. 

{\it Strict $2$--groups}

We provide now the definition of strict $2$--group. 

\hspace{.5cm} A strict $2$--group (in delooped form)
consists of the following set of data: 
\begin{enumerate}

\item a set of $1$-cells $V_1$;

\item a composition law of $1$--cells $\circ: V_1\times V_1\rightarrow V_1$;

\item a inversion law of $1$--cells ${}^{-1_\circ}: V_1\rightarrow V_1$;

\item a distinguished unit $1$--cell $1\in V_1$;

\item for each pair of $1$--cells $a,b\in V_1$, a set of $2$--cells $V_2(a,b)$;

\item for each quadruple of $1$--cells $a,b,c,d\in V_1$, a horizontal composition law of $2$--cells
$\circ:V_2(a,c)\times V_2(b,d)\rightarrow V_2(b\circ a,d\circ c)$;

\item for each pair of $1$--cells $a,b\in V_1$, 
a horizontal inversion law of $2$--cells ${}^{-1_\circ}: V_2(a,b)\rightarrow V_2(a^{-1_\circ},b^{-1_\circ})$;

\item for each triple of $1$--cells $a,b,c\in V_1$, a vertical composition law of $2$--cells
$\bfdot:V_2(a,b)\times V_2(b,c)\rightarrow V_2(a,c)$;

\item for each pair of $1$--cells $a,b\in V_1$, 
a vertical inversion law of $2$--cells ${}^{-1_\bfdot}\!\!: V_2(a,b)\rightarrow V_2(b,a)$;

\item for each $1$--cell $a$, a distinguished %vertical 
unit $2$--cell $1_a\in V_2(a,a)$.
\end{enumerate}
 These are required to satisfy the following axioms. %\pagebreak 
\begin{subequations}
\label{twogr1}
\begin{align}
&(c\circ b)\circ a=c\circ(b\circ a),
\vphantom{\Big]}
\label{twogr1a}
\\
&a^{-1_\circ}\circ a=a\circ a^{-1_\circ}=1,
\vphantom{\Big]}
\label{twogr1b}
\\
&a\circ 1=1\circ a=a,
\vphantom{\Big]}
\label{twogr1c}
%\\
\end{align}
\begin{align}
&(C\circ B)\circ A=C\circ(B\circ A),
\vphantom{\Big]}
\label{twogr1d}
\\
&A^{-1_\circ}\circ A=A\circ A^{-1_\circ}=1_1,
\vphantom{\Big]}
\label{twogr1e}
\\
&A\circ 1_1=1_1\circ A=A,
\vphantom{\Big]}
\label{twogr1f}
\\
&(C\bfdot B)\bfdot A=C\bfdot(B\bfdot A),
\vphantom{\Big]}
\label{twogr1g}
\\
&A^{-1_\bfdot}\!\bfdot A=1_a,\qquad A\bfdot A^{-1_\bfdot}=1_b,
\vphantom{\Big]}
\label{twogr1h}
\\
&A\bfdot 1_a=1_b\bfdot A=A,
\vphantom{\Big]}
\label{twogr1i}
\\
&(D\bfdot C)\circ(B\bfdot A)=(D\circ B)\bfdot(C\circ A).
\vphantom{\Big]}
\label{twogr1j}
\end{align}
\end{subequations}%
Here and in the following, $a,b,c,\dots\in V_1$, $A,B,C,\dots\in V_2$, where 
$V_2$ denotes the set of all $2$-cells. For clarity, we often denote $A\in V_2(a,b)$
as $A:a\Rightarrow b$. 
%In our conventions, if $F:a\Rightarrow c$ and $G:b\Rightarrow d$, 
%then $G\circ F:b\circ a\Rightarrow d\circ b$ and, if $F:a\Rightarrow b$ and $G:b\Rightarrow c$, 
%then $G\bfdot F:a\Rightarrow c$. 
All identities involving the vertical composition and inversion hold whenever defined.  
Relation \eqref{twogr1j} is called interchange law. 
In the following, we shall denote a $2$--group such as the above as $V$ or $(V_1,V_2)$
or $(V_1,V_2,\circ,{}^{-1_\circ},\bfdot,{}^{-1_\bfdot},1_-)$ to emphasize the underlying structure.

$V$ is in fact a one--object
strict $2$--category in which all $1$--morphisms are invertible  and all $2$--morphisms
are both horizontal and vertical invertible, a one--object strict $2$--groupoid. 

If $(V_1,V_2,\circ,{}^{-1_\circ},\bfdot,{}^{-1_\bfdot},1_-)$ is a strict $2$--group, then 
$(V_1,\circ,{}^{-1_\circ},1)$ is an ordinary group and $(V_1,V_2,\bfdot,{}^{-1_\bfdot},1_-)$ is a groupoid. 
Viewing this as a category $V$ having $V_1$, $V_2$  as its collection of objects and morphisms, 
$\circ:V\times V\rightarrow V$ and ${}^{-1_\circ}:V\rightarrow V$ are both functors and 
$V$ turns out to be a strict monoidal category in which every morphism is invertible 
and every object has a strict inverse. 

%\vfil\eject

\subsection{\normalsize \textcolor{blue}{Strict $2$--groups and crossed modules}}\label{sec:crossed}

\hspace{.5cm}  Strict $2$--groups are intimately related to crossed modules. 
A {\it crossed module} \cite{Whitehead:1946} 
consists in the following elements.
\begin{enumerate}

\item a pair of groups $G$, $H$;

\item a group morphism $t:H\rightarrow G$;

\item a group morphism $m:G\rightarrow\Aut(H)$, where
$\Aut(H)$ is the group of automorphisms of $H$.

\end{enumerate}
Further, the following conditions are met.
\begin{subequations}
\label{twogr3}
\begin{align}
&t(m(a)(A))=a t(A)a^{-1},
\vphantom{\Big]}
\label{twogr3a}
\\
&m(t(A))(B)=ABA^{-1},
\vphantom{\Big]}
\label{twogr3b}
\end{align}
\end{subequations}
where $a\in G$, $A,B\in H$.
We shall denote a crossed module such as this
by $(G,H)$ or $(G,H,t,m)$ to explicitly indicate its underlying structure. 

Crossed modules are just another way of describing strict $2$--groups. 
There is in fact a one--to--one 
correspondence between the former and the latter \cite{Brown:1976}.
With any crossed module $(G,H)$, there is associated a strict $2$--group $V$
as follows. 
\begin{enumerate}

\item  $V_1=G$;

\item  %for $a,b\in V_1$, 
$b\circ a=ba$;

\item  %for $a\in V_1$, 
$a^{-1_\circ}=a^{-1}$;

\item $1=1_G$;

\item  %for $a,b\in V_1$, a $2$--cell from $a$ to $b$ is a
$V_2(a,b)$ is the set of pairs $(a,A) \in G\times H$ 
such that $b = t(A)a$;

\item %for $(a,A)\in V_2(a,c)$, $(b,B)\in V_2(b,d)$, 
$(b,B)\circ(a,A)=(ba, Bm(b)(A))$;

\item  %for $(a,A)\in V_2(a,b)$, 
$(a,A)^{-1_\circ}=(a^{-1},m(a^{-1})(A^{-1}))$;

\item  for composable %$(a,A)\in V_2(a,c)$, $(b,B)\in V_2(b,d)$, 
$(a,A)$, $(b,B)$, $(b,B)\bfdot(a,A)=(a, BA)$;

\item  %for $(a,A)\in V_2(a,b)$, 
$(a,A)^{-1_\bfdot}=(t(A)a,A^{-1})$;

\item %for $a\in V_1$, 
$1_a=(a,1_H)$.

\end{enumerate}
%Above, in each expression, the operations of the left hand side refer to 
%$(V_1,V_2)$ and those of the right hand side to $(G,H)$. 
Conversely, with any strict $2$--group $V$ there is associated a crossed module $(G,H)$, 
as follows. 
\begin{enumerate}

\item $G=V_1$;

\item %for $a,b\in G$, 
$ba=b\circ a$;

\item %for $a\in G$, 
$a^{-1}=a^{-1_\circ}$;

\item $1_G=1$;

\item $H$ is the set of all $2$--cells of the form $A:1\Rightarrow a$
for some $a$; %\in V_1$. 

\item %for $A,B\in H$, 
$BA=B\circ A$;

\item %for $A\in H$, 
$A^{-1}=A^{-1_\circ}$;

\item $1_H=1_1$;

\item %for $A\in H$, 
$t(A)=a$ if $A:1\Rightarrow a$. 

\item %for $a\in G$, $A\in A$, 
$m(a)(A) = 1_a \circ A\circ 1_{a^{-1_\circ}}$.
\end{enumerate}
%Above, in each expression, the operations of the left hand side refer to 
%$(G,H)$ and those of the right hand side to $(V_1,V_2)$.

%\vfil\eject

\subsection{\normalsize \textcolor{blue}{Lie $2$--algebras}}\label{sec:linfty}

\hspace{.5cm} In this appendix, we review the notion of Lie $2$--algebra, which is basic in 
the present work. Again, Lie $2$--algebras have an elegant categorical formulation \cite{Baez:2003fs}. 
Here, we shall present them as $2$--term $L_\infty$ algebras, which is an equivalent computationally efficient
description. 

{\it Ordinary Lie algebras} %$\mathfrak{v}$ 

A Lie $2$--algebra consists of the following set of data: 
\begin{enumerate}

\item a vector space $\mathfrak{g}$;

\item a linear map $[\cdot,\cdot]:\mathfrak{g}\wedge \mathfrak{g}\rightarrow \mathfrak{g}$;

\end{enumerate}
This is required to satisfy the following axiom:
\begin{equation}
3[\pi,[\pi,\pi]]=0, 
\label{jacobi}
\end{equation}
where $\pi$ is given by
\begin{equation}
\pi=\pi^a\otimes e_a,
\label{gammadef0}
\end{equation}
$\{e_a\}$ being a basis of $\mathfrak{g}$
and $\{\pi^a\}$ being the basis of $\mathfrak{g}^\vee[1]$
dual to $\{e_a\}$. Here, $\mathfrak{g}^\vee[1]$ is 
the $1$ step degree shifted dual of $\mathfrak{g}$, 
assumed to have degree $0$.
It is immediately verified that \eqref{jacobi} is equivalent to
the familiar Jacobi identity.

{\it Lie algebra Chevalley--Eilenberg cohomology} 

The  Chevalley--Eilenberg algebra 
$\mathrm{CE}(\mathfrak{g})$ of $\mathfrak{g}$ is the graded commutative algebra 
$S(\mathfrak{g}^\vee[1])\simeq \bigwedge^*\mathfrak{g}^\vee$ 
generated by $\mathfrak{g}^\vee[1]$, the $1$ step degree shifted dual of $\mathfrak{g}$.
The Chevalley--Eilenberg differential $\mathcal{Q}_{\mathrm{CE}(\mathfrak{g})}$ 
is the degree $1$ differential defined by
\begin{equation}
\mathcal{Q}_{\mathrm{CE}(\mathfrak{g})}\pi=-\frac{1}{2}[\pi,\pi].
\label{x2tlinalgQ}
\end{equation}
It is immediately verified that $\mathcal{Q}_{\mathrm{CE}(\mathfrak{g})}$ is nilpotent, %\hphantom{xxxxxxxxxxxxxxxxxxxx}
\begin{equation}
\mathcal{Q}_{\mathrm{CE}(\mathfrak{g})}{}^2=0,
\label{xQ2=0}
\end{equation}
as a consequence of \eqref{jacobi}. 
$(\mathrm{CE}(\mathfrak{g}),\mathcal{Q}_{\mathrm{CE}(\mathfrak{g})})$ is so a cochain complex. 
Its cohomology $H_{CE}{}^*(\mathfrak{g})$ is the  Chevalley--Eilenberg cohomology, 
also known as {\it Lie algebra cohomology}, of $\mathfrak{g}$. 

{\it Lie $2$--algebras} %$\mathfrak{v}$ 

A Lie $2$--algebra consists of the following set of data: 
\begin{enumerate}

\item a pair of vector spaces on the same field
$\mathfrak{v}_0,\mathfrak{v}_1$;

\item a linear map $\partial:\mathfrak{v}_1\rightarrow\mathfrak{v}_0$;
%$\xymatrix{0\ar[r]&\mathfrak{v}_1\ar[r]^{\partial}&\mathfrak{v}_0\ar[r]&0}$\!.

\item a linear map $[\cdot,\cdot]:\mathfrak{v}_0\wedge \mathfrak{v}_0\rightarrow \mathfrak{v}_0$;

\item a linear map $[\cdot,\cdot]:\mathfrak{v}_0\otimes \mathfrak{v}_1\rightarrow \mathfrak{v}_1$;

\item a linear map 
$[\cdot,\cdot,\cdot]:\mathfrak{v}_0\wedge \mathfrak{v}_0\wedge \mathfrak{v}_0\rightarrow \mathfrak{v}_1$
\footnote{$\vphantom{\bigg[}$ We denote by $[\cdot,\cdot]$ both 
$2$--argument brackets. It will be clear from context which is which.}.

\end{enumerate}
These are required to satisfy the following axioms:
\begin{subequations}
\label{2tlinalg}
\begin{align}&[\pi,\partial\varPi]-\partial[\pi,\varPi]=0,
\vphantom{\Big]}
\label{2tlinalga}
\\
&[\partial \varPi,\varPi]=0, 
\vphantom{\Big]}
\label{2tlinalgb}
\\
&3[\pi,[\pi,\pi]]-\partial[\pi,\pi,\pi]=0,
\vphantom{\Big]}
\label{2tlinalgc}
\\
%\end{align}
%\eject\noindent
%\begin{align}
&2[\pi,[\pi,\varPi]]-[[\pi,\pi],\varPi]-[\pi,\pi,\partial\varPi]=0,
\vphantom{\Big]}
\label{2tlinalgd}
\\
&4[\pi,[\pi,\pi,\pi]]-6[\pi,\pi,[\pi,\pi]]=0.
\vphantom{\Big]}
\label{2tlinalge}
\end{align}
\end{subequations}
where $\pi$ and $\Pi$ are given by
\begin{subequations}
\label{gammaCdef}
\begin{align}
\pi&=\pi^a\otimes e_a,
\vphantom{\Big]}
\label{gammadef}
\\
\varPi&=\varPi^\alpha\otimes E_\alpha,
\vphantom{\Big]}
\label{Cdef}
\end{align}
\end{subequations}
$\{e_a\}$, $\{E_\alpha\}$ being bases of $\mathfrak{v}_0$, $\mathfrak{v}_1$
and $\{\pi^a\}$, $\{\varPi^\alpha\}$ being the bases of $\mathfrak{v}_0{}^\vee[1]$, 
$\mathfrak{v}_1{}^\vee[2]$ dual to $\{e_a\}$, $\{E_\alpha\}$, respectively. 
Here, $\mathfrak{v}_0{}^\vee[1]$ and $\mathfrak{v}_1{}^\vee[2]$ are
the $1$ and $2$ step degree shifted duals of $\mathfrak{v}_0$, $\mathfrak{v}_1$
assumed to have degree $0$.
We shall denote a Lie $2$--algebra such as the above 
by $\mathfrak{v}$ or, more explicitly, by $(\mathfrak{v}_0,\mathfrak{v}_1)$ or 
$(\mathfrak{v}_0,\mathfrak{v}_1,\partial,[\cdot,\cdot],[\cdot,\cdot,\cdot])$
%$(\mathfrak{v}_0,\mathfrak{v}_1,\partial_{\mathfrak{v}},
%[\cdot,\cdot]_{\mathfrak{v}},[\cdot,\cdot,\cdot]_{\mathfrak{v}})$
to emphasize its underlying structure. 

{\it Lie $2$--algebra Chevalley--Eilenberg cohomology} 

Similarly to ordinary Lie algebras, the  Chevalley--Eilenberg algebra 
$\mathrm{CE}(\mathfrak{v})$ of $\mathfrak{v}$ is the graded commutative algebra 
$S(\mathfrak{v}_0{}^\vee[1]\oplus\mathfrak{v}_1{}^\vee[2])\simeq \bigwedge^*\mathfrak{v}_0{}^\vee
\otimes\bigvee^*\mathfrak{v}_1{}^\vee$ 
generated by $\mathfrak{v}_0{}^\vee[1]\oplus\mathfrak{v}_1{}^\vee[2]$.
The Chevalley--Eilenberg differential $\mathcal{Q}_{\mathrm{CE}(\mathfrak{v})}$ 
is the degree $1$ differential defined by 
\begin{subequations}
\label{2tlinalgQ}
\begin{align}
\mathcal{Q}_{\mathrm{CE}(\mathfrak{v})}\pi&=-\frac{1}{2}[\pi,\pi]+\partial \varPi,
\vphantom{\Big]}
\label{2tlinalgQa}
\\
\mathcal{Q}_{\mathrm{CE}(\mathfrak{v})}\varPi&=-[\pi,\varPi]+\frac{1}{6}[\pi,\pi,\pi].
\vphantom{\Big]}
\label{2tlinalgQb}
\end{align}
\end{subequations}
$\mathcal{Q}_{\mathrm{CE}(\mathfrak{v})}$ turns out to be nilpotent,
\begin{equation}
\mathcal{Q}_{\mathrm{CE}(\mathfrak{v})}{}^2=0,
\label{wQ2=0}
\end{equation}
in virtue of the relations \eqref{2tlinalg}.
$(\mathrm{CE}(\mathfrak{v}),\mathcal{Q}_{\mathrm{CE}(\mathfrak{v})})$ is a so cochain complex. 
The associated Chevalley--Eilenberg cohomology
$H_{CE}{}^*(\mathfrak{v})$ is the Lie $2$--algebra cohomology of $\mathfrak{v}$
generalizing ordinary Lie algebra cohomology.

%\vfil\eject

\subsection{\normalsize \textcolor{blue}{Strict Lie $2$--algebras and differential Lie crossed modules}}
\label{sec:difliecr}

\hspace{.5cm} 
A Lie $2$--algebra $\mathfrak{v}$ is called {\it strict} if its three-- argument bracket
$[\cdot,\cdot,\cdot]$ vanishes identically. 
From\eqref{2tlinalg}, it follows that then $\mathfrak{v}_0$ is an ordinary Lie algebra,
$\mathfrak{v}_1$ is a $\mathfrak{v}_0$ Lie module and $\partial$ is a Casimir for the latter. 

Just as crossed modules provide an equivalent description of strict $2$--groups, 
differential Lie crossed modules furnish an alternative description of strict Lie $2$--algebras.

%Strict Lie $2$--algebras are just differential Lie crossed modules. 

A {\it differential Lie crossed module} \cite{Gerstenhaber:1964}
consists in the following elements.
\begin{enumerate}

\item A pair of Lie algebras $\mathfrak{g}$, $\mathfrak{h}$.

\item A Lie algebra morphism $\tau:\mathfrak{h}\rightarrow\mathfrak{g}$.

\item A Lie algebra morphism $\mu:\mathfrak{g}\rightarrow\mathfrak{der}(\mathfrak{h})$, where
$\mathfrak{der}(\mathfrak{h})$ is the Lie algebra of %outer 
derivations of $\mathfrak{h}$.

\end{enumerate}
Further, the following conditions are verified,
\begin{subequations}
\label{liecross}
\begin{align}
&\tau(\mu(x)(X))=[x,\tau(X)]_{\mathfrak{g}},
\vphantom{\Big]}
\label{liecrossa}
\\
&\mu(\tau(X))(Y)=[X,Y]_{\mathfrak{h}},
\vphantom{\Big]}
\label{liecrossb}
\end{align}
\end{subequations}
where $x\in\mathfrak{g}$, $X,Y\in\mathfrak{h}$.
We shall denote the Lie crossed module 
by $(\mathfrak{g},\mathfrak{h})$ or $(\mathfrak{g},\mathfrak{h},\tau,\mu)$ to explicitly indicate 
its underlying structure. 

There exists a one--to--one
correspondence between strict Lie $2$--algebras and differential Lie crossed modules.
With any differential Lie crossed module $(\mathfrak{g},\mathfrak{h})$, 
there is associated a strict Lie $2$--algebra $\mathfrak{v}$ as follows.
$\vphantom{\ul{\ul{\ul{x}}}}$
\begin{enumerate}

\item $\mathfrak{v}_0=\mathfrak{g}$;

\item $\mathfrak{v}_1=\mathfrak{h}$;

\item $\partial X=\tau(X)$;

\item $[x,y]=[x,y]_{\mathfrak{g}}$; %\pagebreak 

\item $[x,X]=\mu(x)(X)$;

\item $[x,y,z]=0$.

\end{enumerate}
Conversely, with any strict Lie $2$--algebra $\mathfrak{v}$, 
there is associated a differential Lie crossed module $(\mathfrak{g},\mathfrak{h})$ 
as follows.  
\begin{enumerate}

\item $\mathfrak{g}=\mathfrak{v}_0$;

\item $\mathfrak{h}=\mathfrak{v}_1$;

\item $[x,y]_{\mathfrak{g}}=[x,y]$;

\item $[X,Y]_{\mathfrak{h}}=[\partial X,Y]$;

\item $\tau(X)=\partial X$;

\item $\mu(x)(X)=[x,X]$.

\end{enumerate}

\vspace{.5cm}
%$x,y,\ldots\in\mathfrak{g}$, $X,Y,\dots\in\mathfrak{h}$.

%\vfil\eject

\subsection{\normalsize \textcolor{blue}{Strict Lie $2$--groups and their algebras}}
\label{subsec:strict}

\hspace{.5cm} A group $G$ is Lie if the set of $1$--cells $G$ is a smooth 
manifold and the multiplication and inversion of $G$ are smooth functions.

With any Lie group $G$, there is associated a Lie algebra 
$\mathfrak{g}$. $\mathfrak{g}$ is the tangent space to $G$ at $1$. 
The brackets of $\mathfrak{g}$ are defined by the relations 
\begin{equation}
[x,y]=\frac{\partial}{\partial s}\Big(\frac{\partial}{\partial t}a(s)^{-1_\circ}
\circ b(t)^{-1_\circ}\circ a(s) \circ b(t)\Big|_{t=0}\Big)\Big|_{s=0},
\label{strbraa0}
\end{equation}
where $x,y\in\mathfrak{g}$ and $a(t)$, $b(t)$ are curves in $G$ such that 
$a(0)=1$, $da(0)/dt=x$, $b(0)=1$, $db(0)/dt=y$. 
There is a natural exponential map $\exp:\mathfrak{g}\rightarrow G$. 

Similarly, a strict $2$--group $V$ is Lie if the sets of $1$-- and $2$--cells $V_1$, $V_2$ are smooth 
manifolds and the multiplication and inversion of $V_1$ and the horizontal and vertical 
multiplication and inversion of $V_2$ as well as the source and target maps of $V_2$ 
are all smooth functions.

With any strict Lie $2$--group $V$, there is associated a strict Lie $2$--algebra 
$\mathfrak{v}$ as follows. $\mathfrak{v}_0$ is the tangent space to $V_1$ at $1$;
$\mathfrak{v}_1$ is the tangent space to $V_2{}^*=\cup_{a\in V_1}V_2(1,a)$ at $1_1$. 
The brackets and the boundary map of $\mathfrak{v}$ are defined by the relations 
\begin{subequations}
\label{strbra}
\begin{align}
&[x,y]=\frac{\partial}{\partial s}\Big(\frac{\partial}{\partial t}a(s)^{-1_\circ}
\circ b(t)^{-1_\circ}\circ a(s) \circ b(t)\Big|_{t=0}\Big)\Big|_{s=0},
\vphantom{\Big]}
\label{strbraa}
\\
&[x,X]=\frac{\partial}{\partial s}\Big(\frac{\partial}{\partial t}1_{a(s)}\circ A(t) \circ 
1_{a(s)^{-1_\circ}}\Big|_{t=0}\Big)\Big|_{s=0},
\vphantom{\Big]}
\label{strbrab}
\\
&\partial X=\frac{d}{ds}t(A(s))\Big|_{s=0},
\vphantom{\Big]}
\label{strbrac}
\\
&[x,y,z]=0. 
\vphantom{\Big]}
\label{strbrad}
\end{align}
\end{subequations} 
where $x,y\in\mathfrak{v}_0$ and $X\in\mathfrak{v}_1$, 
$a(t)$, $b(t)$ are curves in $V_1$ such that $a(0)=1$, $da(0)/dt=x$, $b(0)=1$, $db(0)/dt=y$
and $A(t)$ is a curve in $V_2{}^*$ such that $A(0)$ $=1_1$, $dA(0)/dt=X$ and $t$ is the target map
of $V_2$. 

The relation between a strict Lie $2$--group $V$ and and its strict Lie $2$--algebra $\mathfrak{v}$ 
can be phrased in more conventional Lie theoretic terms if we view $V$ as a Lie crossed module $(G,H)$
(cf. app. \ref{sec:crossed}).  
Then, $\mathfrak{v}$ can correspondingly 
be viewed the differential Lie crossed module $(\mathfrak{g},\mathfrak{h})$ (cf. app. \ref{sec:difliecr}), where 
$\mathfrak{g}$, $\mathfrak{h}$ are the Lie algebras of $G$, $H$, respectively, and 
\begin{align}
&\tau(X)%=\dot t(X)
=\frac{d t(C(v))}{dv}\Big|_{v=0},
\vphantom{\Big]}
\label{st2grst2lie1}
\\
&\mu(x)(X)%=\frac{d\dot m(c(u)))(X)}{du}\Big|_{u=0},
=\frac{\partial}{\partial u}\Big(\frac{\partial m(c(u))(C(v))}{\partial v}\Big|_{v=0}\Big)|_{u=0},
\vphantom{\Big]}
\label{st2grst2lie3}
\end{align} 
where $x\in\mathfrak{g}$, $X\in\mathfrak{h}$, $c(u)$ is any curve in $G$ 
such that $c(u)\big|_{u=0}=1_G$ and $dc(u)/du\big|_{u=0}=x$ and  $C(v)$ is any curve 
in $H$ such that $C(v)\big|_{v=0}=1_H$ and $dC(v)/dv\big|_{v=0}=X$. 
A natural exponential map $\exp:\mathfrak{v}\rightarrow V$ is defined
in terms of the customary exponential maps 
$\exp:\mathfrak{g}\rightarrow G$, $\exp:\mathfrak{h}\rightarrow H$.

%\vfil\eject 

\subsection{\normalsize \textcolor{blue}{The Lie $2$--algebra
automorphism group }}\label{sec:linftyauto}

\hspace{.5cm} 
Automorphisms of a Lie algebra or a Lie $2$--algebra 
provide structural information and play a basic role in gauge and semistrict 
higher gauge theory as formulated in this paper.

\hspace{.5cm} {\it Automorphisms of an ordinary Lie algebra}

Let $\mathfrak{g}$ be a Lie algebra.
A Lie algebra automorphism of $\mathfrak{g}$ consists of the following datum:
\begin{enumerate}

\item a vector space automorphism  $\phi:\mathfrak{g}\rightarrow\mathfrak{g}$;

\end{enumerate}
which is required to satisfy the following relation:
\begin{equation}
\phi([\pi,\pi])-[\phi(\pi),\phi(\pi)]=0.
\vphantom{\Big]}
\label{mor2tlinalgb0}
\end{equation}
The set $\Aut(\mathfrak{g})$ of all automorphisms of $\mathfrak{g}$ is a group for 
the operations and unit
\begin{subequations}
\label{mor3tlinalg0}
\begin{align}
&\psi\circ\phi(\pi)=\psi\phi(\pi),
\vphantom{\Big]}
\label{mor3tlinalga0}
\\
&\phi^{-1_\circ}(\pi)=\phi^{-1}(\pi), 
\vphantom{\Big]}
\label{mor3/2tlinalgd0}
\\
&\id(\pi)=\pi.
\vphantom{\Big]}
\label{mor3tlinalgg0}
\end{align}
\end{subequations} 
$\Aut(\mathfrak{g})$ is a Lie group. 

{\it Automorphisms of a Lie $2$--algebra}

Let $\mathfrak{v}$ be a Lie $2$--algebra. A Lie $2$--algebra
$1$--automorphism of $\mathfrak{v}$ consists of the following data:
\begin{enumerate}

\item a vector space automorphism  $\phi_0:\mathfrak{v}_0\rightarrow\mathfrak{v}_0$;

\item a vector space automorphism $\phi_1:\mathfrak{v}_1\rightarrow\mathfrak{v}_1$; 

\item a vector space morphism $\phi_2:\mathfrak{v}_0\wedge\mathfrak{v}_0\rightarrow\mathfrak{v}_1$. 

\end{enumerate}
These are required to satisfy the following relations:
\begin{subequations}
\label{mor2tlinalg}
\begin{align}
&\phi_0(\partial \varPi)-\partial\phi_1(\varPi)=0,
\vphantom{\Big]}
\label{mor2tlinalga}
\\
&\phi_0([\pi,\pi])-[\phi_0(\pi),\phi_0(\pi)]-\partial\phi_2(\pi,\pi)=0,
\vphantom{\Big]}
\label{mor2tlinalgb}
\\
&\phi_1([\pi,\varPi])-[\phi_0(\pi),\phi_1(\varPi)]-\phi_2(\pi,\partial \varPi)=0,
\vphantom{\Big]}
\label{mor2tlinalgc}
\\
&3[\phi_0(\pi),\phi_2(\pi,\pi)]+3\phi_2(\pi,[\pi,\pi])
\vphantom{\Big]}
\label{mor2tlinalgd}
\\
&\hspace{3cm}+[\phi_0(\pi),\phi_0(\pi),\phi_0(\pi)]-\phi_1([\pi,\pi,\pi])=0.  
\vphantom{\Big]}
\nonumber
\end{align}
\end{subequations}
%where $x,y,z\in\mathfrak{v}_0$, $X\in\mathfrak{v}_1$. 
In the following, we shall denote a $1$--morphism 
such as the above one by $\phi$ or, more explicitly, by 
$(\phi_0,\phi_1,\phi_2)$ to emphasize its constituent components. 
We shall denote the set of all $1$--automorphisms of $\mathfrak{v}$ 
by $\Aut_1(\mathfrak{v})$.

For any two Lie $2$--algebra $1$--automorphisms $\phi,\psi$, 
a {\it Lie $2$--algebra $2$--auto\-morphism} from $\phi$ to $\psi$ 
consists of a single datum:
\begin{enumerate}

\item a linear map $\varPhi:\mathfrak{v}_0\rightarrow\mathfrak{v}_1$. 

\end{enumerate}
This must satisfy the following relations
\begin{subequations}
\label{mor0tlinalg}
\begin{align}
&\phi_0(\pi)-\psi_0(\pi)-\partial\varPhi(\pi)=0,
\vphantom{\Big]}
\label{mor0tlinalga}
\\
&\phi_1(\varPi)-\psi_1(\varPi)-\varPhi(\partial \varPi)=0,
\vphantom{\Big]}
\label{mor0tlinalgb}
\\
&\phi_2(\pi,\pi)-\psi_2(\pi,\pi)+[\phi_0(\pi)+\psi_0(\pi),\varPhi(\pi)]
%+[\partial\varPhi(\pi),\varPhi(\pi)]
-\varPhi([\pi,\pi])=0.
\vphantom{\Big]}
\label{mor0tlinalgc}
\end{align}
\end{subequations}  %
We shall write a $2$--automorphism such as this as $\varPhi$ or as $\varPhi:\phi\Rightarrow\psi$
to emphasize its source and target.  
We shall denote the set of all $2$--automorphisms 
$\varPhi:\phi\Rightarrow\psi$ by $\Aut_2(\mathfrak{v})(\phi,\psi)$
and the set of all $2$--automorphisms $\varPhi$ by $\Aut_2(\mathfrak{v})$. 

$\Aut_1(\mathfrak{v})$, $\Aut_2(\mathfrak{v})$ are the sets of $1$-- and $2$--cells
of a strict $2$--group $\Aut(\mathfrak{v})$ for the operations and units 
\begin{subequations}
\label{mor3tlinalg}
\begin{align}
&\psi\circ \phi_0(\pi)=\psi_0\phi_0(\pi),
\vphantom{\Big]}
\label{mor3tlinalga}
\\
&\psi\circ \phi_1(\varPi)=\psi_1\phi_1(\varPi),
\vphantom{\Big]}
\label{mor3tlinalgb}
\\
&\psi\circ \phi_2(\pi,\pi)=\psi_1\phi_2(\pi,\pi)+\psi_2(\phi_0(\pi),\phi_0(\pi)),
\vphantom{\Big]}
\label{mor3tlinalgc}
\\
&\phi^{-1_\circ}{}_0(\pi)=\phi_0{}^{-1}(\pi), \hspace{4.05cm}
\vphantom{\Big]}
\label{mor3/2tlinalgd}
\\
&\phi^{-1_\circ}{}_1(\varPi)=\phi_1{}^{-1}(\varPi), 
\vphantom{\Big]}
\label{mor3/2tlinalge}
\\
&\phi^{-1_\circ}{}_2(\pi,\pi)=-\phi_1{}^{-1}\phi_2(\phi_0{}^{-1}(\pi),\phi_0{}^{-1}(\pi)).
\vphantom{\Big]}
\label{mor3/2tlinalgf}
\\
&\id_0(\pi)=\pi,
\vphantom{\Big]}
\label{mor3tlinalgg}
\\
&\id_1(\varPi)=\varPi,
\vphantom{\Big]}
\label{mor3tlinalgh}
\\
&\id_2(\pi,\pi)=0,
\vphantom{\Big]}
\label{mor3tlinalgi}
\\
&\varPsi\circ \varPhi(\pi)=\varPsi\lambda_0(\pi)+\psi_1\varPhi(\pi)=\varPsi\mu_0(\pi)+\phi_1\varPhi(\pi),
\vphantom{\Big]}
\label{mor4tlinalga}
\\
&\varPhi^{-1_\circ}(\pi)=-\lambda_1{}^{-1}\varPhi\mu_0{}^{-1}(\pi)=-\mu_1{}^{-1}\varPhi\lambda_0{}^{-1}(\pi),
\vphantom{\Big]}
\label{mor4/1tlinalgb}
\\
&\varLambda\bfdot \varTheta(\pi)=\varTheta(\pi)+\varLambda(\pi),
\vphantom{\Big]}
\label{mor4tlinalgb}
\\
&\varTheta^{-1_\bfdot}(\pi)=-\varTheta(\pi),
\vphantom{\Big]}
\label{mor4/1tlinalgd}
\\
&\mathrm{Id}_\phi(\pi)=0.
\vphantom{\Big]}
\label{mor4tlinalgc}
\end{align}
\end{subequations} 
where $\varPhi:\lambda\Rightarrow\mu$, $\varPsi:\phi\Rightarrow\psi$, 
$\varTheta:\rho\Rightarrow\sigma$, $\varLambda:\sigma\Rightarrow\tau$.

The strict $2$--group $\Aut(\mathfrak{v})$ can be described as a crossed module. 
The two groups underlying it are $\Aut_1(\mathfrak{v})$, $\Aut_2{}^*(\mathfrak{v})=
\cup_{\phi\in\Aut_1(\mathfrak{v})}\Aut_2(\mathfrak{v})(\id,\phi)=\{\varPhi\,|\,\varPhi\in\Hom(\mathfrak{v}_0,
\mathfrak{v}_1),\,\text{with}\,1_{\mathfrak{v}_0}-\partial\varPhi\in \GL(\mathfrak{v}_0),\,
1_{\mathfrak{v}_1}-\varPhi\partial\in \GL(\mathfrak{v}_1)\}$.
The crossed module operations are as follows, \pagebreak 
\begin{subequations}
\label{cmmor3tlinalg}
\begin{align}
&\psi\circ \phi_0(\pi)=\psi_0\phi_0(\pi),
\vphantom{\Big]}
\label{cmmor3tlinalga}
\\
&\psi\circ \phi_1(\varPi)=\psi_1\phi_1(\varPi),
\vphantom{\Big]}
\label{cmmor3tlinalgb}
\\
&\psi\circ \phi_2(\pi,\pi)=\psi_1\phi_2(\pi,\pi)+\psi_2(\phi_0(\pi),\phi_0(\pi)),
\vphantom{\Big]}
\label{cmmor3tlinalgc}
\\
&\phi^{-1_\circ}{}_0(\pi)=\phi_0{}^{-1}(\pi), \hspace{4.05cm}
\vphantom{\Big]}
\label{cmmor3/2tlinalgd}
\\
&\phi^{-1_\circ}{}_1(\varPi)=\phi_1{}^{-1}(\varPi), 
\vphantom{\Big]}
\label{cmmor3/2tlinalge}
\\
&\phi^{-1_\circ}{}_2(\pi,\pi)=-\phi_1{}^{-1}\phi_2(\phi_0{}^{-1}(\pi),\phi_0{}^{-1}(\pi)).
\vphantom{\Big]}
\label{cmmor3/2tlinalgf}    
\\
&\id_0(\pi)=\pi,
\vphantom{\Big]}
\label{cmmor3tlinalgg}
\\
&\id_1(\varPi)=\varPi,
\vphantom{\Big]}
\label{cmmor3tlinalgh}
\\
&\id_2(\pi,\pi)=0,
\vphantom{\Big]}
\label{cmmor3tlinalgi}
\\
&\varPsi\circ\varPhi(\pi)=\varPsi(\pi)+\varPhi(\pi)-\varPsi\partial\varPhi(\pi),
\vphantom{\Big]}
\label{cmmor4tlinalga}
\\
&\varPhi^{-1_\circ}(\pi)=-\varPhi(1_{\mathfrak{v}_0}-\partial\varPhi)^{-1}(\pi)
=-(1_{\mathfrak{v}_1}-\varPhi\partial)^{-1}\varPhi(\pi),
\vphantom{\Big]}
\label{cmmor4tlinalgb}
\\
&\mathrm{Id}_\phi(\pi)=0.   
\vphantom{\Big]}
\label{cmmor4tlinalgc}
\\
&t(\varPhi)_0(\pi)=(1_{\mathfrak{v}_0}-\partial\varPhi)(\pi)
\vphantom{\Big]}
\label{cmmor4tlinalgd}
\\
&t(\varPhi)_1(\varPi)=(1_{\mathfrak{v}_1}-\varPhi\partial)(\varPi)
\vphantom{\Big]}
\label{cmmor4tlinalge}
\\
&t(\varPhi)_2(\pi,\pi)=2[\pi,\varPhi(\pi)]-[\partial\varPhi(\pi),\varPhi(\pi)]-\varPhi([\pi,\pi]),
\vphantom{\Big]}
\label{cmmor4tlinalgf}
\\
&m(\phi)(\varPhi)(\pi)=\phi_1\varPhi\phi_0{}^{-1}(\pi).
\vphantom{\Big]}
\label{cmmor4tlinalgg}
\end{align}
\end{subequations} 

$\Aut(\mathfrak{v})$ is a strict Lie $2$--group. 

%\vfil\eject

\subsection{\normalsize \textcolor{blue}{The derivation  Lie $2$--algebra}}
\label{sec:linftyinfaut}

\hspace{.5cm} Derivations of a Lie algebra or a Lie $2$--algebra play an important role 
because of the structural information they provide and the constructive applications they have. 

{\it The derivation Lie algebra}

Let $\mathfrak{g}$ be an ordinary Lie algebra.
An element $\alpha$ of $\mathfrak{aut}(\mathfrak{g})$, a derivation  
of $\mathfrak{g}$,  is 

\begin{enumerate}

\item a vector space morphism $\alpha:\mathfrak{g}\rightarrow\mathfrak{g}$

\end{enumerate}
with the property that 
\begin{equation}
\alpha([\pi,\pi])-[\alpha(\pi),\pi]-[\pi,\alpha(\pi)]=0,
\label{mor5tlinalgb0}
\end{equation}
With the brackets %of $\mathfrak{aut}(\mathfrak{g})$ 
\begin{equation}
[\alpha,\beta]_\circ (\pi)=\alpha\beta(\pi)-\beta\alpha(\pi), 
\label{mor5tlinalgb0/1}
\end{equation}
$\mathfrak{aut}(\mathfrak{g})$ is the Lie algebra, in fact that associated with the Lie group $\Aut(\mathfrak{g})$
of $\mathfrak{g}$--automorphisms, as suggested by the notation (cf. subsect. \ref{sec:linftyauto}). 

{\it Lie algebra adjoint action} 

For any $x\in\mathfrak{g}$,
the mapping
\begin{equation}
\ad x(\pi)=[x,\pi]
\label{ad}
\end{equation}
defines a derivation $\ad x\in \mathfrak{aut}(\mathfrak{g})$, the adjoint of $x$.

{\it Lie algebra exponential map}

The exponential map $\exp_\circ :\mathfrak{aut}(\mathfrak{g})\rightarrow \Aut(\mathfrak{g})$ is defined 
as expected. For $\alpha\in \mathfrak{aut}(\mathfrak{g})$, $\exp_\circ (\alpha)\in \Aut(\mathfrak{g})$ 
is given by
\begin{equation}
\exp_\circ (\alpha)(\pi)=\exp(\alpha)(\pi).
\label{exp00}
\end{equation}
If $G$ is a Lie group with Lie algebra $\mathfrak{g}$, we have 
\begin{equation}
\exp_\circ (\ad x)(\pi)=\Ad\exp(x)(\pi)
\label{exp00/1}
\end{equation}
for $x\in\mathfrak{g}$, where in the right hand side $\exp:\mathfrak{g}\rightarrow G$ 
is the usual Lie theoretic exponential map.

{\it The derivation Lie $2$--Lie algebra}

Let $\mathfrak{v}$ be a Lie $2$--algebra.
The derivation strict Lie $2$--Lie algebra $\mathfrak{aut}(\mathfrak{v})$ 
of $\mathfrak{v}$ is described as follows. 

An element of $\alpha$
of $\mathfrak{aut}_0(\mathfrak{v})$, a $1$--derivation,  consists of three mappings.
\begin{enumerate}

\item a vector space morphism $\alpha_0:\mathfrak{v}_0\rightarrow\mathfrak{v}_0$;

\item a vector space morphism $\alpha_1:\mathfrak{v}_1\rightarrow\mathfrak{v}_1$;

\item a vector space morphism $\alpha_2:\mathfrak{v}_0\wedge\mathfrak{v}_0\rightarrow\mathfrak{v}_1$.

\end{enumerate} 
These must satisfy the following relations:
\begin{subequations}
\label{mor5tlinalg}
\begin{align}
&\alpha_0(\partial \varPi)-\partial\alpha_1(\varPi)=0,
\vphantom{\Big]}
\label{mor5tlinalga}
\\
&\alpha_0([\pi,\pi])-[\alpha_0(\pi),\pi]-[\pi,\alpha_0(\pi)]-\partial\alpha_2(\pi,\pi)=0,
\hspace{1cm}
\vphantom{\Big]}
\label{mor5tlinalgb}
\\
&\alpha_1([\pi,\varPi])-[\alpha_0(\pi),\varPi]-[\pi,\alpha_1(\varPi)]-\alpha_2(\pi,\partial \varPi)=0,
\vphantom{\Big]}
\label{mor5tlinalgc}
\\
&3[\pi,\alpha_2(\pi,\pi)]+3\alpha_2(\pi,[\pi,\pi])
\vphantom{\Big]}
\label{mor5tlinalgd}
\\
&\hspace{4cm}+3[\pi,\pi,\alpha_0(\pi)]-\alpha_1([\pi,\pi,\pi])=0.
\vphantom{\Big]}
\nonumber
\end{align}
\end{subequations}
%where $x,y,z\in\mathfrak{v}_0$, $\varPi\in\mathfrak{v}_1$. 
%As usual, we shall conveniently denote a Lie algebra ele\-ment like this
%by $\alpha$ or, more explicitly, by $(\alpha_0,\alpha_1,\alpha_2)$.

An element of $\varGamma$ of 
$\mathfrak{aut}_1(\mathfrak{v})$, a $2$--derivation, consists of a single mapping. 
\begin{enumerate}

\item a vector space morphism $\varGamma:\mathfrak{v}_0\rightarrow\mathfrak{v}_1$.

\end{enumerate} 
No restrictions are imposed on it. 

The boundary map and the brackets of $\mathfrak{aut}(\mathfrak{v})$ 
%$[\alpha,\beta]_\circ $ of $\alpha,\beta\in \mathfrak{aut}(\mathfrak{v})$ 
are given by the expressions 
\begin{subequations}
\label{mor7tlinalg}
\begin{align}
&\partial_\circ \varGamma_0(\pi)=-\partial\varGamma(\pi), \hspace{6.1cm}
\vphantom{\Big]}
\label{mor7tlinalgx}
\\
&\partial_\circ \varGamma_1(\varPi)=-\varGamma(\partial \varPi),
\vphantom{\Big]}
\label{mor7tlinalgy}
\\
&\partial_\circ \varGamma_2(\pi,\pi)=2[\pi,\varGamma(\pi)]-\varGamma([\pi,\pi]),
\vphantom{\Big]}
\label{mor7tlinalgz}
\\
&[\alpha,\beta]_{\circ 0}(\pi)=\alpha_0\beta_0(\pi)-\beta_0\alpha_0(\pi), 
\vphantom{\Big]}
\label{mor7tlinalga}
\\
&[\alpha,\beta]_{\circ 1}(\varPi)=\alpha_1\beta_1(\varPi)-\beta_1\alpha_1(\varPi),
\vphantom{\Big]}
\label{mor7tlinalgb}
\\
&[\alpha,\beta]_{\circ 2}(\pi,\pi)=\alpha_1\beta_2(\pi,\pi)+2\alpha_2(\beta_0(\pi),\pi)
\vphantom{\Big]}
\label{mor7tlinalgc}
\\
&\qquad\qquad\qquad\qquad\qquad -\beta_1\alpha_2(\pi,\pi)-2\beta_2(\alpha_0(\pi),\pi),
\vphantom{\Big]}
\nonumber
\\
&[\alpha,\varGamma]_\circ (\pi)=\alpha_1\varGamma(\pi)-\varGamma\alpha_0(\pi),
\vphantom{\Big]}
\label{mor7tlinalgv}
\\
&[\alpha,\beta,\gamma]_\circ (\pi)=0.
\vphantom{\Big]}
\label{mor7tlinalgw}
\end{align}
\end{subequations}
%where $x,y\in\mathfrak{v}_0$, $\varPi\in\mathfrak{v}_1$. 
Relations  \eqref{mor5tlinalg} ensure that the basic relations \eqref{2tlinalg}
are satisfied by the above boundary and brackets.

The strict Lie $2$--algebra $\mathfrak{aut}(\mathfrak{v})$ can be described as a differential Lie 
crossed module. The two Lie algebras underlying it are $\mathfrak{aut}_0(\mathfrak{v})$, 
$\mathfrak{aut}_1(\mathfrak{v})$.
The differential Lie crossed module operations are as follows,
\begin{subequations}
\label{dlcmmor7tlinalg}
\begin{align}
&[\alpha,\beta]_{\circ 0}(\pi)=\alpha_0\beta_0(\pi)-\beta_0\alpha_0(\pi), 
\vphantom{\Big]}
\label{dlcmmor7tlinalga}
\\
&[\alpha,\beta]_{\circ 1}(\varPi)=\alpha_1\beta_1(\varPi)-\beta_1\alpha_1(\varPi),
\vphantom{\Big]}
\label{dlcmmor7tlinalgb}
\\
&[\alpha,\beta]_{\circ 2}(\pi,\pi)=\alpha_1\beta_2(\pi,\pi)+2\alpha_2(\beta_0(\pi),\pi)
\vphantom{\Big]}
\label{dlcmmor7tlinalgc}
\\
&\qquad\qquad\qquad\qquad\qquad -\beta_1\alpha_2(\pi,\pi)-2\beta_2(\alpha_0(\pi),\pi),
\vphantom{\Big]}
\nonumber
\\
&[\varGamma,\varDelta]_\circ (\pi)=-\varGamma\partial\varDelta(\pi)+\varDelta\partial\varGamma(\pi),
\vphantom{\Big]}
\label{dlcmmor7tlinalgw}
\\
&\tau_\circ (\varGamma)_0(\pi)=-\partial\varGamma(\pi), \hspace{6.1cm}
\vphantom{\Big]}
\label{dlcmmor7tlinalgx}
\\
&\tau_\circ (\varGamma)_1(\varPi)=-\varGamma(\partial \varPi),
\vphantom{\Big]}
\label{dlcmmor7tlinalgy}
\\
&\tau_\circ (\varGamma)_2(\pi,\pi)=2[\pi,\varGamma(\pi)]-\varGamma([\pi,\pi]),
\vphantom{\Big]}
\label{dlcmmor7tlinalgz}
\\
&\mu_\circ (\alpha)(\varGamma)(\pi)=\alpha_1\varGamma(\pi)-\varGamma\alpha_0(\pi),
\vphantom{\Big]}
\label{dlcmmor7tlinalgv}
\end{align}
\end{subequations}
$\mathfrak{aut}(\mathfrak{v})$ is the strict Lie $2$--algebra associated with the strict Lie $2$--group 
$\Aut(\mathfrak{g})$ of $\mathfrak{v}$--automorphisms, as indicated  by the notation (cf. subsect. \ref{sec:linftyauto}). 

For any Lie $2$--algebra, $\Aut(\mathfrak{v})$ is a strict Lie $2$--group. 
Its associated strict Lie $2$--algebra
is $\mathfrak{aut}(\mathfrak{v})$ (cf. subsect. \ref{subsec:strict}).

{\it Lie $2$--algebra adjoint action}

For any $x\in\mathfrak{v}_0$, the mappings 
\begin{subequations}
\label{ad0}
\begin{align}
&\ad x_0(\pi)=[x,\pi],
\vphantom{\Big]}
\label{ada}
\\
&\ad x_1(\varPi)=[x,\varPi],
\vphantom{\Big]}
\label{adb}
\\
&\ad x_2(\pi,\pi)=[x,\pi,\pi]
\vphantom{\Big]}
\label{adc}
\end{align}
\end{subequations} 
define an element $\ad x\in\mathfrak{aut}_0(\mathfrak{v})$, the adjoint of $x$.
Furthermore, for any $x,y\in\mathfrak{v}_0$ and any $X\in\mathfrak{v}_1$, the mappings 
\begin{subequations}
\label{ad1}
\begin{align}
&\ad x\wedge y(\pi)=[x,y,\pi],
\vphantom{\Big]}
\label{ade}
\\
&\ad X(\pi)=[\pi,X]
\vphantom{\Big]}
\label{add}
\end{align}
\end{subequations} 
define two elements $\ad x\wedge y,\,\ad X\in\mathfrak{aut}_1(\mathfrak{v})$, the adjoints of $x,y$ and $X$.

{\it Lie $2$--algebra exponential map}
 
The exponential map $\exp_\circ 
:\mathfrak{aut}(\mathfrak{v})\rightarrow \Aut(\mathfrak{v})$ can be described rather explicitly.
For $\alpha\in\mathfrak{aut}_0(\mathfrak{v})$, $\varGamma\in \mathfrak{aut}_1(\mathfrak{v})$, 
$\exp_\circ (\alpha)\in \Aut_1(\mathfrak{v})$, 
$\exp_\circ (\varGamma)\in \Aut_2{}^*(\mathfrak{v})$ are given 
by the expressions
\begin{subequations}
\label{exp0}
\begin{align}
&\exp_\circ (\alpha)_0(\pi)=\exp(\alpha_0)(\pi),
\vphantom{\Big]}
\label{exp0a}
\\
&\exp_\circ (\alpha)_1(\varPi)=\exp(\alpha_1)(\varPi),
\vphantom{\Big]}
\label{exp0b}
\\
&\exp_\circ (\alpha)_2(\pi,\pi)
\vphantom{\Big]}
\label{exp0c}
\\
&\hspace{2cm}
=\int_0^1dt\,\exp((1-t)\alpha_1)\alpha_2\big(\exp(t\alpha_0)(\pi),\exp(t\alpha_0)(\pi)\big),
\vphantom{\Big]}
\nonumber
\\
&\exp_\circ (\varGamma)(\pi)
=\frac{1_{\mathfrak{v}_1}-\exp(-\varGamma\partial)}{\varGamma\partial}\,\varGamma(\pi)
=\varGamma\,\frac{1_{\mathfrak{v}_0}-\exp(-\partial\varGamma)}{\partial\varGamma}(\pi)
%=\Ein'(\varGamma\partial)\varGamma(\pi)=\varGamma\Ein'(\partial\varGamma)(\pi).
\vphantom{\Big]}
\label{exp0d}
\end{align}
\end{subequations}
%where $\Ein'(t)$ is the entire function $(1-\exp(-t))/t$.
%The expression in the right hand side is defined by the Taylor series of the numerical function
%$e(t)=(1-\exp(-t))/t$ with $t$ replaced by the endomorphism $\varGamma\partial$. 

The above expressions can be made more explicit in the case where $\mathfrak{v}$ 
is a strict Lie $2$--algebra corresponding to the differential Lie crossed module 
$(\mathfrak{g},\mathfrak{h})$ of a Lie crossed module $(G,H)$ (cf. app. \ref{subsec:strict}), 
\begin{subequations}
\label{crexp0}
\begin{align}
&\exp_\circ (\ad x)_0(\pi)=\Ad\exp(x)(\pi),
\vphantom{\Big]}
\label{crexp0a}
\\
&\exp_\circ (\ad x)_1(\varPi)=\dot m(\exp(x))(\varPi),
\vphantom{\Big]}
\label{crexp0b}
\\
&\exp_\circ (\ad x)_2(\pi,\pi)=0
\vphantom{\Big]}
\label{crexp0c}
\\
&\exp_\circ (\ad X)(\pi)=Q(\pi,\exp(X))
\vphantom{\Big]}
\label{crexp0d}
\end{align}
\end{subequations}
for $x\in\mathfrak{g}$, $X\in\mathfrak{h}$, where,  
for $a\in G$, $A\in H$, $x\in\mathfrak{g}$, $X\in\mathfrak{h}$,
$\dot m(a)(X)\in\mathfrak{h}$ and $Q(x,A)\in\mathfrak{h}$ 
are defined by 
%\begin{subequations}
%\label{crexp1}
\begin{align}
&\dot m(a)(X)=\frac{d}{dv}m(a)(C(v))\Big|_{v=0}
\vphantom{\Big]}
\label{crexp1a}
\\
&Q(x,A)=\frac{d}{du}m(c(u))(A)A^{-1}\Big|_{u=0}, 
\vphantom{\Big]}
\label{crexp1b}
\end{align}
%\end{subequations}
with $c(u)$ being a curve in $G$ 
such that $c(u)\big|_{u=0}=1_G$ and $dc(u)/du\big|_{u=0}=x$ and  $C(v)$ being a curve 
in $H$ such that $C(v)\big|_{v=0}=1_H$ and $dC(v)/dv\big|_{v=0}=X$. 

%\vfil\eject

\subsection{\normalsize \textcolor{blue}{Balanced Lie $2$--algebras}}\label{sec:linftybal}

\hspace{.5cm} Balanced Lie $2$--algebras play a major role in the construction 
higher Chern--Simons theory. 
The notion of balancement has non counterpart in ordinary Lie algebra theory.

{\it Balanced Lie $2$--algebras}

A Lie $2$--algebra $\mathfrak{v}$ is said balanced if $\dim\mathfrak{v}_0=\dim\mathfrak{v}_1$.

For any non balanced Lie $2$--algebra $\mathfrak{v}$, there exists a
balanced Lie $2$--algebra $\mathfrak{v}^\sim$ minimally extending
$\mathfrak{v}$. By this, we mean:
\begin{enumerate}

\item $\mathfrak{v}$ is contained in $\mathfrak{v}^\sim$;

\item $\dim\mathfrak{v}^\sim$ is minimal;

\item $\mathfrak{v}^\sim$ is as trivial as possible outside $\mathfrak{v}$.

\end{enumerate}

In more precise terms, the following propositions hold.

Let $\mathfrak{v}$ be a Lie $2$--algebra such that 
$\dim\mathfrak{v}_0<\dim\mathfrak{v}_1$. Then, there is a balanced 
Lie $2$--algebra with the following properties.

\begin{enumerate}

\item $\mathfrak{v}^\sim{}_0=\mathfrak{v}_0\oplus\mathfrak{w}$, wehere
$\mathfrak{w}$ is a vector space such that $\dim\mathfrak{w}=
\dim\mathfrak{v}_1-\dim\mathfrak{v}_0$, and $\mathfrak{v}^\sim{}_1=\mathfrak{v}_1$.

\item For $x,y,z\in\mathfrak{v}_0$, $a,b,c\in\mathfrak{w}$, $X\in\mathfrak{v}_1$, \pagebreak 
\begin{subequations}
\label{linftybal1,2,3,4}
\begin{align}
&\partial^\sim X=\partial X\oplus 0,
\vphantom{\Big]}
\label{linftybal1}
\\
&[x\oplus a,y\oplus b]^\sim=[x,y]\oplus 0,
\vphantom{\Big]}
\label{linftybal2}
\\
&[x\oplus a,X]^\sim=[x,X],
\vphantom{\Big]}
\label{linftybal3}
\\
&[x\oplus a,y\oplus b,z\oplus c]^\sim=[x,y,z].
\vphantom{\Big]}
\label{linftybal4}
\end{align}
\end{subequations}
\end{enumerate}
Further, $\mathfrak{v}^\sim$ is unique up to (non canonical) isomorphism.

Let $\mathfrak{v}$ be a Lie $2$--algebra such that 
$\dim\mathfrak{v}_0>\dim\mathfrak{v}_1$. Then, there is a balanced 
Lie $2$--algebra with the following properties.

\begin{enumerate}

\item $\mathfrak{v}^\sim{}_0=\mathfrak{v}_0$ and $\mathfrak{v}^\sim{}_1=\mathfrak{v}_1\oplus\mathfrak{f}$,
wehere $\mathfrak{f}$ is a vector space such that $\dim\mathfrak{f}=
\dim\mathfrak{v}_0-\dim\mathfrak{v}_1$, 

\item For $x,y,z\in\mathfrak{v}_0$, $X\in\mathfrak{v}_1$, $A\in\mathfrak{f}$, 
\begin{subequations}
\label{linftybal5.6.7.8}
\begin{align}
&\partial^\sim(X\oplus A)=\partial X,
\vphantom{\Big]}
\label{linftybal5}
\\
&[x,y]^\sim=[x,y],
\vphantom{\Big]}
\label{linftybal6}
\\
&[x,X\oplus A]^\sim=[x,X]\oplus 0,
\vphantom{\Big]}
\label{linftybal7}
\\
&[x,y,z]^\sim=[x,y,z]\oplus 0.
\vphantom{\Big]}
\label{linftybal8}
\end{align}
\end{subequations}
\end{enumerate}
Further, $\mathfrak{v}^\sim$ is unique up to (non canonical) isomorphism.

Using the above results, we can always assume that the Lie $2$--algebra 
$\mathfrak{v}$ we are dealing with is balanced.

%\vfil\eject

\subsection{\normalsize \textcolor{blue}{Balanced Lie $2$--algebras
with invariant form}}\label{sec:linftyform} %%%ooooooo

\hspace{.5cm}  Balanced Lie $2$--algebras  are the basic data in 
higher Chern--Simons theory.

{\it Invariant forms on Lie algebras}

Let $\mathfrak{g}$ be a Lie algebra. An invariant form 
on $\mathfrak{g}$ is a non singular symmetric bilinear mapping 
$(\cdot,\cdot):\mathfrak{g}\times \mathfrak{g}\rightarrow \mathbb{R}$
such that 
\begin{equation}
(x,[\pi,y])+(y,[\pi,x])=0
\label{linftyform20}
\end{equation}
for any $x,y\in\mathfrak{g}$. 

We assume below that $\mathfrak{g}$ is a Lie algebra with invariant form
$(\cdot,\cdot)$. 

{\it The orthogonal automorphisms of a Lie algebra with invariant form} 

A automorphism $\phi\in\Aut(\mathfrak{g})$ is said orthogonal if 
\begin{equation}
(\phi(x),\phi(y))=(x,y),
\label{linftyform40}
\end{equation}
for any $x,y\in\mathfrak{g}$. 
We shall denote by $\OAut(\mathfrak{g})$ the subset of all orthogonal elements
$\phi\in\Aut(\mathfrak{g})$. $\OAut(\mathfrak{g})$ is a Lie subgroup of the Lie group 
$\Aut(\mathfrak{g})$.

{\it The orthogonal derivations of a Lie algebra with invariant form} 

A derivation $\alpha\in\mathfrak{aut}(\mathfrak{g})$ is said orthogonal if 
\begin{equation}
(\alpha(x),y)+(x,\alpha(y))=0,
\vphantom{\Big]}
\label{linftyform80}
\end{equation}
for any $x,y\in\mathfrak{g}$. 
We shall denote by $\mathfrak{oaut}(\mathfrak{g})$ the subset of all orthogonal elements
$\alpha\in\mathfrak{aut}(\mathfrak{g})$. $\mathfrak{oaut}(\mathfrak{g})$ is a Lie subalgebra of the 
Lie algebra $\mathfrak{aut}(\mathfrak{g})$. $\mathfrak{oaut}(\mathfrak{g})$ is the Lie algebra
of the Lie group $\OAut(\mathfrak{g})$. 

{\it Adjoint action and orthogonality in Lie algebras with invariant form}

For any $x\in\mathfrak{g}$, the derivation $\ad x\in\mathfrak{aut}(\mathfrak{g})$
is orthogonal, $\ad x\in\mathfrak{oaut}(\mathfrak{g})$ (cf. eq. \eqref{ad}). 
This is an immediate consequence of \eqref{linftyform20}. 

{\it Exponential map and orthogonality in Lie algebras with invariant form}

The exponential map $\exp_\circ:\mathfrak{oaut}(\mathfrak{g})\rightarrow\OAut(\mathfrak{g})$ 
of $\mathfrak{oaut}(\mathfrak{g})$ is simply the restriction of the exponential map $\exp_\circ 
:\mathfrak{aut}(\mathfrak{g})\rightarrow \Aut(\mathfrak{g})$ of $\mathfrak{aut}(\mathfrak{g})$ 
to $\mathfrak{oaut}(\mathfrak{g})$. In particular, the orthogonal exponential 
is still computed by the expression \eqref{exp00}. 

{\it Invariant forms on balanced Lie $2$--algebras}

Let $\mathfrak{v}$ be a balanced Lie $2$--algebra. An invariant form 
on $\mathfrak{v}$ is a non singular bilinear mapping 
$(\cdot,\cdot):\mathfrak{v}_0\times \mathfrak{v}_1\rightarrow \mathbb{R}$
enjoying the following properties.
\begin{subequations}
\label{linftyform1,2,3}
\begin{align}
&(\partial X,Y)-(\partial Y,X)=0, 
\vphantom{\Big]}
\label{linftyform1}
\\
&([\pi,x],X)+(x,[\pi,X])=0,
\vphantom{\Big]}
\label{linftyform2}
\\
&(x,[\pi,\pi,y])+(y,[\pi,\pi,x])=0,
\vphantom{\Big]}
\label{linftyform3}
\end{align}
\end{subequations}
for any $x,y\in\mathfrak{v}_0$, $X,Y\in \mathfrak{v}_1$. 

We assume below that $\mathfrak{v}$ is a balanced Lie $2$--algebra 
equipped with an invariant form $(\cdot,\cdot)$. 

{\it The orthogonal automorphisms of a balanced algebra with invariant form} %strict $2$--group 

A $1$--automorphism $\phi\in\Aut_1(\mathfrak{v})$ is said orthogonal if 
\begin{subequations}
\label{linftyform4,5}
\begin{align}
&(\phi_0(x),\phi_1(X))=(x,X),
\vphantom{\Big]}
\label{linftyform4}
\\
&(\phi_0(x),\phi_2(y,z))+(\phi_0(z),\phi_2(y,x))=0,
\vphantom{\Big]}
\label{linftyform5}
\end{align}
\end{subequations}
for any $x,y,z\in\mathfrak{v}_0$, $X\in \mathfrak{v}_1$. 
We shall denote by $\OAut_1(\mathfrak{v})$ the set of all orthogonal elements
$\phi\in\Aut_1(\mathfrak{v})$. 

A $2$--automorphism $\varPhi\in\Aut_2(\mathfrak{v})(\phi,\psi)$, $\phi,\psi\in\Aut_1(\mathfrak{v})$
being two $1$--auto\-morphism,
is said orthogonal if both $\phi,\,\psi$ are. For any $\phi,\psi\in\OAut_1(\mathfrak{v})$,
we shall set $\OAut_2(\mathfrak{v})(\phi,\psi)=\Aut_2(\mathfrak{v})(\phi,\psi)$. We further set
$\OAut_2(\mathfrak{v})=\bigcup_{\phi,\psi\in\OAut_1(\mathfrak{v})}$ $\Aut_2(\mathfrak{v})(\phi,\psi)$. 

The following theorem holds true. 
$\OAut(\mathfrak{v})=(\OAut_1(\mathfrak{v}),\OAut_2(\mathfrak{v}))$ 
is a Lie $2$--subgroup of the strict Lie $2$--group
$\Aut(\mathfrak{v})=(\Aut_1(\mathfrak{v}),\Aut_2(\mathfrak{v}))$, 
by which we mean that $\OAut(\mathfrak{v})$ is closed under all operations of the strict $2$--group  
$\Aut(\mathfrak{v})$ (cf. app. \ref{sec:linftyauto}).

$\OAut(\mathfrak{v})$ can be described as a crossed module. The two groups underlying it are 
$\OAut_1(\mathfrak{v})$ and $\OAut_2{}^*(\mathfrak{v})=\bigcup_{\phi\in\OAut_1(\mathfrak{v})}\Aut_2(\mathfrak{v})(\id,\phi)$.
$\OAut_2{}^*(\mathfrak{v})$ can be characterized as the set of 
the elements $\varPhi\in\Aut_2{}^*(\mathfrak{v})$ with the property that
\begin{subequations}
\label{linftyform6,7}
\begin{align}
&(\partial\varPhi(x),X)+(x,\varPhi(\partial X))-(\partial\varPhi(x),\varPhi(\partial X))=0,
\vphantom{\Big]}
\label{linftyform6}
\\
&(y,[x,\varPhi(z)]+[z,\varPhi(x)])+(x-\partial\varPhi(x),\varPhi([y,z]))
\vphantom{\Big]}
\label{linftyform7}
\\
&\hspace{5cm}
+(z-\partial\varPhi(z),\varPhi([y,x]))=0,
\vphantom{\Big]}
\nonumber
\end{align}
\end{subequations}
for $x,y,z\in\mathfrak{v}_0$, $X\in\mathfrak{v}_1$. (cf. app. \ref{sec:linftyauto}).
In this description, as expected, $\OAut(\mathfrak{v})$ is a Lie crossed submodule of the 
Lie crossed module $\Aut(\mathfrak{v})$ (cf. app. \ref{sec:linftyauto}).

{\it The orthogonal derivations of a balanced algebra with invariant form} %strict Lie $2$--algebra}

A $1$--derivation $\alpha\in\mathfrak{aut}_0(\mathfrak{v})$ 
is said orthogonal if 
\begin{subequations}
\label{linftyform8,9}
\begin{align}
&(\alpha_0(x),X)+(x,\alpha_1(X))=0,
\vphantom{\Big]}
\label{linftyform8}
\\
&(x,\alpha_2(y,z))+(z,\alpha_2(y,x))=0,
\vphantom{\Big]}
\label{linftyform9}
\end{align}
\end{subequations}
for any $x,y,z\in\mathfrak{v}_0$, $X\in \mathfrak{v}_1$. 
We shall denote by $\mathfrak{oaut}_0(\mathfrak{v})$ the subset of all orthogonal elements
$\alpha\in\mathfrak{aut}_0(\mathfrak{v})$.

A $2$--derivation $\varGamma\in\mathfrak{aut}_1(\mathfrak{v})$ 
is said orthogonal if, for $x,y,z\in\mathfrak{v}_0$, $X\in\mathfrak{v}_1$,  
\begin{subequations}
\label{linftyform10,11}
\begin{align}
&(\partial\varGamma(x),X)+(x,\varGamma(\partial X))=0,
\vphantom{\Big]}
\label{linftyform10}
\\
&(y,[x,\varGamma(z)]+[z,\varGamma(x)])+(x,\varGamma([y,z]))+(z,\varGamma([y,x]))=0. 
\vphantom{\Big]}
\label{linftyform11}
\end{align}
\end{subequations}
We shall denote by $\mathfrak{oaut}_1(\mathfrak{v})$ the subset of all orthogonal elements
$\varGamma\in\mathfrak{aut}_1(\mathfrak{v})$.

The following theorem holds true. 
$\mathfrak{oaut}(\mathfrak{v})=(\mathfrak{oaut}_0(\mathfrak{v}),\mathfrak{oaut}_1(\mathfrak{v}))$ 
is a strict Lie $2$--subalgebra of 
$\mathfrak{aut}(\mathfrak{v})=(\mathfrak{aut}_0(\mathfrak{v}),\mathfrak{aut}_1(\mathfrak{v}))$, 
by which we mean that $\mathfrak{oaut}(\mathfrak{v})$
is closed under all operations of the strict Lie $2$--algebra  
$\mathfrak{aut}(\mathfrak{v})$.

For any Lie $2$--algebra $\mathfrak{v}$ with invariant form, $\OAut(\mathfrak{v})$ 
is a strict Lie $2$--group having precisely $\mathfrak{oaut}(\mathfrak{v})$ as its associated strict 
Lie $2$--algebra (cf. subsect. \ref{subsec:strict}).

{\it Adjoint action and orthogonality in balanced algebras with invariant form}

For any $x\in\mathfrak{v}_0$, the $1$-derivation $\ad x\in\mathfrak{aut}_0(\mathfrak{v})$
is orthogonal, $\ad x\in\mathfrak{oaut}_0(\mathfrak{v})$ (cf. eqs. \eqref{ada}--\eqref{adc}). 
Likewise, for and $x,y\in\mathfrak{v}_0$
and any $X\in\mathfrak{v}_1$, the $2$--derivations $\ad x\wedge y,\ad X\in\mathfrak{aut}_1(\mathfrak{v})$
are orthogonal, $\ad x\wedge y,\ad X\in\mathfrak{oaut}_1(\mathfrak{v})$
(cf. eqs. \eqref{ade}, \eqref{add}). This is an immediate consequence of \eqref{linftyform1,2,3}. 

{\it Exponential map and orthogonality in balanced algebras with invariant form}

The exponential map $\exp_\circ:\mathfrak{oaut}(\mathfrak{v})\rightarrow\OAut(\mathfrak{v})$ 
of $\mathfrak{oaut}(\mathfrak{v})$ is simply the restriction of the exponential map $\exp_\circ 
:\mathfrak{aut}(\mathfrak{v})\rightarrow \Aut(\mathfrak{v})$ of $\mathfrak{aut}(\mathfrak{v})$ 
to $\mathfrak{oaut}(\mathfrak{v})$. In particular, the orthogonal exponential 
is still computed by the expressions \eqref{exp0}. 

\vspace{1cm}

\textcolor{blue}{Acknowledgements.} We thank J. A. de Azcarraga, S. Baseilhac, D. Calaque, N. Cantarini, 
A. Cattaneo, E. Getzler, V. Ginzburg, C. Imbimbo,
J.M. Izquierdo, E. Latini, D. Lejay, M. Penkava, V. Schlegel  and M. Zambon 
for helpful discussions and suggestions.

\vfil\eject

\end{document}